\documentclass[pra,twocolumn,showpacs,showkeys,preprintnumbers,amsmath,amssymb,floatfix,aps,longbibliography,superscriptaddress]{revtex4-1}

\usepackage{graphicx}
\usepackage{bm}

\usepackage{physics}
\usepackage{placeins}
\usepackage{graphics}
\usepackage{color}
\usepackage{hyperref}
\usepackage{multirow}
\usepackage{blindtext}

\usepackage{pdfpages}

\makeatletter
\AtBeginDocument{\let\LS@rot\@undefined}
\makeatother

\usepackage{xcolor}
\hypersetup{
    colorlinks,
    linkcolor={red!50!black},
    citecolor={blue!50!black},
    urlcolor={blue!80!black}
}

\begin{document}


\title{Twist- and gate-tunable proximity spin-orbit coupling, spin relaxation anisotropy, and charge-to-spin conversion  in heterostructures of graphene and transition-metal dichalcogenides}

\author{Klaus Zollner}
\email{klaus.zollner@physik.uni-regensburg.de}
\affiliation{Institute for Theoretical Physics, University of Regensburg, 93040 Regensburg, Germany}

\author{Simão M. João}
\affiliation{Department of Materials, Imperial College London, South Kensington Campus, London SW7 2AZ, United Kingdom}

\author{Branislav K. Nikoli\'{c}}
\affiliation{Department of Physics and Astronomy, University of Delaware, Newark, DE 19716, USA}

\author{Jaroslav Fabian}
\affiliation{Institute for Theoretical Physics, University of Regensburg, 93040 Regensburg, Germany}

\begin{abstract}
Proximity-induced phenomena in van der Waals heterostructures have emerged as a platform to tailor the electronic, spin, optical, and topological properties in two dimensional materials. A crucial degree of freedom, which has only recently been recognized, is the relative twist angle between the monolayers. 
While partial results exist in the literature, we present here a comprehensive first-principles based investigation of the twist-angle dependent proximity spin-orbit coupling (SOC) in graphene in contact with, or encapsulated by, monolayer transition metal dichalcogenides (TMDCs)
MoS$_2$, MoSe$_2$, WS$_2$, and WSe$_2$. Crucially, our commensurate supercells comprise monolayers
with strains of less than 2.5\%, minimizing band-offset artifacts. We confirm earlier DFT results that for Mo-based TMDCs the proximity valley-Zeeman SOC exhibits a maximum 
at around 15--20$^{\circ}$, and vanishes at 30$^{\circ}$ for symmetry reasons. Although such a maximum was also predicted by tight-binding simulations 
for W-based TMDCs, we find an almost linear decrease of proximity valley-Zeeman SOC in graphene/WSe$_2$ and graphene/WS$_2$ when twisting from 0$^{\circ}$ to 30$^{\circ}$.
We also refine previous DFT simulations and show that the induced Rashba SOC 
is rather insensitive to twisting, while acquiring a nonzero Rashba phase
angle $\varphi$ which measures the deviation of the electron spin from in-plane transverse direction to the momentum, for twist angles different from 0$^{\circ}$ and 30$^{\circ}$. The Rashba phase
angle $var\phi$ varies from $-20^{\circ}$ to 40$^{\circ}$, with the largest variation (40$^{\circ}$) found for MoS$_2$ at a twist angle of 20$^{\circ}$. 
This finding contradicts earlier tight-binding predictions that the Rashba angle can be 90$^{\circ}$ in the studied systems.   
In addition, we study the influence of a transverse electric field, vertical and lateral shifts, and TMDC encapsulation on the proximity SOC for selected twist angles. Within our investigated electric field limits of $\pm 2$~V/nm, mainly the Rashba SOC can be tuned by about 50\%. The interlayer distance provides a giant tunability, since the proximity-induced SOC can be increased by a factor of 2--3, when reducing the distance by only about 10\%. When encapsulating graphene between two TMDCs, both twist angles are important to control the interference of the individual proximity-induced SOCs, allowing to precisely tailor the proximity-induced valley-Zeeman SOC in graphene, while the Rashba SOC becomes suppressed. Finally, based on our effective Hamiltonians with fitted parameters to low-energy \emph{ab initio} band structures, we calculate experimentally measurable quantities such as spin lifetime anisotropy and charge-to-spin conversion efficiencies. 
The spin lifetime anisotropy---being the ratio between out-of-plane and in-plane spin lifetimes---can become giant (up to 100), depending on the TMDC, twist angle, transverse electric field, and the interlayer distance. 
The charge-to-spin conversion can be divided into three components which are due to spin-Hall and Rashba-Edelstein effects with non-equilibrium spin-density polarizations that are perpendicular and parallel to the applied charge current. All conversion efficiencies are highly tunable by the twist angle and the Fermi level. 
\end{abstract}

\pacs{}
\keywords{spintronics, transition-metal dichalcogenides, heterostructures, proximity spin-orbit coupling}
\maketitle

\section{Introduction}

Van der Waals (vdW) heterostructures based on two-dimensional (2D) materials are emerging as an important platform for investigating novel solid state phenomena \cite{Sierra2021:NN,Zutic2019:MT,Gibertini2019:NN,Briggs2019:2DM,Novoselov2016:Sci,Burch2018:Nat,Duong2017:ACS,Bora2021:JPM}. While 2D materials exhibit extraordinary physical properties on the atomic scale, we can combine different monolayers to form artificial vdW crystals with customized electronic, optical, magnetic, or topological properties \cite{Geim2013:Nat,Novoselov2016:Sci,Zutic2019:MT,Sierra2021:NN}. The prime example are heterostructures based on 
monolayer graphene, where proximity interactions, such as spin-orbit coupling (SOC)~\cite{Gmitra2015:PRB,Gmitra2016:PRB,Szalowski2023:2DM,Naimer2021:arxiv,Zollner2019:PRR,Zollner2021:PSSB,Zollner2019:PRB,Herlin2020:APL,Safeer2019:NL,Fulop2021:arxiv, Khokhriakov2020:NC,Zihlmann2018:PRB,Song2018:NL,Garcia2018:CSR,Khoo2017:NL,Omar2017:PRB, Omar2018:PRB,Tiwari2022:2DM}, exchange coupling~\cite{Zollner2018:NJP,Zollner2019:PRR,Zollner2021:arXiv2,Dyrdal2017:2DM,Hallal2017:2DM,Cardoso2018:PRL,Karpiak2019:arxiv,Zollner2016:PRB,Zhang2015:PRB,Zhang2018:PRB,Yang2013:PRL,Song2017:JPD, Haugen2008:PRB, Zhang2015:SR, Su2017:PRB,Cardoso2018:PRL,Singh2017:PRL, Swartz2012:ACS}, and superconductivity~\cite{Moriya2020:PRB} can be induced via neighboring layers. Important, the proximity-induced interactions can be controlled by gating, doping, straining, lateral stacking, and twisting.

Particularly interesting for spintronics \cite{Zutic2004:RMP} are graphene/transition-metal dichalcogenide (TMDC) bilayers \cite{Gmitra2015:PRB,Gmitra2016:PRB, Wang2015:NC, Island2019:Nat}. First-principles calculations \cite{Gmitra2015:PRB} and experiments \cite{Ghiasi2019:NL,Benitez2020:NM,Khokhriakov2020:NC,Hoque2021:CP,Amann2021:arxiv} on graphene/TMDC structures have already demonstrated that proximity SOC can be tuned by the application of a transverse electric field. Recent DFT simulations show
a potential tunability via controlled alloying of the TMDC \cite{Khatibi2022:arxiv}; this should be 
experimentally realizable given the impressive progress in TMDC growth techniques \cite{Nugera2022:S}.
Since proximity effects are short-ranged and originate from the wavefunction overlap of different layers, also the vdW distance plays an important role. Recent experiments have shown that external pressure, which reduces the interlayer distance, can significantly boost proximity interactions~\cite{Fulop2021:arxiv,Fulop2021:arxiv2}. 
The proximity coupling of graphene with TMDCs has already lead to fascinating experimental findings, 
such as optical spin injection \cite{Avsar2017:ACS,Gmitra2015:PRB,Luo2017:NL} , gate tunable charge-to-spin conversion~\cite{Khokhriakov2020:NC,Ingla2022:2DM,Benitez2020:NM,Ghiasi2019:NL}, giant spin relaxation anisotropy \cite{Cummings2017:PRL, Zihlmann2018:PRB, Ghiasi2017:NL, Offidani2018:PRB, Leutenantsmeyer2018:PRL, Omar2019:arxiv}, and field-effect spin transistor operation~\cite{Ingla2021:PRL}.

Recently, the relative twist angle between the monolayers has emerged as another important control knob. In general, vdW heterostructures composed of twisted monolayers 
\cite{Carr2017:PRB,Hennighausen2021:ES,Ribeiro2018:SC,Carr2020:NRM}
promise great tunability of electronic, optical, and magnetic properties.
For example, magic-angle twisted bilayer graphene exhibits magnetism and superconductivity due to strong correlations \cite{Cao2018:Nat,Cao2018a:Nat,Arora2020:arxiv,Stepanov2020:Nat,Lu2019:Nat,Sharpe2019:SC, Saito2021:Nat,Serlin2020:S,Nimbalkar2020:NML,Bultinck2020:PRL,Repellin2020:PRL,Choi2019:NP,Lisi2021:NP,Balents2020:NP,Wolf2019:PRL}. In twisted TMDCs, a strong trapping potential for excitons can arise due to the emerging moir\'e pattern~\cite{Gobato2022:NL,Lin2023:NL}.
In graphene/Cr$_2$Ge$_2$Te$_6$ bilayers, twisting allows to reverse the proximity-induced exchange splitting of the Dirac bands \cite{Zollner2021:arXiv2}.
Finally, gating and twisting are two efficient control knobs to tune the valley splitting in TMDC/CrI$_3$ heterostructures~\cite{Zollner2023:PRB}.
All the above demonstrates that the twist angle has a highly non-trivial influence on physical observables.

There have already been theoretical \cite{Pezo2021:2DM,David2019:arxiv,Li2019:PRB,Naimer2021:arxiv,Lee2022:arxiv, Wang2022:JAP} and experimental \cite{ Rockinger:PC} studies investigating the impact of twisting on the electronic properties and proximity-induced SOC in graphene/TMDC heterostructures \cite{Rockinger:PC}. 
Tight-binding studies have predicted that the relative rotation of the monolayers can greatly enhance the proximity SOC, with an expected maximum at around 15--20$^{\circ}$, for graphene in contact with MoS$_2$, MoSe$_2$, WS$_2$, and WSe$_2$ \cite{David2019:arxiv,Li2019:PRB}. 
However, tight-binding calculations have to rely on some input parameters. For example, the position of the Dirac point within the TMDC band gap seems is rather crucial for predicting  twist-angle dependent proximity SOC \cite{David2019:arxiv}. 

In a systematic DFT investigation, Naimer \textit{et al.} \cite{Naimer2021:arxiv} showed
that strain (the study used up to 10\% of strain in graphene) in twisted graphene/TMDC supercells affects the proximity effects due to strain-induced band offsets, prompting the application of a transverse displacement field to remove these artifacts. This \emph{ad hoc} procedure has produced qualitatively similar results as the aforementioned tight-binding studies for Mo-based TMDCs, but has found that the valley-Zeeman proximity coupling for W-based TMDCs decreases with increasing the twist angle from 0$^{\circ}$ to 30$^{\circ}$, not exhibiting a global maximum. This DFT study \cite{Naimer2021:arxiv} also found specific values for the Rashba phase angles, predicted on symmetry grounds to be different from zero (the reference angle at which the in-plane spin is perpendicular to the momentum) away from 0$^{\circ}$ to 30$^{\circ}$ \cite{David2019:arxiv,Li2019:PRB}. 
Also Pezo \textit{et al.} \cite{Pezo2021:2DM} considered large-scale supercells of graphene on strained (up to 3.5\%) MoTe$_2$ and WSe$_2$, employing twist angles around 0$^{\circ}$, 15$^{\circ}$, and 30$^{\circ}$, predicting strong variations of the proximity SOC, although 
the limited set of twist angles was insufficient to uncover systematic trends.  
Finally, Lee \textit{et al.} \cite{Lee2022:arxiv} performed DFT investigations of twisted 
graphene/WSe$_2$ heterostructures with small strain (less than 2\%) finding a nearly constant valley-Zeeman SOC up to about 18$^{\circ}$, followed by a linear decrease to 30$^{\circ}$; the Rashba SOC was found to be  nearly constant for all the investigated twist angles. There is already evidence from weak antilocalization experiments \cite{Rockinger:PC} on twisted graphene/WSe$_2$ structures showing small ($\sim0.05$~meV) valley-Zeeman and finite ($\sim 0.5$~meV) Rashba SOC at 30$^{\circ}$, in agreement with theory. In contrast, samples with 15$^{\circ}$ twist angle show larger SOC values, with Rashba $\sim1.5$~meV and valley-Zeeman $\sim0.4$~meV.

In this paper, we aim to provide a comprehensive DFT-based picture of proximity SOC in twisted graphene/TMDC heterostructures by considering only small-strain supercells (less than 2.5\% of strain in graphene and zero strain in TMDCs) for all four semiconducting TMDC monolayers  MoS$_2$, MoSe$_2$, WS$_2$, and WSe$_2$. In addition to providing 
systematic dependencies of the valley-Zeeman and Rashba SOC on the twist angles, we also 
address the effects of a transverse electric field, encapsulation, and lateral and vertical shifts.
We confirm earlier DFT studies that upon twisting from 0$^{\circ}$ to 30$^{\circ}$, the induced valley-Zeeman SOC decreases almost linearly to zero for W-based TMDCs, while for Mo-based TMDCs it exhibits a maximum at around 15--20$^{\circ}$. The induced Rashba SOC stays rather constant upon twisting, and acquires a phase angle $\varphi \neq 0$, due to symmetry breaking, for twist angles different from 0$^{\circ}$ and 30$^{\circ}$.
For WSe$_2$ our results also correspond to the findings of Ref.~\cite{Lee2022:arxiv}, but we additionally cover the twist angle behavior for graphene on MoS$_2$, MoSe$_2$, and WS$_2$.
Within our investigated electric field limits of $\pm 2$~V/nm, mainly the Rashba SOC can be tuned by about 50\%. The interlayer distance, correlating to external pressure in experiments \cite{Fulop2021:arxiv,Fulop2021:arxiv2}, provides a giant tunability, since the proximity-induced SOC can be increased by a factor of 2--3, when reducing the distance by only about 10\%. When encapsulating graphene between two TMDCs, both twist angles are important to control the interference of the individual proximity-induced SOCs, allowing to precisely tailor the valley-Zeeman SOC, while the Rashba SOC becomes suppressed. More precisely, when the twist angles of the encapsulating TMDC layers are equal, say both are $0^{\circ}$, the induced valley-Zeeman SOC is roughly doubled, since the layer-resolved proximity effect is additive on the graphene sublattices. In contrast, when the twist angles differ by $60^{\circ}$, the sublattices are effectively exchanged and the effective valley-Zeeman SOC becomes suppressed. The Rashba SOC is always suppressed due to the nearly restored $z$-mirror symmtery in encapsulated structures.

Finally, combining the first-principles calculations, low energy model Hamiltonian, fitted parameters, and real-space transport calculations, we make specific predictions for experimentally measurable quantities such as spin lifetime anisotropy and charge-to-spin conversion efficiency. We find that the spin lifetime anisotropy---the ratio between out-of-plane and in-plane spin lifetimes---can become giant, up to 100, especially in graphene on MoS$_2$ and WS$_2$ as the valley-Zeeman dominates over the Rashba SOC, pinning the spin to the out-of-plane direction. Our calculated anisotropies are in agreement with experiments \cite{Ghiasi2017:NL, Benitez2018:NP,Zihlmann2018:PRB} and further tunability is provided by twisting, an external electric field, and the interlayer distance. The real-space transport calculations reveal that twisted heterostructures provide a tunable charge-to-spin conversion via spin-Hall and Rashba-Edelstein effects. With gating and twisting, it is possible to tailor not only the magnitude but also the direction of the non-equilibrium spin-density, making graphene/TMDC heterostructures a versatile platform for creating and detecting spin polarized currents without the need of conventional ferromagnets.

The manuscript is organized as follows. In Sec.~\ref{Sec:Geometry}, we first address the structural setup and summarize the calculation details for obtaining the electronic structures of the twisted graphene/TMDC bilayers. In Sec.~\ref{Sec:Model}, we introduce the model Hamiltonian that captures the proximitized Dirac bands, which is used to fit the first-principles results.
In Sec.~\ref{Sec:FPR}, we show and discuss exemplary calculated electronic structures, along with the model Hamiltonian fits. We also address the influence of the twist-angle, transverse electric field, and the interlayer distance on the proximity SOC. In Sec.~\ref{Sec:Encap}, we briefly discuss TMDC-encapsulated graphene structures, where proximity SOC can be enhanced or suppressed due to interference of the encapsulating layers. In Sec.~\ref{Sec:Discuss}, we address some open questions and discuss the origin of our findings in more detail. 
In Sec.~\ref{Sec:Aniso} and Sec.~\ref{Sec:SCC} we analyze experimentally relevant quantities, which are the twist-angle and gate tunability of the spin-lifetime anisotropy and charge-to-spin conversion efficiencies.
Finally, in Sec.~\ref{Sec:Summary} we conclude the manuscript.

\section{Geometry Setup and Computational Details}
\label{Sec:Geometry}

    \begin{figure}[htb]
     \includegraphics[width=.95\columnwidth]{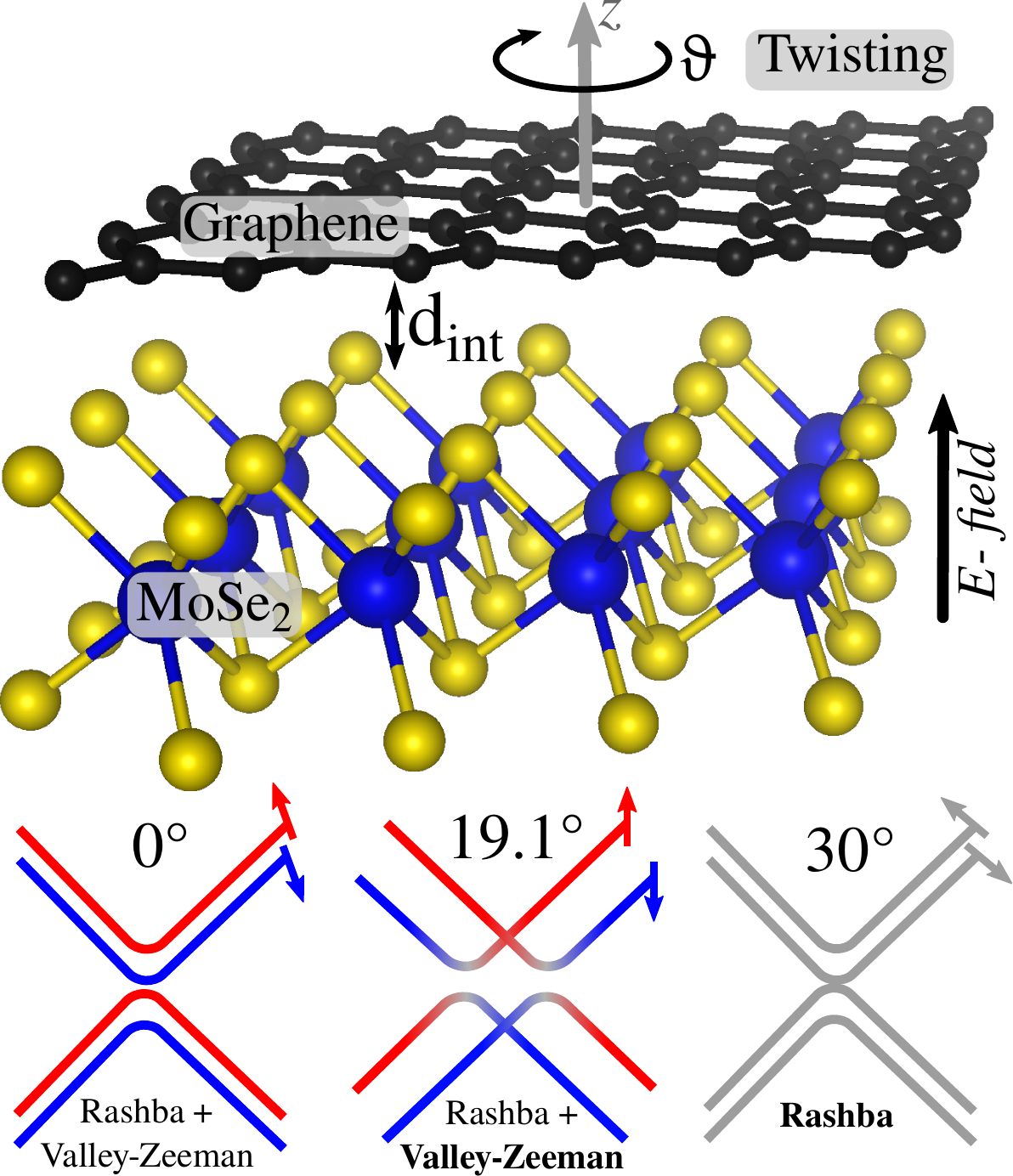}
     \caption{3D view of graphene above a TMDC (MoSe$_2$), where we define the interlayer distance, $\mathrm{d}_{\mathrm{int}}$. We twist graphene by an angle $\vartheta$ around the $z$ axis with respect to the TMDC.
     The twist-angle evolution of the proximitized Dirac states is sketched. Red (blue) bands are polarized spin up (spin down), while grey bands are in-plane polarized.
     At 0$^{\circ}$, the proximity-induced SOC in the Dirac states, is of Valley-Zeeman and Rashba type. At around 19.1$^{\circ}$, the Valley-Zeeman SOC is maximized leading to a band inversion. At 30$^{\circ}$, Valley-Zeeman SOC vanishes and only Rashba SOC remains.
    }\label{Fig:Structure}
    \end{figure}

    The graphene/TMDC heterostructures, for which we consider several twist angles between the two monolayers, are set-up with the {\tt atomic simulation environment (ASE)} \cite{ASE} and the {\tt CellMatch} code \cite{Lazic2015:CPC}, implementing the coincidence lattice method \cite{Koda2016:JPCC,Carr2020:NRM}.
    Within this method, a graphene/TMDC heterostructure contains a ($n$,$m$) graphene supercell and a ($n'$,$m'$) TMDC supercell, where integers $n,m,n'$ and $m'$ define the corresponding supercell lattice vectors. 
    Monolayers of graphene and TMDCs are based on hexagonal unit cells, with experimental lattice constants \cite{Baskin1955:PR, Wakabayashi1975:PRB, Schutte1987:JSSC, James1963:AC} of 
    $a = 2.46$~\AA~(graphene), $a = 3.288$~\AA~(MoSe$_2$), $a = 3.282$~\AA~(WSe$_2$), $a = 3.15$~\AA~(MoS$_2$), and $a = 3.153$~\AA~(WS$_2$), which additionally need to be strained in the twisted heterostructures, in order to form commensurate supercells for periodic density functional theory (DFT) calculations. Since MoSe$_2$ and WSe$_2$ have nearly the same lattice constant, we set them to 3.28~\AA~in the following. The same we do for MoS$_2$ and WS$_2$, where we use 3.15~\AA.
    In Table~S1 and Table~S2 we summarize the main structural information for the twist angles we consider. 
    In total, we investigate 12 different angles between 0$^{\circ}$ and 30$^{\circ}$, for each graphene/TMDC heterostructure. Especially these angles are suitable for DFT calculations, since strain applied to the monolayers is below 2.5\%. 
    We already know that biaxial strain strongly influences the band gap of monolayer TMDCs \cite{Zollner2019:strain} and therefore we leave them nearly unstrained in the heterostructures. 
    The residual strain is applied to the graphene lattice, which mainly influences the Fermi velocity of Dirac states \cite{Naimer2021:arxiv}. 
    In addition, the number of atoms is kept below 250. Otherwise, also
    other angles could be investigated, but beyond reasonable strain limits and above a computationally feasible number of atoms in the structure.

	The electronic structure calculations and structural relaxations of the graphene/TMDC heterostructures
	are performed by DFT~\cite{Hohenberg1964:PRB} 
	with {\tt Quantum ESPRESSO}~\cite{Giannozzi2009:JPCM}.
	Self-consistent calculations are carried out with a $k$-point sampling of 
	$n_k\times n_k\times 1$. The number $n_k$ is listed in Table~S1 and Table~S2 for all twist angles and depends on the number of atoms in the heterostructure.
	In addition, $n_k$ is limited by our computational power. Nevertheless, for large supercells the heterostructure Brillouin Zone is small and only few $k$-points are necessary to get converged results. 
	
	We use an energy cutoff for charge density of $560$~Ry and
	the kinetic energy cutoff for wavefunctions is $70$~Ry for the fully relativistic pseudopotentials
	with the projector augmented wave method~\cite{Kresse1999:PRB} with the 
	Perdew-Burke-Ernzerhof exchange correlation functional~\cite{Perdew1996:PRL}.
	Spin-orbit coupling (SOC) is included in the calculations.
	For the relaxation of the heterostructures, we add DFT-D2 vdW corrections~\cite{Grimme2006:JCC,Grimme2010:JCP,Barone2009:JCC} and use 
	quasi-Newton algorithm based on trust radius procedure. 
	Dipole corrections \cite{Bengtsson1999:PRB} are also included to get correct band offsets and internal electric fields.
	In order to simulate quasi-2D systems, we add a vacuum of about $20$~\AA~to avoid interactions between periodic images in our slab geometry. To get proper interlayer distances and to capture possible moir\'{e} reconstructions, we allow all atoms to move freely within the heterostructure geometry during relaxation. Relaxation is performed until every component of each force is reduced below $5\times10^{-4}$~[Ry/$a_0$], where $a_0$ is the Bohr radius.
	
    After relaxation of the graphene/TMDC heterostructures, we calculate the mean interlayer distances, $d_{\textrm{int}}$, and the standard deviations, $\Delta z_{\textrm{grp}}$, from the $z$ coordinates of the C atoms of graphene. The standard deviations represent the amount of rippling of graphene. The results are summarized in Table~S1 and Table~S2. 
	The interlayer distances are nearly independent of the twist angle and range from about 3.3 to 3.4~\AA.
	The graphene itself stays nearly flat, as the rippling stays below about 3~pm. 
	In Fig.~\ref{Fig:Structure}, we show the general structural setup of our graphene/TMDC heterostructures, where the graphene resides above the TMDC.
    When we apply the transverse electric field (modeled by a zigzag potential), a positive field points along $z$ direction from the TMDC towards graphene.

\section{Model Hamiltonian}
\label{Sec:Model}
From our first-principles calculations we obtain the low energy Dirac band structure of the spin-orbit proximitized graphene. We then extract realistic parameters for an effective Hamiltonian describing graphene's low energy Dirac bands. The Hamiltonian together with the fitted parameters provide an effective description for the low-energy physics, which is relevant for studying transport \cite{Fulop2021:arxiv, Veneri2022:arxiv, Lee2022:arxiv,Offidani2017:PRL,Karpiak2019:arxiv, Zollner2019:PRR}, topology \cite{Hogl2020:PRL,Frank2018:PRL}, or spin relaxation \cite{Cummings2017:PRL, Zollner2021:PRB, Zollner2019:PRB, Offidani2018:PRB}. Due to the short-range nature of the proximity effects in van der Waals heterostructures, the effective model parameters are transferable and can be employed for bilayer and trilayer graphene heterostructures \cite{Zollner2020:PRL,Zollner2021:arxiv,Zollner2022:PRB}.

The band structure of spin-orbit proximitized graphene can be modeled by symmetry-derived Hamiltonians \cite{Kochan2017:PRB}. For graphene in
heterostructues with $C_{3}$ symmetry, the effective low energy Hamiltonian is
\begin{flalign}
\label{Eq:Hamiltonian}
&\mathcal{H} = \mathcal{H}_{0}+\mathcal{H}_{\Delta}+\mathcal{H}_{\textrm{I}}+\mathcal{H}_{\textrm{R}}+E_D,\\
&\mathcal{H}_{0} = \hbar v_{\textrm{F}}(\tau k_x \sigma_x - k_y \sigma_y)\otimes s_0, \\
&\mathcal{H}_{\Delta} =\Delta \sigma_z \otimes s_0,\\
&\mathcal{H}_{\textrm{I}} = \tau (\lambda_{\textrm{I}}^\textrm{A} \sigma_{+}+\lambda_{\textrm{I}}^\textrm{B} \sigma_{-})\otimes s_z,\\
&\mathcal{H}_{\textrm{R}} = -\lambda_{\textrm{R}}\textrm{e}^{-\textrm{i}\varphi\frac{s_z}{2}}(\tau \sigma_x \otimes s_y + \sigma_y \otimes s_x)\textrm{e}^{\textrm{i}\varphi\frac{s_z}{2}}
\end{flalign}
Here $v_{\textrm{F}}$ is the Fermi velocity and the in-plane wave vector 
components $k_x$ and $k_y$ are measured from $\pm$K, corresponding to the valley index $\tau = \pm 1$.
The Pauli spin matrices are $s_i$, acting on spin space ($\uparrow, \downarrow$), and $\sigma_i$ are pseudospin matrices, acting on sublattice space (C$_\textrm{A}$, C$_\textrm{B}$), with $i = \{ 0,x,y,z \}$ and $\sigma_{\pm} = \frac{1}{2}(\sigma_z \pm \sigma_0)$.
The staggered potential gap is $\Delta$, arising from sublattice asymmetry.
The parameters $\lambda_{\textrm{I}}^\textrm{A}$ and $\lambda_{\textrm{I}}^\textrm{B}$ describe the sublattice-resolved intrinsic SOC and $\lambda_{\textrm{R}}$ stands for the Rashba SOC.
In addition, a phase angle $\varphi$ can be present in the usual Rashba term, which leads to a rotation of the spin-orbit field around the $z$-axis \cite{Li2019:PRB,David2019:arxiv}. 
When the intrinsic SOC parameters satisfy $\lambda_{\textrm{I}}^\textrm{A} = -\lambda_{\textrm{I}}^\textrm{B}$, it is also called valley-Zeeman or Ising type SOC, while in the case of $\lambda_{\textrm{I}}^\textrm{A} = \lambda_{\textrm{I}}^\textrm{B}$, it is called Kane-Mele type SOC~\cite{Kane2005:PRL}.
Charge transfer between the monolayers in the DFT calculation is captured by the Dirac point energy, $E_D$, which adjusts the Dirac point with respect to the Fermi level.
The basis states are $\ket{\Psi_{\textrm{A}}, \uparrow}$, $\ket{\Psi_{\textrm{A}}, \downarrow}$, $\ket{\Psi_{\textrm{B}}, \uparrow}$, and $\ket{\Psi_{\textrm{B}}, \downarrow}$, resulting in four eigenvalues $\varepsilon_{1/2}^{\textrm{CB/VB}}$.
For each considered heterostructure, we calculate the proximitized low energy Dirac bands in the vicinity of the K point.
To extract the fit parameters from the first-principles data, we employ a least-squares routine \cite{lmfit}, taking into account band energies, splittings, and spin expectation values.

\section{First-Principles Results and Discussion}	
\label{Sec:FPR}
\subsection{Twist angle dependence of proximity SOC}

    \begin{figure*}[htb]
     \includegraphics[width=.99\textwidth]{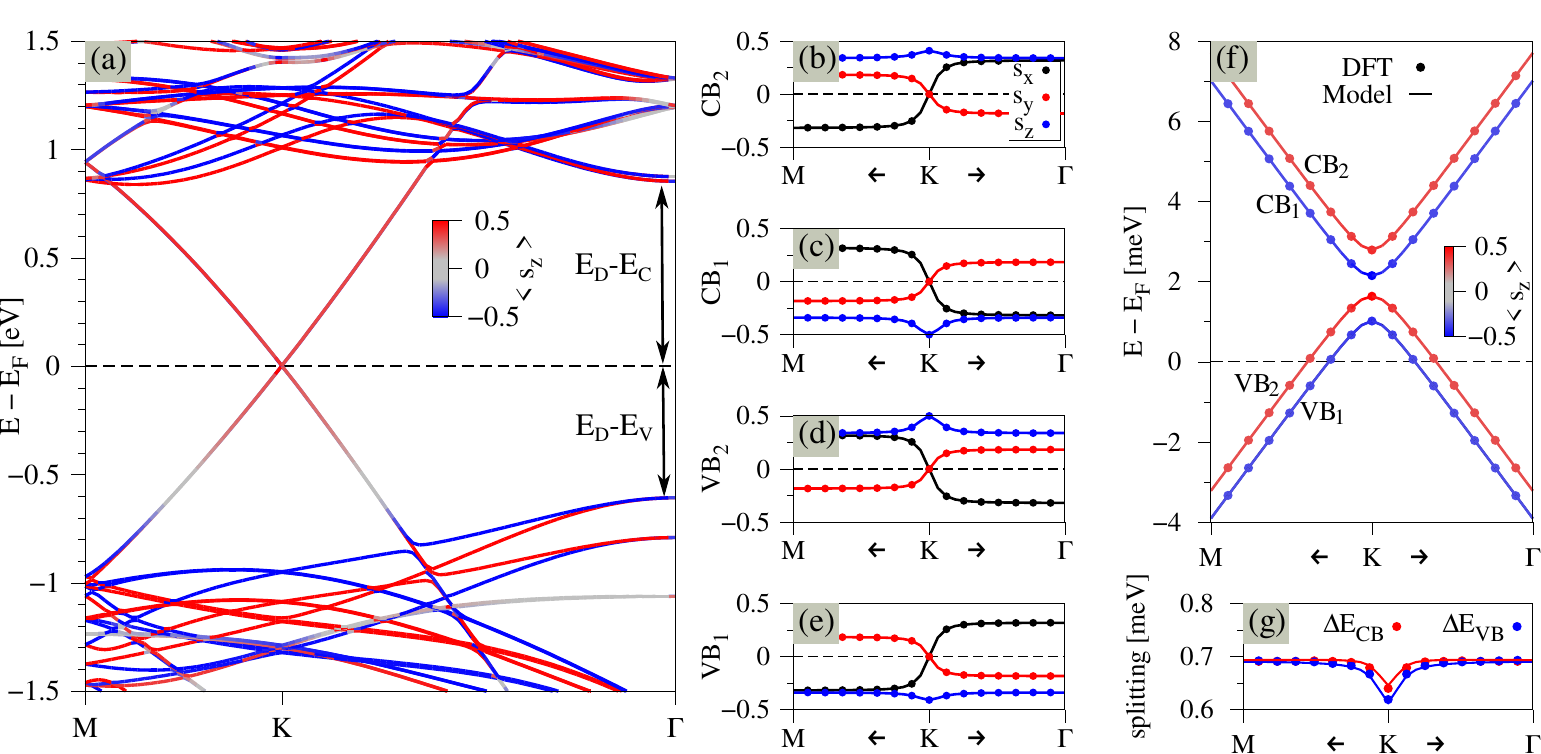}
     \caption{(a) DFT-calculated band structure of the graphene/MoSe$_2$ heterostructure along the high-symmetry path M-K-$\Gamma$ for a twist angle of 0$^{\circ}$. The color of the lines corresponds to the $s_z$ spin expectation value. We also indicate the position of the Dirac point with respect to the TMDC valence (conduction) band edge, $E_D-E_V$ ($E_D-E_C$).
     (b)-(e) The spin expectation values of the 4 low-energy bands as labeled in (f). (f) Zoom to the calculated low-energy bands (symbols) near the Fermi level around the $K$ point, corresponding to the band structure in (a), with a fit to the model Hamiltonian (solid lines). (g) The energy splitting of the low energy Dirac bands.
     }\label{bands_MoSe2_0deg}
    \end{figure*}

In Fig.~\ref{bands_MoSe2_0deg}(a), we show the calculated global band structure of the graphene/MoSe$_2$ heterostructure for a twist angle of 0$^{\circ}$, as an exemplary case. The Dirac states of graphene are nicely preserved within the band gap of the TMDC, and are located about $0.61$~eV ($-0.85$~eV) above (below) the relevant $K$ point valence (conduction) band egde of the TMDC, see Table~S3. 
Actually in Fig.~\ref{bands_MoSe2_0deg}(a), the conduction band edge of the TMDC is located close to the $M$ point. However, we note that we use a lattice constant of 3.28~\AA~for MoSe$_2$, and not the exact experimental one of 3.288~\AA. Already at such small tensile strain, MoSe$_2$ becomes an indirect band gap semiconductor, with the conduction band edge at the $Q$ side valley \cite{Zollner2019:strain}. In addition, the relevant $K$ points of TMDC band edges are backfolded to the $\Gamma$ point due to the $3\times 3$ MoSe$_2$ supercell we use for the 0$^{\circ}$ case. 

In Figs.~\ref{bands_MoSe2_0deg}(b)-(g), we summarize the low-energy band properties of the graphene Dirac states near the Fermi level.
Due to proximity-induced SOC, the Dirac bands split into four states, $\varepsilon_{1/2}^{\textrm{CB/VB}}$. The magnitude of the splitting is on the order of 0.7~meV. By fitting the low-energy Dirac dispersion to our model Hamiltonian, we find that proximity-induced intrinsic SOCs are of valley-Zeeman type, $\lambda_{\textrm{I}}^\textrm{A} \approx -\lambda_{\textrm{I}}^\textrm{B} \approx 0.23$~meV. In addition, a Rashba SOC is present, $\lambda_{\textrm{R}} \approx 0.25$~meV, being of the same magnitude. The obtained SOC parameters are giant compared to the intrinsic SOC of pristine graphene, being about $20$--$40~\mu$eV \cite{Gmitra2009:PRB,Sichau2019:PRL}. In addition, Dirac states display an orbital gap, which results from the potential asymmetry of the sublattices (connected to the rippling of graphene), characterized by parameter $\Delta$. 
The Dirac states, band splittings, and spin expectation values are perfectly reproduced by our model Hamiltonian employing the parameters in Table~\ref{tab:fit}. 
The results for 0$^{\circ}$ are in good agreement to earlier calculations of proximity SOC in graphene/TMDC heterostructures \cite{Gmitra2016:PRB}.

\begin{table*}[!htb]
\begin{ruledtabular}
\caption{\label{tab:fit} Fit parameters of the model Hamiltonian, Eq.~\eqref{Eq:Hamiltonian},
for the graphene/TMDC heterostructures for different twist angles $\vartheta$.
We summarize the Fermi velocity $v_{\textrm{F}}$, the staggered potential gap $\Delta$, the sublattice-resolved intrinsic SOC parameters $\lambda_{\textrm{I}}^\textrm{A}$ and $\lambda_{\textrm{I}}^\textrm{B}$, the
Rashba SOC parameter $\lambda_{\textrm{R}}$, the phase angle $\varphi$, and the position of the Dirac point, $E_{\textrm{D}}$, with respect to the Fermi level. }
\begin{tabular}{lcccccccc}
TMDC & $\vartheta$ [°] & $\Delta$~[meV] & $v_{\textrm{F}}/10^5 [\frac{\textrm{m}}{\textrm{s}}]$ & $\lambda_{\textrm{I}}^\textrm{A}$~[meV] & $\lambda_{\textrm{I}}^\textrm{B}$~[meV] &  $\lambda_{\textrm{R}}$~[meV] & $\varphi$ [°] & $E_{\textrm{D}}$ [meV] \\
\hline
MoSe$_2$ & 0.0000 & 0.4917 & 8.2538 & 0.2422 & -0.2258 & 0.2550 & 0 & 1.8970 \\
& 2.6802 & 0.4346 & 8.2382 & 0.2213 & -0.2120 & 0.2664 & -2.2919 & 0.0024 \\
& 3.8858 & -0.3121 & 8.1250 & -0.1860 & 0.1954 & 0.2859 & -4.1254 & -0.0311 \\
& 5.2087 & -1.1162 & 8.5072 & -0.2920 & 0.2166 & 0.2448 & -1.3751 & 1.9400 \\
& 8.2132 & -0.6569 & 8.3124 & -0.3046 & 0.2434 & 0.2613 & -2.8076 & 0.0046 \\
& 12.2163 & -0.7117 & 8.4028 & -0.5062 & 0.3877 & 0.2136 & 2.8190 & 0.1276 \\
& 14.3916 & 0.4097 & 8.0799 & 0.3838 & -0.4240 & 0.3247 & -7.9644 & 0.0592 \\
& 19.1066 & 0.1163 & 8.0073 & 0.5627 & -0.5827 & 0.3326 & 4.7156 & 1.0680 \\
& 22.4987 & -0.0826 & 8.2585 & -0.5181 & 0.5041 & 0.2912 & 31.8860 & -0.1366 \\
& 25.2850 & -0.0173 & 7.9727 & -0.3393 & 0.3320 & 0.3110 & 29.5139 & 0.0445 \\
& 30.0000 & 0.0040 & 8.3109 & 0.0013 & -0.0055 & 0.2398 & 0 & 0.2514 \\ \hline
WSe$_2$ & 0.0000 & 0.5878 & 8.2500 & 1.1722 & -1.1572 & 0.5303 & 0 & 1.2931\\
& 2.6802 & 0.5438 & 8.2687 & 1.0775 & -1.0650 & 0.5475 & -1.3522 & -0.0502\\
& 3.8858 & -0.4079 & 8.2968 & -0.9045 & 0.9120 & 0.5592 & -3.1055 & -0.0509\\
& 5.2087 & -1.3110 & 8.3911 & -1.1868 & 1.0555 & 0.5979 & -1.3293 & 1.6139\\
& 8.2132 & -0.8307 & 8.3230 & -1.0482 & 0.9122 & 0.6210 & -3.4092 & 1.0818\\
& 12.2163 & -0.8494 & 8.4755 & -1.2914 & 0.9973 & 0.6129 & -1.8794 & -0.0278\\
& 14.3916 & 0.4444 & 8.0440 & 0.6371 & -0.7484 & 0.8339 & -17.3382 & 0.0158\\
& 19.1066 & 0.0876 & 7.8914 & 0.5899 & -0.6420 & 0.8215 & -19.6129 & 2.2178\\
& 22.4987 & -0.0813 & 8.2654 & -0.7106 & 0.6654 & 0.6441 & 3.8985 & -0.0464\\
& 25.2850 & -0.0037 & 7.9577 & -0.2522 & 0.2382 & 0.5237 & 18.6102 & 0.0107\\
& 30.0000 & -0.0093 & 8.3185 & -0.0165 & 0.0128 & 0.6197 & 0 & 1.1670\\ \hline
MoS$_2$ & 1.0445 & -0.7794 & 8.3275 & -0.2990 & 0.2672 & 0.0737 & 6.1881 & -0.1036 \\
& 6.5868 & 0.4420 & 8.0126 & 0.2445 & -0.2647 & 0.0854 & 21.1428 & 0.2847 \\
& 8.9483 & 0.3782 & 7.9692 & 0.2244 & -0.2460 & 0.0953 & 17.5330 & -0.0681 \\
& 12.8385 & -0.2796 & 7.9358 & -0.2393 & 0.2140 & 0.1106 & 8.2508 & -0.0696 \\
& 14.4649 & 0.3765 & 8.1134 & 0.3053 & -0.3565 & 0.1245 & 15.0692 & 1.1699 \\
& 16.1021 & -0.3058 & 8.2297 & -0.4126 & 0.3517 & 0.1287 & 14.3244 & 0.0450 \\
& 22.4109 & -0.0546 & 8.0486 & -0.1347 & 0.1216 & 0.0718 & 37.4152 & 0.0025 \\
& 27.6385 & -0.0002 & 8.1439 & -0.0410 & 0.0373 & 0.0843 & 32.8887 & 0.1104 \\
& 29.2649 & 0.0011 & 8.0021 & 0.0027 & -0.0049 & 0.0395 & 18.4498 & 0.0020 \\ \hline
WS$_2$  & 1.0445 & -0.9678 & 8.1209 & -1.1390 & 1.0407 & 0.2131 & 5.3688 & -0.0787\\
& 6.5868 & 0.6485 & 8.0248 & 0.7849 & -0.8638 & 0.2337 & 16.8970 & 1.6459\\
& 8.9483 & 0.5615 & 7.9988 & 0.6581 & -0.7354 & 0.2705 & 9.8609 & 0.5747\\
& 12.8385 & -0.3525 & 7.9563 & -0.5200 & 0.4531 & 0.3206 & -4.9620 & 0.0493\\
& 14.4649 & 0.4676 & 8.1248 & 0.5635 & -0.6826 & 0.3678 & -1.3236 & 0.3962\\
& 16.1021 & -0.3602 & 8.1780 & -0.6841 & 0.5536 & 0.3956 & -4.8474 & 0.0075\\
& 22.4109 & -0.0472 & 8.0434 & -0.0158 & -0.0082 & 0.1777 & 2.4793 & 0.3277\\
& 27.6385 & 0.0025 & 8.2009 & 0.0059 & -0.0113 & 0.2410 & 18.7310 & 1.8203\\
& 29.2649 & -0.0007 & 8.0090 & -0.0212 & 0.0194 & 0.1462 & 9.0129 & 0.3090
\end{tabular}
\end{ruledtabular}
\end{table*}

    \begin{figure}[htb]
     \includegraphics[width=.99\columnwidth]{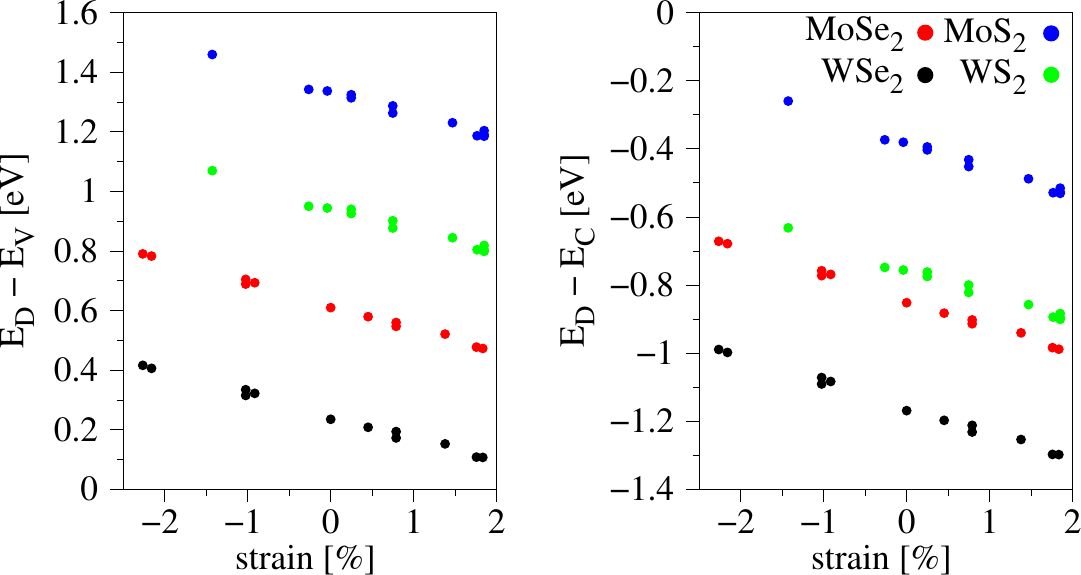}
     \caption{The calculated position of the Dirac point with respect to the TMDC valence (conduction) band edge, $E_D-E_V$ ($E_D-E_C$), as function of the biaxial strain in graphene for the different TMDCs. 
     The data are summarized in Table~S3.
    }\label{Fig:bandoffs}
    \end{figure}

    \begin{figure*}[htb]
     \includegraphics[width=.9\textwidth]{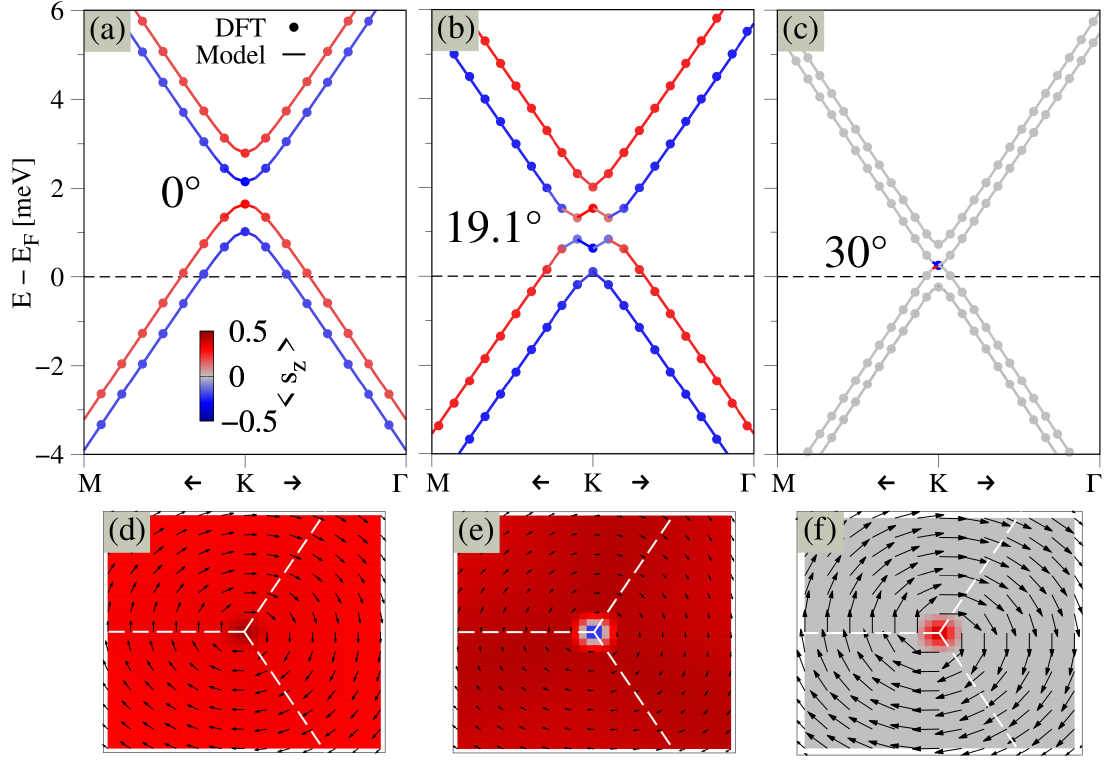}
     \caption{(a) Zoom to the calculated low-energy bands (symbols) of the graphene/MoSe$_2$ heterostructure near the Fermi level around the $K$ point, for a twist angle of 0$^{\circ}$ and with a fit to the model Hamiltonian (solid lines). The color of the lines/points corresponds to the $s_z$ spin expectation value.
     (b) and (c) The same as (a), but for twist angles of 19.1$^{\circ}$ and 30$^{\circ}$. 
     (d) The calculated spin-orbit field, in the vicinity of the $K$ point, of the spin-up valence band from the low-energy dispersion shown in (a). The color represents the $s_z$ spin expectation value, while the arrows represent $s_x$ and $s_y$ spin expectation values. The dashed white lines represent the edges of the hexagonal Brillouin zone, with the $K$ point at the center. (e) and (f) The same as (d), but for twist angles of 19.1$^{\circ}$ and 30$^{\circ}$.
     }\label{low_energy_bands_MoSe2}
    \end{figure*}
    
Before we show and discuss the twist-angle dependence of proximity SOC, we first want to address how strain affects the dispersion. Since the lattice constant of the TMDC is fixed for all twist angles, the main changes are in the graphene Dirac states and band offsets. From literature, we know that the Dirac states of graphene are quite robust against biaxial strain~\cite{Si2016:NS,Choi2010:PRB}, apart from a renormalization of the Fermi velocity. From recent studies \cite{Zollner2021:arXiv2,Naimer2021:arxiv}, we already know that band offsets are tunable by strain. 
In Fig.~\ref{Fig:bandoffs}, we plot the position of the Dirac point with respect to the TMDC valence (conduction) band edge, $E_D-E_V$ ($E_D-E_C$), as defined in Fig.~\ref{bands_MoSe2_0deg}(a), as function of the strain applied to graphene. The different twist angles provide different strain, and the plotted information are summarized in Tables S1, S2, and S3. 
We find a linear dependence of the band offsets with respect to the graphene strain as in a previous study \cite{Naimer2021:arxiv}. 
In experiment, one can expect that both graphene and the TMDCs are nearly unstrained due to weak vdW bonding and only the zero strain band offsets are relevant.
For our exemplary case of MoSe$_2$, we find the Dirac cone roughly in the middle of the TMDC band gap. 
From Fig.~\ref{Fig:bandoffs} we can extract the zero strain band offsets and the rates $\gamma$ at which the band offsets change via straining, by fitting the data with a linear dependence. The extrapolated values are summarized in Table~\ref{tab:bandoffs_slopes}. We find that for lighter (heavier) elements in the TMDC, the Dirac cone is located closer to the conduction (valence) band edge, as is the case for MoS$_2$ (WSe$_2$).
Especially the zero strain band offsets should be also useful for tight-binding models of graphene/TMDC bilayers ~\cite{David2019:arxiv,Li2019:PRB}, where the position of the Dirac point within the TMDC band gap enters as an unknown parameter. 
In addition, despite the strain in graphene is kept below $\pm$2.5\% in our heterostructure calculations, we observe variations in the band offsets of several hundreds of meV. The reason is that the rates $\gamma \approx -80$~meV/\% are quite large, but similar for all TMDCs, and band offsets can be massively tuned by straining. 
In particular, tensile (compressive) strain will shift the Dirac states closer to the TMDC valence (conduction) band edge. 
Our calculated zero strain band offsets show that the Dirac cone is clearly located within the TMDC band gap, which is in agreement to experiments \cite{Pierucci2016:NL,Kim2015:ACS}.
The tunability of the band offset with straining graphene is expected, since the individual workfunctions of the layers determine the band alignment, and the workfunction of graphene shows a significant strain dependence within our strain limits~\cite{Grassano2020:PRB}. In particular, the workfunction of graphene increases (decreases) with positive (negative) strain~\cite{Grassano2020:PRB}, shifting the Dirac point towards more negative (positive) energy, which is consistent with our observations in Fig.~\ref{Fig:bandoffs}.

In contrast to Ref.~\cite{Naimer2021:arxiv}, our heterostructures have smaller strain so we do not compensate the strain-related band offsets with an electric field. Also, we perform structural relaxation at each 
twist angle which leads to rippling and twist-dependent interlayer distance. As we show, both effects
influence the proximity induced SOC, so that electric-field compensation would not necessarily make the
results more representative. We demonstrate this by comparing 0$^{\circ}$ graphene/MoSe$_2$ and graphene/WSe$_2$ heterostructures with different strains and setup conditions~\footnotemark[1].
We believe that the field correction as in Ref.~\cite{Naimer2021:arxiv} makes sense to be applied only in the scenario of a flat graphene layer and fixed interlayer distance, to extract the bare twist-angle dependence while
disregarding other effects. Otherwise all these effects: band offset, rippling, and interlayer distance, which 
are in some way connected to strain and which affect proximity SOC, would be difficult to disentangle.

\begin{table}[!htb]
\begin{ruledtabular}
\caption{\label{tab:bandoffs_slopes} Zero strain band offsets $E_D-E_{V}$ and $E_D-E_{C}$ and the rates $\gamma$ at which the band offsets change via straining, extrapolated by fitting the data in Fig.~\ref{Fig:bandoffs} with linear functions. }
\begin{tabular}{lcccc}
TMDC & $E_D-E_V$ [eV] & $E_D-E_C$ [eV] & $\gamma$ [meV/\%] \\ \hline
MoS$_2$ & 1.3360 & -0.3817 & -78.95 \\
WS$_2$ & 0.9473 & -0.7531 & -77.04 \\
MoSe$_2$ & 0.6159 & -0.8458 & -77.35 \\
WSe$_2$ & 0.2446 & -1.1606 & -75.72 \\
\end{tabular}
\end{ruledtabular}
\end{table}

Now we turn to the most important result, which is the twist-angle dependence of proximity-induced SOC. In Fig.~\ref{low_energy_bands_MoSe2}, we show the calculated low energy Dirac states for the graphene/MoSe$_2$ heterostructure for three different twist angles, 0$^{\circ}$, 19.1$^{\circ}$, and 30$^{\circ}$, as exemplary cases. As already mentioned, the Dirac states are split due to proximity SOC. In the case of 0$^{\circ}$, the splitting is moderate, caused by nearly equal valley-Zeeman and Rashba SOC ($\lambda_{\textrm{I}}^\textrm{A} \approx -\lambda_{\textrm{I}}^\textrm{B} \approx 0.23$~meV, $\lambda_{\textrm{R}} \approx 0.25$~meV). 
This can be also seen in the calculated spin-orbit field of one of the Dirac bands.
Overall, spins have an out-of-plane component due to intrinsic SOCs, while Rashba SOC is responsible for the vortex-like in-plane components. Both components are nearly equal away from the $K$ point, see also Fig.~\ref{bands_MoSe2_0deg}. 
For 19.1$^{\circ}$, the splitting is maximized, a band inversion can be obtained, and valley-Zeeman SOC dominates over the Rashba one ($\lambda_{\textrm{I}}^\textrm{A} \approx -\lambda_{\textrm{I}}^\textrm{B} \approx 0.57$~meV, $\lambda_{\textrm{R}} \approx 0.33$~meV). The band inversion is due to the fact that the sublattice potential asymmetry $\Delta$ is small compared to the magnitude of the intrinsic SOCs. The spin-orbit field shows almost only an out-of-plane component, while in-plane components are suppressed.   
For 30$^{\circ}$, the splitting is minimal, valley-Zeeman SOC vanishes and Rashba SOC dominates ($\lambda_{\textrm{I}}^\textrm{A} \approx -\lambda_{\textrm{I}}^\textrm{B} \approx 0$~meV, $\lambda_{\textrm{R}} \approx 0.24$~meV). 
In fact, the valley-Zeeman SOC should completely vanish at 30$^{\circ}$, due to a mirror plane symmetry, restoring the sublattice symmetry~\cite{Lee2022:arxiv}. However, due to the small rippling in graphene from structural relaxations, this symmetry is not fully restored and small, but finite, intrinsic SOCs arise even at 30$^{\circ}$. 
The spin-orbit field almost solely shows vortex-like in-plane components, while an out-of-plane component is only present right at the $K$ point.
Such a twist-angle tunability of SOC and the corresponding spin-orbit fields will have a huge impact on spin transport and relaxation \cite{Cummings2017:PRL}, as we will discuss later.

For all the investigated twist angles and the different TMDCs, our model Hamiltonian can faithfully describe the low-energy Dirac states, with the fit parameters summarized in Table \ref{tab:fit}. For structures from Tables S1 and S2, which satisfy $n-m = 3\cdot l $, $l \in \mathbb{Z}$, the Dirac states of graphene from both $K$ and $K^{\prime}$ fold back to the $\Gamma$ point. Consequently, we cannot apply our fitting routine employing the model Hamiltonian, Eq. \eqref{Eq:Hamiltonian}, for some twist angles, which are then absent in Table \ref{tab:fit}.

Note that, when graphene sublattices (C$_\textrm{A}$ and C$_\textrm{B}$) are interchanged in the geometry, the parameter $\Delta$ changes sign, while parameters $\lambda_{\textrm{I}}^\textrm{A}$ and $\lambda_{\textrm{I}}^\textrm{B}$ are interchanged as well. Such an exchange of sublattices corresponds to an additional 
60$^{\circ}$ twist applied to graphene above the TMDC. 
Therefore twist angles $\vartheta$ and $\vartheta+60^{\circ}$ cannot be distinguished from the geometries. 
In Table \ref{tab:fit}, the fit parameters show such a sign change for the investigated twist angles. This is connected to the setup of the heterostructure supercells for different angles, since 1) the starting point stacking of the non-rotated layers is arbitrary, 2) the origin of the rotation axis can be chosen randomly, 3) the lattice vectors, defining the periodic heterostructure supercell, can be imposed differently on the moir\'{e} structure from the twisted layers. 
Consequently, one would have to consider several structures for each twist angle to obtain well justified results (in terms of value and sign). Considering subsequent lateral shifts (see below) is particularly helpful to see how the proximity SOC changes for different atomic registries.  However, it is enough to consider only angles between 0$^{\circ}$ and 30$^{\circ}$, since the parameters for the other angles can be obtained by symmetry considerations \cite{Naimer2021:arxiv}. 

    \begin{figure}[htb]
     \includegraphics[width=.99\columnwidth]{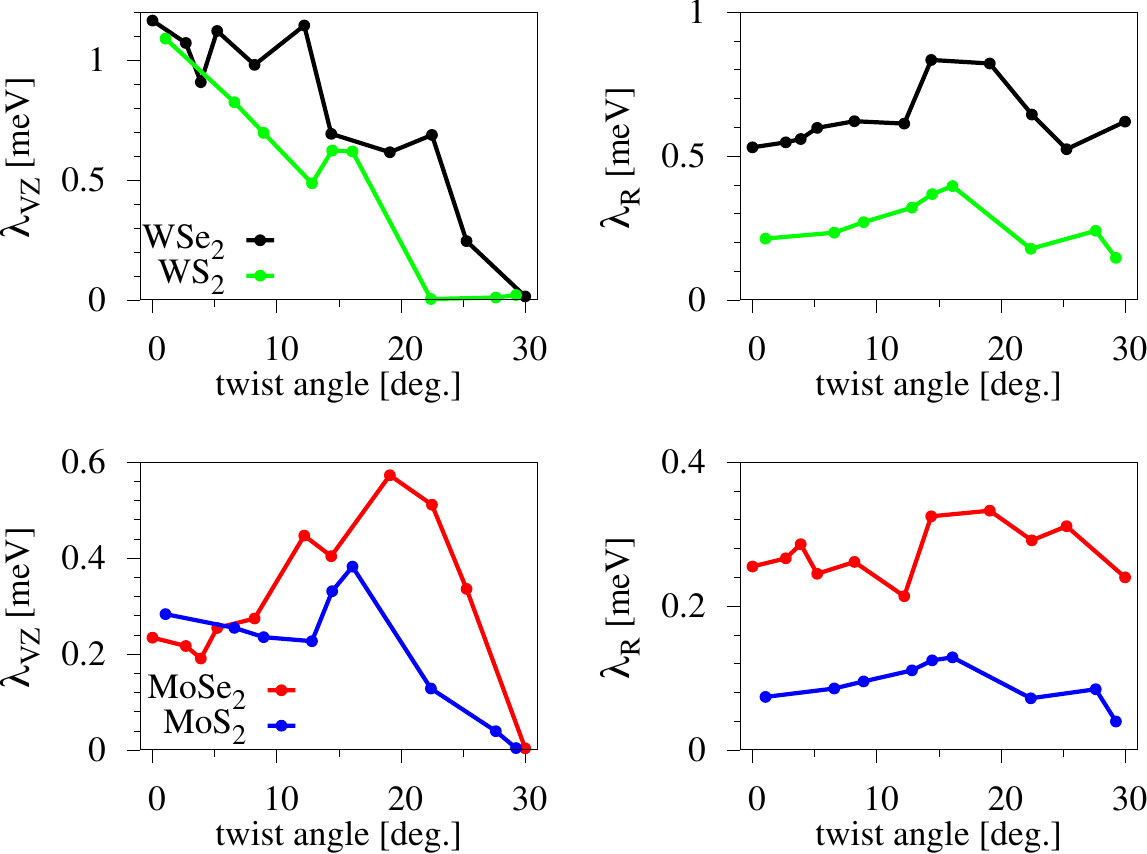}
     \caption{Calculated twist-angle dependence of the valley-Zeeman and Rashba SOC for the different TMDCs. The data are summarized in Table~\ref{tab:fit}.
    }\label{Fig:SOC_params}
    \end{figure}

From the experimental point of view, e. g., in spin transport or spin-charge conversion experiments, that consider twisted graphene/TMDC heterostructures, only the magnitude and type of proximity SOC plays a role, since a well-defined manufacturing process with atomically precise control of stacking and twisting of two different monolayers is not yet possible. 
Due to this and the mentioned sign issue from the DFT results, in Fig.~\ref{Fig:SOC_params} we plot the absolute values of valley-Zeeman and Rashba SOC as function of the twist angle for all TMDs, as summarized in Table~\ref{tab:fit}. 
Note that the valley-Zeeman SOC is defined as $\lambda_{\textrm{VZ}} = (\lambda_{\textrm{I}}^\textrm{A}-\lambda_{\textrm{I}}^\textrm{B})/2$. 
We find a clear and strong twist-angle dependence of the proximity-induced SOC. 
The heavier the elements in the TMDC, the larger is the proximity SOC. For untwisted structures (0$^{\circ}$), both valley-Zeeman and Rashba SOC are finite. At 30$^{\circ}$, the valley-Zeeman SOC vanishes and Rashba SOC dominates, independent of the TMDC. 
While the Rashba SOC stays rather constant upon twisting, the valley-Zeeman SOC shows a marked twist-angle dependence, different for Mo- and W-based TMDCs. 
For WS$_2$ and WSe$_2$, the valley-Zeeman SOC gradually decreases when twisting from 0$^{\circ}$ to 30$^{\circ}$. This finding is consistent with Ref.~\cite{Lee2022:arxiv}. In contrast, for MoS$_2$ and MoSe$_2$, the valley-Zeeman SOC exhibits a maximum at around 15$^{\circ}$ to 20$^{\circ}$.

\subsection{Influence of vertical and lateral shifts}

How sensitive is the proximity-induced SOC with respect to the atomic registry (stacking) and the interlayer distance?
Recent experiments have shown that one can tune proximity SOC by external pressure, thereby reducing the interlayer distance between graphene and the TMDC \cite{Fulop2021:arxiv,Fulop2021:arxiv2}. In particular, by applying external pressure of about 1.8~GPa to a graphene/WSe$_2$ heterostructure and diminishing the interlayer distance by about 9\%, leads to a 2-fold enhancement of the proximity-induced Rashba SOC, as found by magnetotransport experiments \cite{Fulop2021:arxiv}.
In this section, we study how variations of the interlayer distance influence proximity SOC. For selected twist angles we vary $d_{\textrm{int}}$ in steps of 0.1~\AA, starting from the relaxed equilibrium distances listed in Tables~S1 and S2, keeping the rest of the geometry (rippling of graphene and the TMDC) fixed. 
In addition, we study how lateral shifts, which essentially change the exact stacking of graphene above the TMDC, influence proximity SOC. For the lateral shifts, we use crystal coordinate notation, i.~e., we shift graphene above the TMDC by fractions $x$ and $y$ of the supercell lattice vectors. We perform structural relaxations in the case of lateral shifts before we calculate the proximitized low energy Dirac bands, since the stacking may influence the graphene rippling and the interlayer distance.

    \begin{figure}[htb]
     \includegraphics[width=.99\columnwidth]{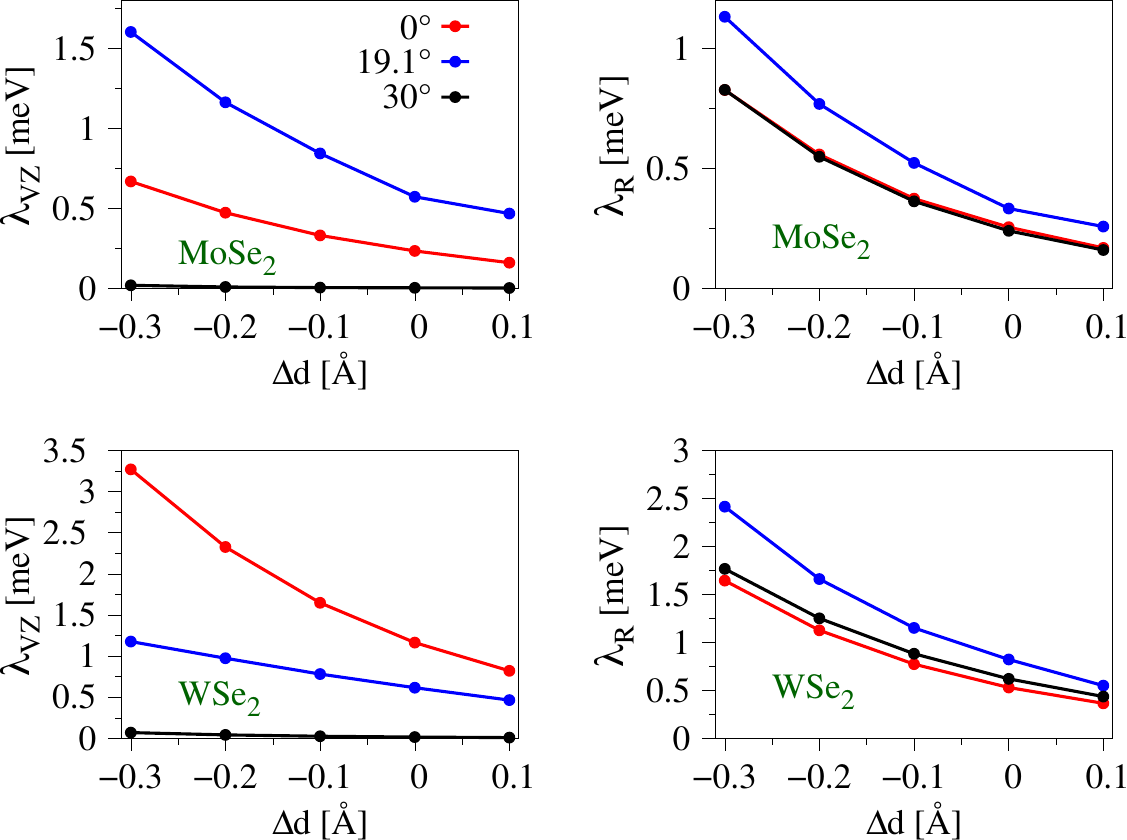}
     \caption{Calculated interlayer distance dependence of the valley-Zeeman and Rashba SOC for MoSe$_2$ and WSe$_2$ structures for selected twist angles. The data are summarized in Table~S4 and Table~S5.
    }\label{Fig:Shifts}
    \end{figure}

Since Mo- and W-based TMDCs produce different trends in the twist-angle dependence of proximity SOC, we focus on MoSe$_2$ and WSe$_2$ only. In addition, we consider only three selected twist angles, namely 0$^{\circ}$, 19.1$^{\circ}$ and 30$^{\circ}$.
In Table~S4 and Table~S5 we summarize the fit results, when tuning the interlayer distance or changing the stacking.
By reducing the interlayer distance, we find that Dirac states are pushed towards the TMDC valence band edge. In addition, the sublattice asymmetry, represented by the staggered potential $\Delta$ increases, when decreasing the distance. 
Most important, the induced valley-Zeeman and Rashba SOC depends strongly on the distance, as summarized in Fig.~\ref{Fig:Shifts}. By reducing the interlayer distance, the SOC can be heavily increased, in agreement with experiments \cite{Fulop2021:arxiv,Fulop2021:arxiv2}. 
In particular, the proximity-induced SOC can be increased by a factor of 2--3, when reducing the distance by only about 10\%. The only exception is the valley-Zeeman SOC for the 30$^{\circ}$ structures, which is absent (or at least very small in our case due to rippling) due to symmetry. 
In contrast, the precise atomic registry (stacking) has negligible influence on the magnitude of proximity SOC in graphene/TMDC heterostructures.
This results probably from the fact that the considered heterostructure supercells are large compared to the monolayer unit cells, such that an averaging effect takes place.

\subsection{Gate tunability of proximity SOC}

In experiment, gating is a tool to further control and tailor the proximity SOC in graphene-based heterostructures \cite{Ghiasi2019:NL,Benitez2020:NM,Khokhriakov2020:NC,Hoque2021:CP}. 
For example, in Ref. \cite{Hoque2021:CP} it has been shown that a gate voltage can be employed to control the spin-charge conversion efficiency in graphene/MoTe$_2$ heterostructures. 
We wish to answer the question: How does a transverse electric field affect proximity SOC for different twist angles? Again, we focus only on MoSe$_2$ and WSe$_2$ and twist angles of 0$^{\circ}$, 19.1$^{\circ}$ and 30$^{\circ}$. The positive field direction is indicated in Fig.~\ref{Fig:Structure}. 

    \begin{figure}[htb]
     \includegraphics[width=.99\columnwidth]{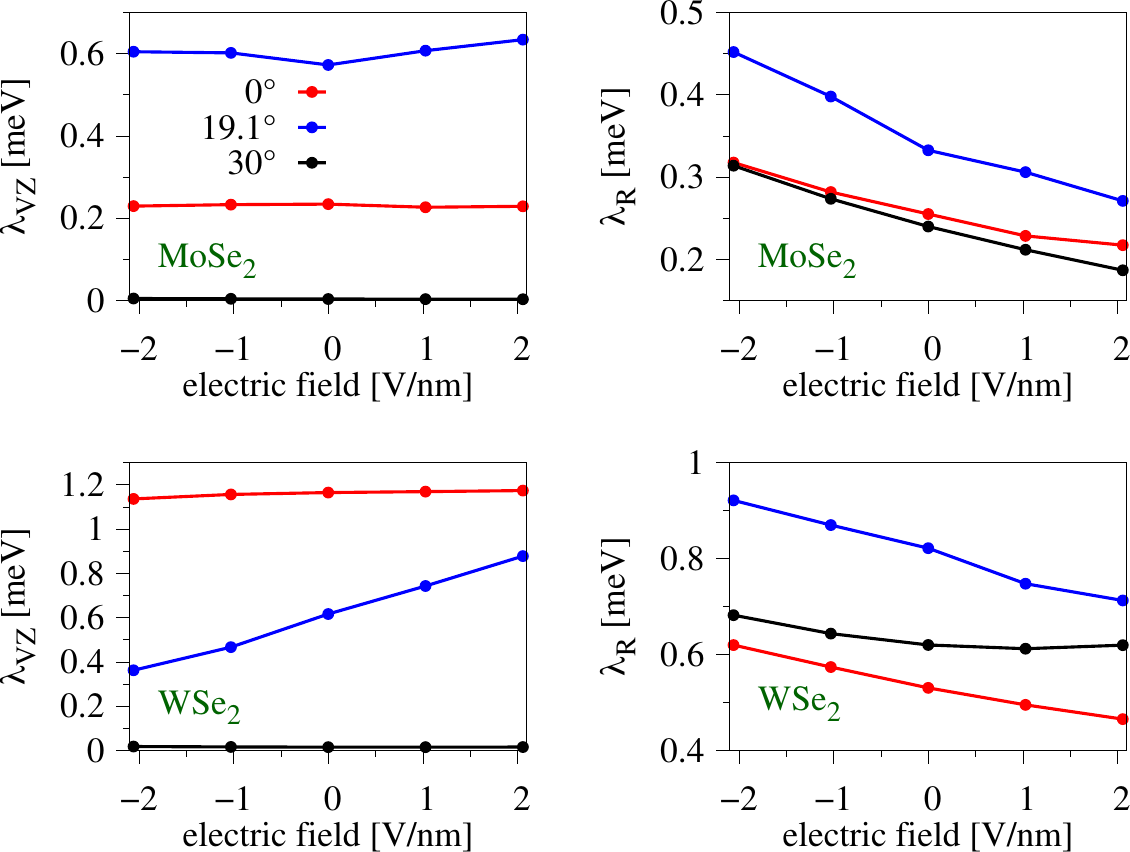}
     \caption{Calculated electric field dependence of the valley-Zeeman and Rashba SOC for MoSe$_2$ and WSe$_2$ structures for selected twist angles. The data are summarized in Table~S6 and Table~S7.
    }\label{Fig:Efield}
    \end{figure}

The fit results are summarized in Tab.~S6 for graphene/MoSe$_2$ and Tab.~S7 for graphene/WSe$_2$ bilayers.
In general, the electric field simply shifts the Dirac cone up or down in energy within the TMDC band gap, as can be seen from the band offsets. The tunability is about 100~meV per V/nm of applied field. Since the band offsets change, also the interlayer coupling along with proximity SOC changes. 
In Fig.~\ref{Fig:Efield} we show how the valley-Zeeman and Rashba SOC are affected by the external transverse electric field. 
We find that for MoSe$_2$, the field barely influences the valley-Zeeman SOC, while the Rashba one can be tuned in a linear fashion, similar for all the different twist angles we consider. More precisely, within our field limits of $\pm 2$~V/nm, the Rashba SOC can be tuned by about 50\%. 
In particular, recalling that the ratio between valley-Zeeman and Rashba SOC determines the spin relaxation anisotropy \cite{Cummings2017:PRL}, the electric field will lead to an enormous tunability of the latter. 

In the case of WSe$_2$, the behaviour is rather similar but the 19.1$^{\circ}$ twist angle is an exception. 
For this angle, also the valley-Zeeman SOC is highly tunable by the field. Moreover, we find that the valley-Zeeman SOC increases, while the Rashba one decreases for positive field amplitudes and vice versa for negative fields.

\section{Encapsulated Geometries}
\label{Sec:Encap}
\begin{figure*}[htb]
     \includegraphics[width=.9\textwidth]{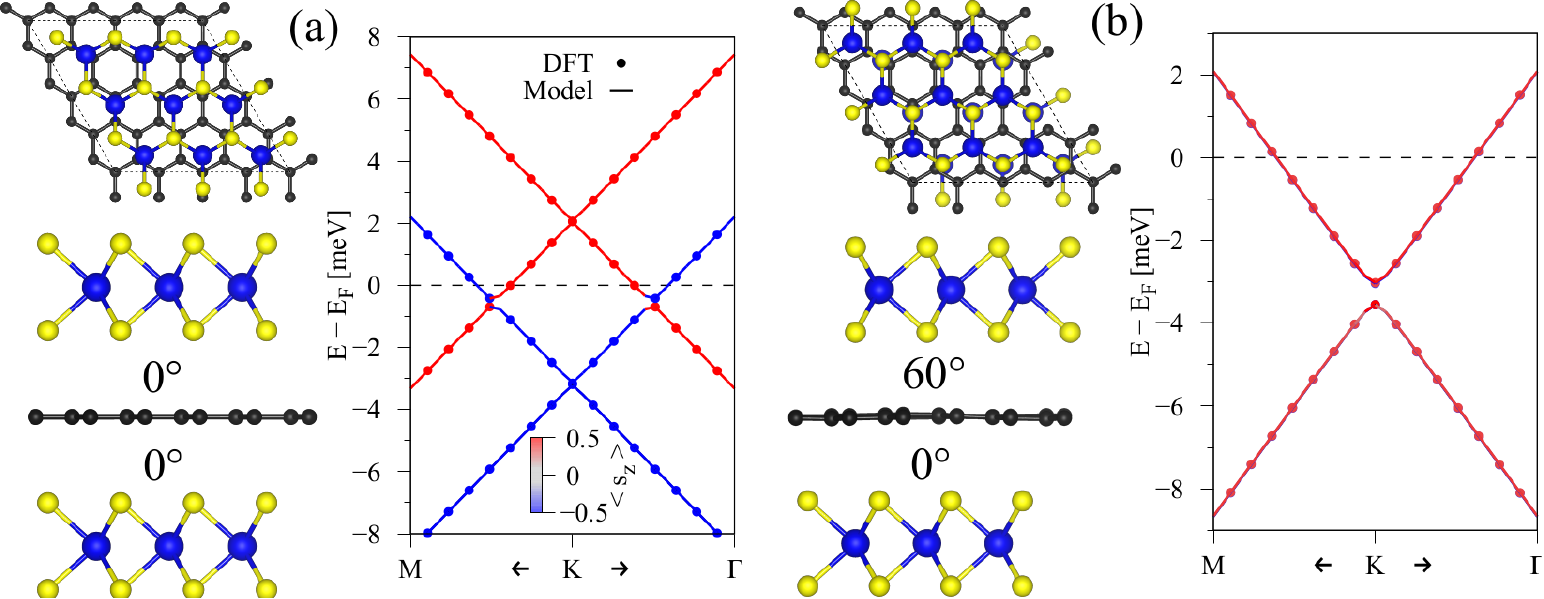}
     \caption{Top and side view of the WSe$_2$-encapsulated graphene and corresponding proximitized low energy Dirac bands for twist angles of (a) $\vartheta_b=0^{\circ}$, $\vartheta_t=0^{\circ}$ and (b) $\vartheta_b=0^{\circ}$, $\vartheta_t=60^{\circ}$
    }\label{Fig:encap_bands}
    \end{figure*}

Maximizing the proximity SOC in graphene is advantageous for example in spin-charge conversion experiments \cite{Benitez2020:NM,Offidani2017:PRL,Ghiasi2019:NL,Khokhriakov2020:NC,Herlin2020:APL,Lee2022:arxiv,Veneri2022:arxiv}.
We have already seen, that proximity-induced SOC is maximized for WSe$_2$ at 0$^{\circ}$ and for MoSe$_2$ at 19.1$^{\circ}$. 
Can we further enhance proximity SOC, by encapsulating graphene between two TMDC monolayers?
We consider the graphene/WSe$_2$ heterostructure with 0$^{\circ}$ twist angle and place another WSe$_2$ monolayer on top. The top WSe$_2$ layer is considered to have a relative twist angle of 0$^{\circ}$ and 0+60$^{\circ}$ with respect to the subjacent graphene/WSe$_2$ bilayer, see Fig.~\ref{Fig:encap_bands}. 
Similarly, we consider the graphene/MoSe$_2$ heterostructure with 19.1$^{\circ}$ twist angle and place another MoSe$_2$ monolayer on top, with a relative twist angle of 19.1$^{\circ}$ and 19.1+60$^{\circ}$.
We also perform a structural relaxation on the encapsulated structures, similar as above, before we proceed to calculate the proximitized Dirac dispersion. 

\begin{table*}[!htb]
\begin{ruledtabular}
\caption{\label{tab:struct_encapsulated} Structural information and calculated band offsets for the TMDC/graphene/TMDC heterostructures. We summarize the relative twist angles $\vartheta_b$ ($\vartheta_t$) of graphene with respect to bottom (top) TMDC layer, the relaxed interlayer distances $d_{\textrm{b}}$ ($d_{\textrm{t}}$), the rippling of the graphene layer $\Delta z_{\textrm{grp}}$, the calculated dipole of the structures, and the position of the Dirac point with respect to the TMDC valence (conduction) band edge, $E_D-E_V$ ($E_D-E_C$). }
\begin{tabular}{lcccccc}
TMDC & $\vartheta_b$ ($\vartheta_t$) [°] & $d_{\textrm{b}}$ ($d_{\textrm{t}}$) [\AA] & $\Delta z_{\textrm{grp}}$ [pm] & dipole [debye] & $E_D-E_V$ [eV] & $E_D-E_C$ [eV] \\
\hline
MoSe$_2$ & 19.1 (19.1) &  3.4114 (3.4152) & 0.0020 & 0.0008 & 0.5196  & -0.9346    \\
         & 19.1 (19.1+60) & 3.4222 (3.4083) & 0.5701 & -0.0057 & 0.5180 & -0.9394   \\\hline
WSe$_2$ & 0.0 (0.0) & 3.3489 (3.3609)  & 0.1847  & 0.0099 & 0.1821 &  1.2108   \\
        & 0.0 (0.0+60) & 3.3410 (3.3419) & 3.7920 & 0.0135 & 0.1739  &  -1.2246   \\
\end{tabular}
\end{ruledtabular}
\end{table*}

\begin{table*}[!htb]
\begin{ruledtabular}
\caption{\label{tab:fit_encapsulated} Fit parameters of the model Hamiltonian, Eq.~\eqref{Eq:Hamiltonian},
for the TMDC/graphene/TMDC heterostructures. We summarize the relative twist angles $\vartheta_b$ ($\vartheta_t$) of graphene with respect to bottom (top) TMDC layer, the Fermi velocity $v_{\textrm{F}}$, the staggered potential gap $\Delta$, the sublattice-resolved intrinsic SOC parameters $\lambda_{\textrm{I}}^\textrm{A}$ and $\lambda_{\textrm{I}}^\textrm{B}$, the
Rashba SOC parameter $\lambda_{\textrm{R}}$, the phase angle $\varphi$, and the position of the Dirac point, $E_{\textrm{D}}$, with respect to the Fermi level. }
\begin{tabular}{lcccccccc}
TMDC & $\vartheta_b$ ($\vartheta_t$) [°] & $\Delta$~[meV] & $v_{\textrm{F}}/10^5 [\frac{\textrm{m}}{\textrm{s}}]$ & $\lambda_{\textrm{I}}^\textrm{A}$~[meV] & $\lambda_{\textrm{I}}^\textrm{B}$~[meV] &  $\lambda_{\textrm{R}}$~[meV] & $\varphi$ [°] & $E_{\textrm{D}}$ [meV] \\
\hline
MoSe$_2$ & 19.1 (19.1) & 0.1049 & 7.8918 & 1.1320 & -1.1357 & 0.0057 & 4.1166 & -0.5327 \\
         & 19.1 (19.1+60) & -0.2099 & 7.8872 & -0.0488 & 0.0066 & -0.0187 & -5.2934 &  -0.7439 \\\hline
WSe$_2$ & 0.0 (0.0) & 0.0399 & 8.1670 & 2.6068 & -2.6201 & 0.0334 & 0 & -0.5580 \\
        & 0.0 (0.0+60) & 0.2623 & 8.1523 & 0.0106 & -0.0002 & 0.0042 & 0 & -3.2908\\
\end{tabular}
\end{ruledtabular}
\end{table*}

The structural information for the encapsulated structures are summarized in Table~\ref{tab:struct_encapsulated}. The relaxed top and bottom graphene/TMDC interlayer distances are nearly identical for the different cases we consider, and coincide with the non-encapsulated geometries. In addition, the intrinsic dipole of the trilayer structure is strongly diminished, but still finite due to a small aysmmetry in the interlayer distances. 
The rippling of the graphene layer is small (large) for symmetric (asymmetric) encapsulation when twist angles are the same for top and bottom monolayers (when the top TMDC monolayer has an additional 60$^{\circ}$ twist). The calculated band offsets are also nearly identical to the non-encapsulated structures. 

We expect that symmetric encapsulation will boost proximity SOC in graphene, while for asymmetric encapsulation the proximity SOC in graphene will nearly vanish. The reason is the valley-Zeeman type of SOC combined with the interchange of the graphene sublattices upon 60$^{\circ}$ rotation. 
For example, the induced SOC from the bottom WSe$_2$ is $\lambda_{\textrm{I}}^\textrm{A} \approx - \lambda_{\textrm{I}}^\textrm{B} \approx 1.2$~meV in the case of 0$^{\circ}$ twist angle. If the top WSe$_2$ layer has the same alignment to graphene as the bottom WSe$_2$ layer, the induced SOC will be the same and we can expect a doubling of valley-Zeeman SOC. However, if the top WSe$_2$ layer is rotated by 60$^{\circ}$ with respect to the underlying graphene/WSe$_2$ bilayer, the graphene sublattices are effectively interchanged with respect to the top WSe$_2$ layer. Hence, bottom and top TMDC layers induce opposite valley-Zeeman SOC, which in total leads to a cancellation. 

In Table~\ref{tab:fit_encapsulated}, we summarize the fit results for the TMDC encapsulated geometries, while in Fig.~\ref{Fig:encap_bands}, we explicitly show the results for WSe$_2$-encapsulated graphene and the different twist angle scenarios.
Indeed, symmetric encapsulation strongly enhances and roughly doubles the proximity-induced intrinsic SOC parameters, compared to non-encapsulated geometries. In contrast, the Rashba SOC is drastically reduced, since TMDC encapsulation nearly restores the $z$-mirror symmetry. Also the dipole (intrinsic electric field) of the structures is almost zero. For asymmetric encapsulation, the proximity-induced intrinsic and Rashba SOC is strongly reduced, as expected.
Actually, for perfectly symmetric encapsulation, the Rashba SOC should exactly vanish. Also the valley-Zeeman SOC should vanish in encapsulated structures where inversion symmetry is restored. However, our heterostructures still show a finite structural asymmetry after atomic relaxation, leading to finite values of proximity SOC. 

In conclusion, TMDC encapsulation will only boost proximity SOC in graphene, if both TMDC layers 
offer the valley-Zeeman SOC in an additive way. In other words, both twist angles are important control knobs to tailor the interference of the individual proximity effects, as also discussed in Ref.~\cite{Peterfalvi2021:arxiv}.

\section{Physics behind the spin-orbit proximity effect}
\label{Sec:Discuss}
There are several open questions related to the presented DFT and simulation results that we wish to address: Why is the proximity-induced SOC of valley-Zeeman (sublattice-odd) and not Kane-Mele (sublattice-even) type? 
What is the exact origin of the proximity-induced SOC? Why is the twist-angle dependence so different for different TMDCs and not as universal as predicted by recent tight-binding studies~\cite{Li2019:PRB,David2019:arxiv}? Which atomic type (transition-metal or chalcogen) contributes most to the proximity-induced SOC? Why is the electric field tunability of valley-Zeeman SOC so pronounced for WSe$_2$ and a twist angle of 19.1$^{\circ}$?

We start by addressing the question about which atomic type contributes most to proximity SOC.
We already know that the different transition-metal and chalcogen atoms provide very different contributions to the TMDC spin splittings~\cite{Zollner2019:strain}, which should also influence proximity effects. 
Therefore, we have turned off SOC on different atoms by employing non-relativistic pseudopotentials, and recalculated the proximitized Dirac bands for different TMDCs and twist angles. The fit results are summarized in the SM~\footnotemark[1]. We find, as expected, that the heavier the element (Mo or W, S or Se), the larger the contribution to the proximity-induced SOC. In particular, the contribution of W, Mo, Se, and S atoms to the proximity-induced valley-Zeeman SOC is roughly 1.2, 0.3, 0.1, and 0.01~meV for small ($0$ to $8^{\circ}$) twist angles. Remarkably, this can be drastically different for other twist angles. For example, at 19.1$^{\circ}$ the contribution of Se atoms to the valley-Zeeman SOC is roughly twice as large as the one from W or Mo atoms. The reason is that the graphene Dirac cone couples to different $k$-points within the TMDC Brillouin zone for different twist angles. At different $k$-points, the TMDC bands have a different atomic and orbital decomposition~\cite{Zollner2019:strain}. Therefore, for different twist angles different atomic contributions and orbitals are involved.

\begin{figure*}[htb]
     \includegraphics[width=.99\textwidth]{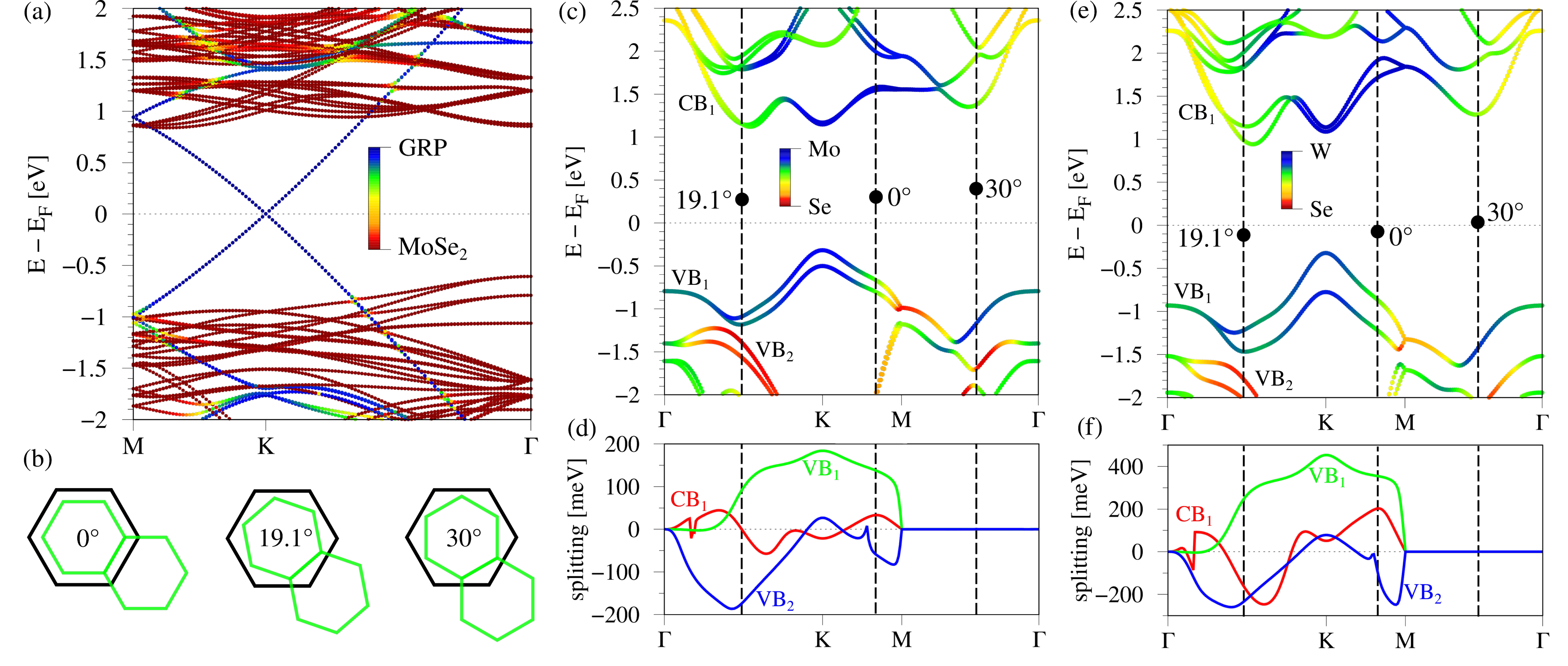}
     \caption{(a) DFT-calculated band structure of the graphene/MoSe$_2$ heterostructure along the high-symmetry path M-K-$\Gamma$ for a twist angle of 0$^{\circ}$. The color code shows the contribution of the individual monolayers to the bands, i. e., the bands appear dark-reddish (dark-blueish) when only MoSe$_2$ (graphene) orbitals contribute. (b) The backfolding of the graphene Dirac point at $K$ for different twist angles. The black (green) hexagon represents the graphene (TMDC) Brillouin zone. (c) DFT-calculated band structure of monolayer MoSe$_2$ with lattice constant of $a = 3.28$~\AA~along the high-symmetry path $\Gamma$-K-M-$\Gamma$. The vertical dashed lines indicate the $k$-points, to which the Dirac states couple to, according to the backfolding in (b). The black dots are the locations of the Dirac point for the different twist angles from Table~S3. (d) The spin splittings $\Delta_{s} =  E_{\uparrow}-E_{\downarrow}$ of the  MoSe$_2$ bands VB$_1$, VB$_2$, and CB$_1$, extracted from the band structure in (c). (e) and (f) are the same as (c) and (d), but for WSe$_2$ monolayer.
    }\label{Fig:Comparison_MoSe2_WSe2}
    \end{figure*}

Why is the proximity SOC of valley-Zeeman type? The graphene Dirac states at $K$ are split as if an external magnetic field would be present, see Fig.~\ref{low_energy_bands_MoSe2}. In particular, for $0^{\circ}$, spin down states are shifted to lower energies compared to spin up, see Fig.~\ref{low_energy_bands_MoSe2}(a), hence a Zeeman-like band splitting. Due to time-reversal symmetry the Dirac states at $K^{\prime}$ are energetically the same, but have the opposite spin. Hence, the charge carriers effectively experience the opposite magnetic field, i.~e., a valley-dependent Zeeman-like spin splitting arises. What causes this splitting in the first place? As we find from the projected band structures for different twist angles, the Dirac states predominantly couple to high-energy TMDC bands, see  for example Fig.~\ref{Fig:Comparison_MoSe2_WSe2}(a) and SM~\footnotemark[1].
Considering a particular twist angle, the Dirac states at $K$ couple differently to the spin up and spin down TMDC band manifolds. For simplicity, imagine that the coupling of Dirac states is only to TMDC conduction band states and the coupling to the spin down manifold is stronger than to the spin up one. According to second order perturbation theory, coupled energy levels repel. When the coupling to spin down is stronger, the spin down Dirac states would be pushed to lower energies compared to spin up, explaining the Zeeman-like splitting for a given valley. Due to time-reversal symmetry, the other valley shows the opposite behavior. Of course, in our heterostructures the coupling is also to TMDC valence bands and there is a delicate balance to the coupling to spin up and spin down manifolds, where one outweighs the other. 
This is similar to recent considerations in twisted graphene/Cr$_2$Ge$_2$Te$_6$ heterostructures~\cite{Zollner2021:arXiv2}.
In particular for $30^{\circ}$ twist angle, the Dirac states of graphene are folded to the $\Gamma$-M high-symmetry line of the TMDC Brillouin zone, see Fig.~\ref{Fig:Comparison_MoSe2_WSe2}, where TMDC bands are spin degenerate, and proximity-induced valley-Zeeman SOC vanishes~\footnotemark[1].

Regarding the electric field tunability of valley-Zeeman SOC for WSe$_2$ and a twist angle of 19.1$^{\circ}$, we first have to consider the location in the TMDC Brillouin zone, where the Dirac cone folds back, see Fig.~\ref{Fig:Comparison_MoSe2_WSe2}(b) and SM~\footnotemark[1]. In particular, the graphene $K$ point folds near the WSe$_2$ $Q$ side-valley, see Fig.~\ref{Fig:Comparison_MoSe2_WSe2}(f), where the spin splitting of the first TMDC conduction band is very large ($\sim 200$~meV). 
Moreover, the electric field results in Table~S7 show that the closer the Dirac point shifts towards the TMDC conduction band, the larger is the proximity-induced valley-Zeeman SOC.  
Considering a coupling of Dirac states to the energetically closest TMDC bands, for this particular twist angle, we come to the conclusion that mainly the first conduction band is responsible for the spin splitting of Dirac states. The contributions from the first two WSe$_2$ valence bands seem to cancel each other, due to opposite spin splittings. 
Another supporting factor is that at the $Q$ valley, the TMDC conduction band wave function is strongly delocalized across the TMDC layer, see Fig.~\ref{Fig:Comparison_MoSe2_WSe2}(e), allowing for a more efficient wavefunction overlap between the layers and an enhanced transfer of the SOC to the graphene layer. 
Therefore, a coupling to the Dirac states should be enhanced, once the energy difference is reduced by applying an external electric field. 
In contrast, for MoSe$_2$ the spin splittings of the relevant bands at the $Q$ valley are very different in magnitude compared to WSe$_2$, see Fig.~\ref{Fig:Comparison_MoSe2_WSe2}(d), and therefore the electric field dependence is not as pronounced for the same twist angle. 

This also relates to the question, why our twist angle results are not universal for all the TMDCs, as the tight-binding studies suggest~\cite{Li2019:PRB,David2019:arxiv}.
Even though the individual TMDCs are very similar, there are profound differences such as atomic and orbital decompositions of bands, leading to different spin splittings across the Brillouin zone. On top of that, our  DFT calculations capture the full picture, including monolayer dispersions, spin-orbit effects, and interlayer interactions. In contrast, the tight-binding description of the heterostructure \cite{Li2019:PRB} employs assumptions for the interlayer interactions and a specific parametrization of the TMDC monolayer dispersion based on first-principles results \cite{Fang2015:PRB}, which does not perfectly reproduce band energies nor spin splittings. 
Anyway, both DFT and the tight-binding descriptions have advantages and drawbacks, but help to gain insights on the physics of proximity-induced SOC in graphene/TMDC heterostructures.

\section{Spin Relaxation Anisotropy}
\label{Sec:Aniso}
\begin{figure*}[htb]
     \includegraphics[width=.99\textwidth]{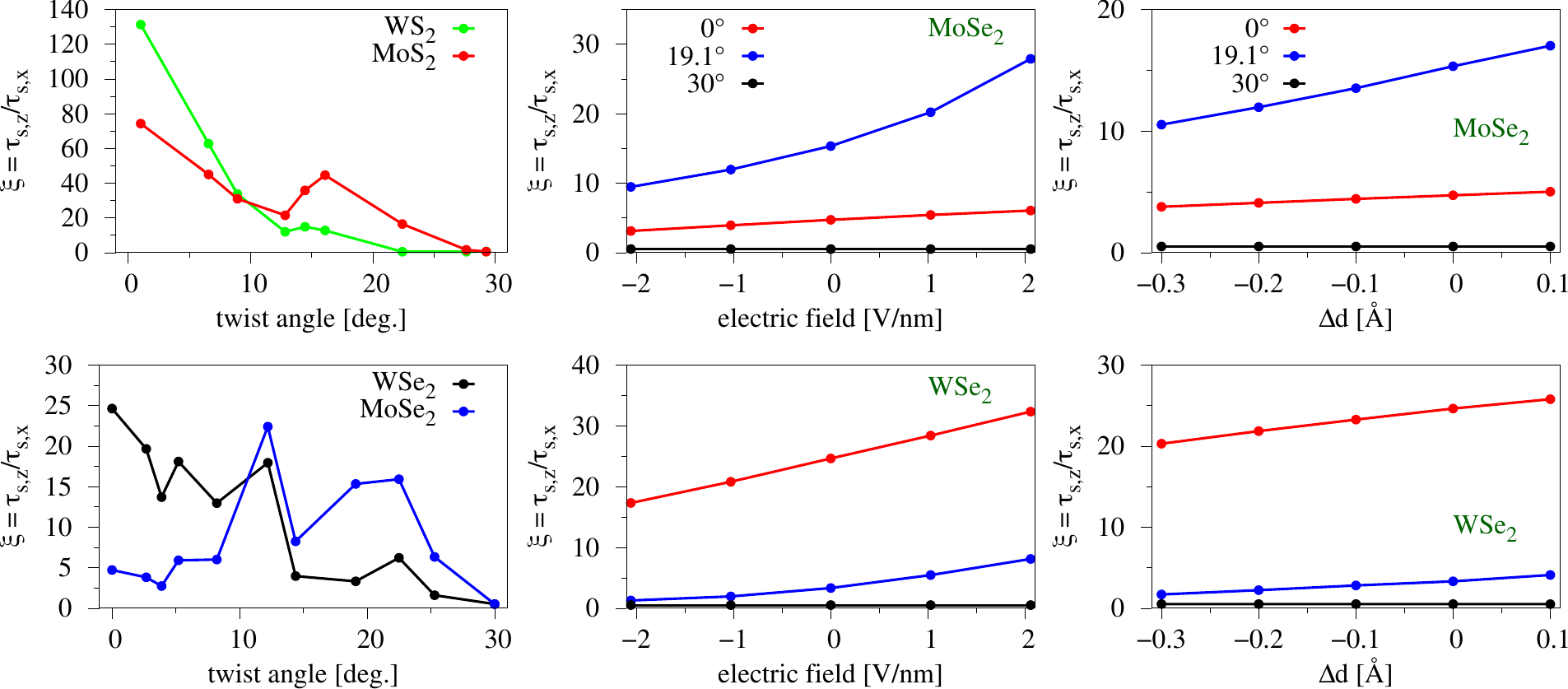}
     \caption{Calculated spin relaxation anisotropy $\xi$, employing Eq.~\eqref{Eq:anisotropy}. Left: Anisotropy as function of the twist angle for the different graphene/TMDC heterostructures, employing the parameters from Table~\ref{tab:fit}. Middle: Anisotropy as function of the transverse electric field for MoSe$_2$ and WSe$_2$ structures for selected twist angles, employing the parameters from Table~S6 and Table~S7.
     Right: Anisotropy as function of the interlayer distance for MoSe$_2$ and WSe$_2$ structures for selected twist angles, employing the parameters from Table~S4 and Table~S5.
    }\label{Fig:anisotropy}
    \end{figure*}

An experimentally verifiable fingerprint of the proximity-induced SOC in graphene/TMDC heterostructures is the anisotropy of the spin lifetimes \cite{Cummings2017:PRL,Ghiasi2017:NL,Leutenantsmeyer2018:PRL,Omar2019:arxiv,Offidani2018:PRB,Zihlmann2018:PRB}. The intrinsic SOC parameters provide a spin-orbit field that points out of the monolayer plane, while the Rashba SOC creates, in the simplest case, a vortex-like in-plane spin-orbit field. 
Depending on the interplay of both SOCs, spins pointing in different directions relax on different timescales, creating a spin lifetime anisotropy.
The spin relaxation anisotropy, $\xi$, which is defined as the ratio between the out-of-plane ($\tau_{s,z}$) and in-plane ($\tau_{s,x}$) spin relaxation times, can be easily calculated from the fitted parameters via \cite{Cummings2017:PRL}
\begin{equation}
\label{Eq:anisotropy}
  \xi  = \frac{\tau_{s,z}}{\tau_{s,x}} = \left(\frac{\lambda_{\textrm{VZ}}}{\lambda_{\textrm{R}}}\right)^{2}\left(\frac{\tau_{\textrm{iv}}}{\tau_{\textrm{p}}}\right)+\frac{1}{2}.
\end{equation}
A similar expression has been derived in Ref.~\cite{Offidani2018:PRB}.
Here, the ratio between the valley-Zeeman and the Rashba SOC strength predominantly determines the anisotropy, but also the ratio between intervalley ($\tau_{\textrm{iv}}$) and momentum ($\tau_{\textrm{p}}$) scattering times play a role. 
In the following, we assume $\tau_{\textrm{iv}}/\tau_{\textrm{p}} = 5$, as in Ref.~\cite{Cummings2017:PRL}.
In Fig.~\ref{Fig:anisotropy}, we summarize the calculated anisotropies as function of the 1) twist angle, 2) the applied electric field, and 3) the interlayer distance, employing the results from above.

The anisotropy is extraordinarily large for WS$_2$ and MoS$_2$ at 0$^{\circ}$, since the valley-Zeeman SOC is giant compared to the Rashba one, pinning the spins to the out-of-plane direction. At 30$^{\circ}$, the anisotropy reduces to $1/2$, i.~e., the Rashba limit, since the valley-Zeeman SOC vanishes independent of the TMDC. In general, the twist angle is an experimental knob to tailor the spin relaxation anisotropy. Once a twist angle is fixed, the proximity SOC can be further tuned by a transverse electric field or pressure engineering of the interlayer distance. Tuning the electric field from $-2$ to $2$~V/nm essentially decreases the Rashba SOC and consequently increases the anisotropy. A strong tunability can be especially observed in WSe$_2$ for 0$^{\circ}$ and for MoSe$_2$ for 19.1$^{\circ}$, where the anisotropies can be increased by a factor of 2--3. In contrast, reducing the interlayer distance both valley-Zeeman and Rashba SOC increase, but at different rates, and the anisotropies decrease.
A particular strong anistoropy can be expected in TMDC-encapsulated graphene, as the Rashba SOC can be suppressed compared to the valley-Zeeman SOC, see Table~\ref{tab:fit_encapsulated}. In particular, considering the WSe$_2$-encapsulated case, and both twist angles to be 0$^{\circ}$, the calculated anisotropy would be gigantic $\xi\approx 3\times10^{4}$.

\section{Spin-Charge Conversion}
\label{Sec:SCC}
Another experimentally verifiable fingerprint of proximity-induced SOC is the possibility to convert between charge and spin currents in proximitized graphene without the need of conventional ferromagnetic electrodes, which is highly desirable for all-2D spintronic devices \cite{Benitez2020:NM,Offidani2017:PRL,Ghiasi2019:NL,Khokhriakov2020:NC,Herlin2020:APL,Lee2022:arxiv,Veneri2022:arxiv,Safeer2022:2DM,Offidani2017:PRL,Monaco2021:PRR,Ferreira2021:JP,Milletari2017:PRL,Dyrdal2014:PRB,Garcia2017:NL,Hoque2021:CP, Ingla2022:2DM, Camosi2022:2DM}. Recent theoretical calculations \cite{Lee2022:arxiv,Veneri2022:arxiv} have already considered the twist angle dependence of the charge-to-spin conversion in graphene/TMDC heterostructures. Remarkably, not only the conventional spin-Hall effect (SHE) and Rashba-Edelstein effect (REE) occur, but also an unconventional REE (UREE) can arise. While for SHE and REE the current-induced non-equilibrium spin density has a polarization perpendicular to the charge current \cite{Ghiasi2019:NL}, for the UREE the spin density polarization is collinear to the applied electric current. 
A similar unconventional charge-to-spin conversion has already been experimentally detected in the semimetals WTe$_2$ \cite{Zhao2020:AM} and MoTe$_2$ \cite{Safeer2019:NL2, Hoque2021:CP,Ontoso2022:arxiv}, and can be attributed to reduced symmetries \cite{Culcer2007:PRL}. 
Recent experiments on graphene/NbSe$_2$~\cite{Ingla2022:2DM}, graphene/WTe$_2$~\cite{Camosi2022:2DM}, and graphene/MoTe$_2$~\cite{Safeer2019:NL2,Ontoso2022:arxiv} heterostructures have demonstrated the spin-to-charge conversion of spins oriented in all three directions. However, in these structures NbSe$_2$, WTe$_2$, and MoTe$_2$ are metallic, contributing directly to the conversion process, along with the proximitized graphene. 

The figure of merit for charge-to-spin conversion for comparing 3D and 2D systems is given by $\alpha \lambda_{\textrm{SF}}$, where $\alpha$ is the conversion efficiency and $\lambda_{\textrm{SF}}$ is the spin diffusion length~\cite{Rojas2019:PRA,Khokhriakov2020:NC,Safeer2019:NL}. Especially $\lambda_{\textrm{SF}}$ can be giant in proximitized graphene ($\sim~\mu$m)~\cite{Khokhriakov2020:NC,Benitez2020:NM,Benitez2018:NP}, much larger than in conventional 3D bulk heavy metals such as Pt or W ($\sim$~nm)~\cite{Rojas2014:PRL,Kim2016:PRL}. Therefore, 2D material heterostructures can outperform 3D systems, even though the conversion efficiencies of, e.~g., Pt (7\%)~\cite{Wang2014:APL} or W (20\%)~\cite{Kim2016:PRL} are sizable.

    \begin{figure}[htb]
     \includegraphics[width=.99\columnwidth]{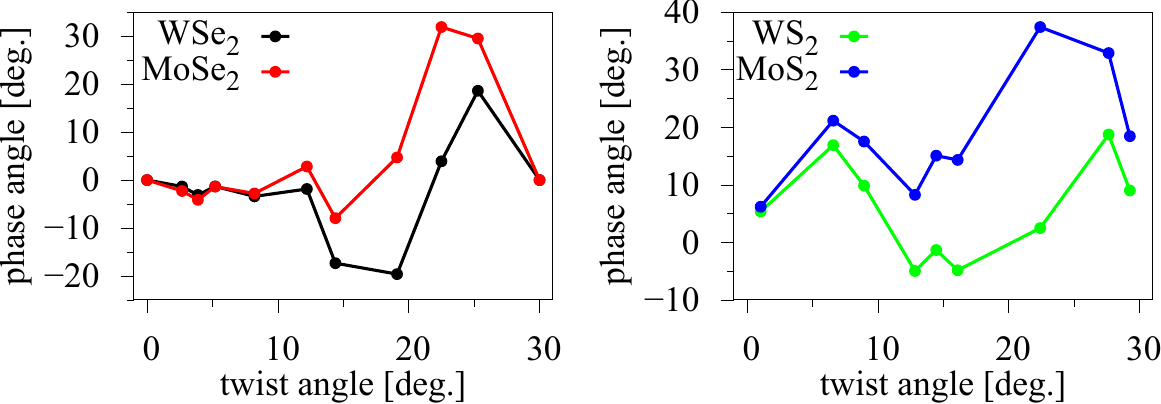}
     \caption{Calculated twist-angle dependence of the Rashba phase angle $\varphi$. The data are summarized in Table~\ref{tab:fit}.
    }\label{Fig:Rashba_phase}
    \end{figure}

The reason for the UREE in graphene/semiconductor-TMDC heterostructures \cite{Lee2022:arxiv,Veneri2022:arxiv} is the Rashba phase angle $\varphi$ of the proximitized Dirac bands. When $\varphi = 0$, no radial in-plane spin-orbit field components arise. In other words, the in-plane spins are always perpendicular to momentum, see for example Fig.~\ref{low_energy_bands_MoSe2}(f), and consequently the generated spin density polarization will also be perpendicular to the applied current direction. However, when $\varphi \neq 0$, also radial spin-orbit field components arise, see for example Fig.~S11, meaning that a current-induced spin density can have a polarization component parallel to the current. Consequently, the UREE will be maximized when $\varphi = 90^{\circ}$. In Fig.~\ref{Fig:Rashba_phase}, we summarize the twist-angle dependence of the Rashba phase angle for our investigated graphene/TMDC structures. 
For our exemplary case of MoSe$_2$, we therefore expect that UREE will be maximized for a twist angle of $\vartheta \approx 23^{\circ}$, where the Rashba phase angle has a maximum of $\varphi \approx 30^{\circ}$.

    \begin{figure*}[htb]
     \includegraphics[width=.95\textwidth]{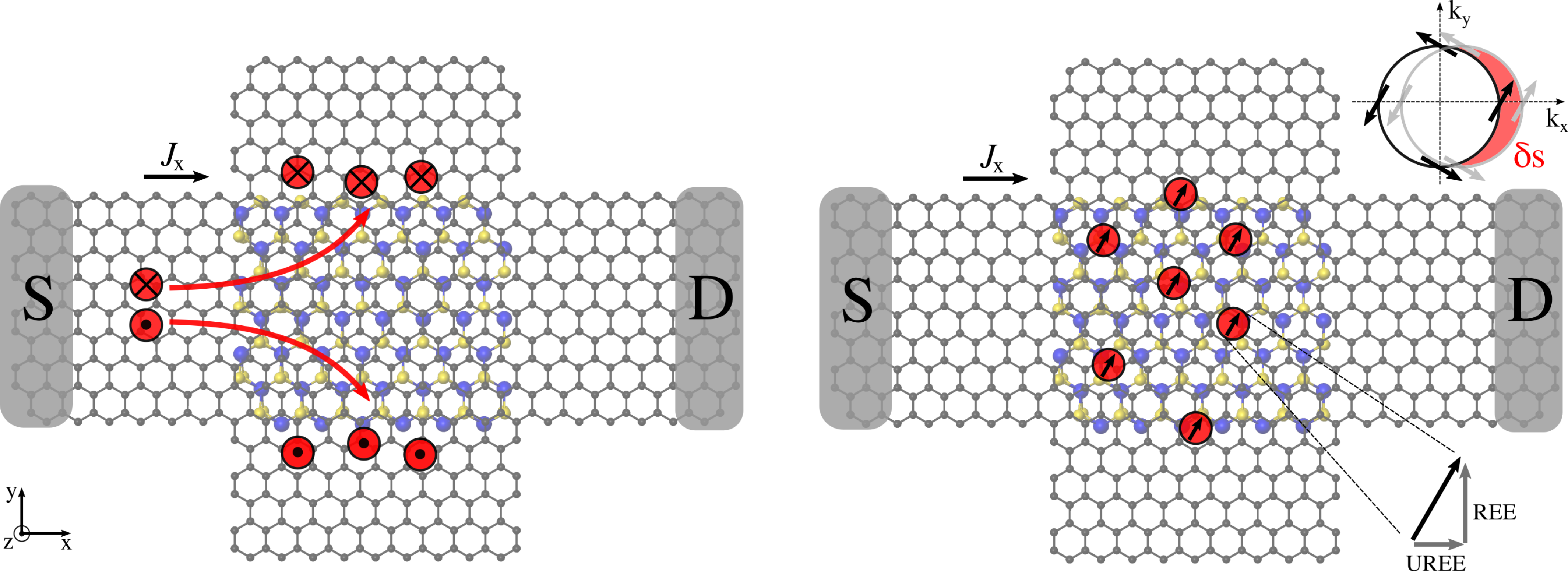}
     \caption{Sketch of the charge-to-spin conversion processes in an experimental setup. Left: A charge current, $J_{x}$, along the $x$ direction results in a spin current flowing along $y$ direction with spins polarized along $z$ due to SHE at the graphene/TMDC region. Right: The charge current shifts the Fermi contour, i.e., the proximitized Dirac bands, and generates a non-equilibrium spin density $\delta s$ at the graphene/TMDC interface. The spin density has components perpendicular (REE) and parallel (UREE) to the charge current, due to the Rashba phase angle $\varphi \neq 0$. 
    }\label{Fig:SCC_sketch}
    \end{figure*}

    \begin{figure*}[htb]
     \includegraphics[width=.99\textwidth]{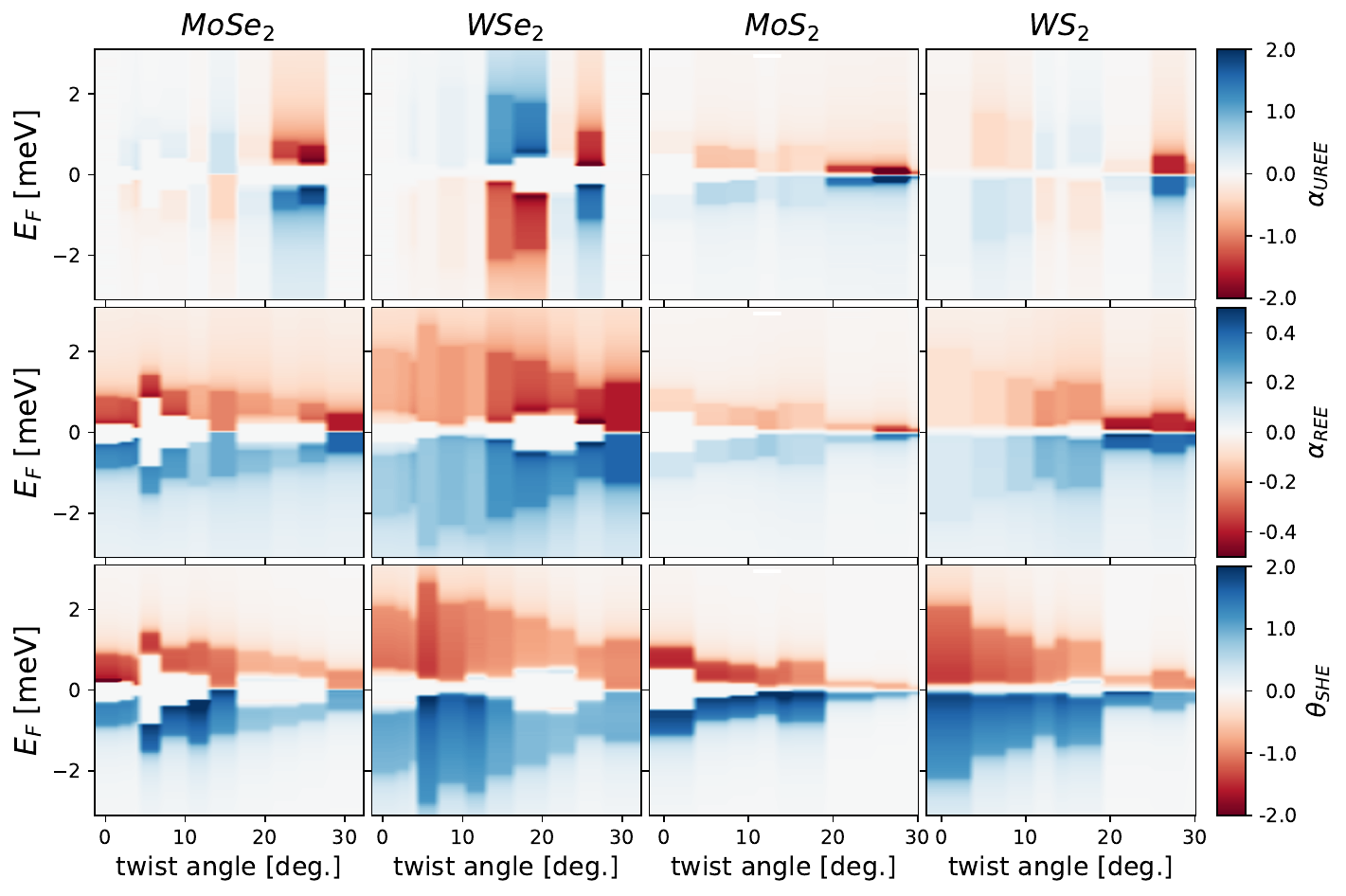}
     \caption{NEGF-computed conversion efficiencies, $\Theta_{\textrm{SHE}}$, $\alpha_{\textrm{REE}}$ and $\alpha_{\textrm{UREE}}$, as function of the twist angle and Fermi level for the different graphene/TMDC heterostructures.
    }\label{Fig:REE_and_UREE}
    \end{figure*}

In Fig.~\ref{Fig:SCC_sketch}, we schematically sketch the different conversion processes in an experimental setup. A charge current along $x$ direction generates a spin current along $y$ with spins polarized along $z$ due to SHE. Similarly, a non-equilibrium spin density $\delta s$ is generated, which is in-plane polarized, due to combined REE and UREE.
In order to get the conversion efficiencies, we have performed real-space quantum transport calculations~\cite{Nikolic2018:book,Nikolic2006,Nikolic2005d}, employing the honeycomb tight-binding version~\cite{Zollner2019:PRR} of the Hamiltonian $\mathcal{H}$, Eq.~\eqref{Eq:Hamiltonian}. The conversion efficiencies $\Theta_{\textrm{SHE}}$, $\alpha_{\textrm{REE}}$, and $\alpha_{\textrm{UREE}}$, are evaluated as
\begin{flalign}
    &\Theta_{\text{SHE}}=\left(2/\hbar\right) J_{y}^{z} / J_{x}\\
    &{\alpha_{\text{REE}}=\left(2ev_{F}/\hbar\right)\delta s_{y}/J_{x}}\\
    &{\alpha_{\text{UREE}}=\left(2ev_{F}/\hbar\right)\delta s_{x}/J_{x}}
\end{flalign}
 where $J_x$ is the charge current along the direction of the applied bias voltage $V_b$ and $\delta s_x$ ($\delta s_y$) is the current-induced nonequilibrium spin density along the $x$ ($y$) axis. Analogously, $J_{y}^{z} = (e/2)\{s_{z},v_{y}\}$ is the Hermitian operator~\cite{Wang2016a,Nikolic2006} of spin current along the $y$-axis  which carries spins oriented alo)ng the $z$-axis. The local spin and charge currents~\cite{Nikolic2006,Wang2016a}, as well as nonequilibrium spin density~\cite{Zollner2019:PRR,Nikolic2005d}, were calculated using the nonequilibrium Green's function formalism (NEGF)~\cite{Stefanucci2013} applied to Landauer geometry~\cite{Wang2016a,Nikolic2018:book} where the central region of finite length is an armchair nanoribbon that is attached to two semi-infinite leads terminating into macroscopic source (S) and drain (D) reservoirs at infinity. The  difference of their electrohemical potentials defines the bias voltage, $\mu_S-\mu_D=eV_b$.  Such clean (i.e., without any impurities) system is then periodically repeated in the transverse direction, which requires carefully checking of convergence in $k_y$ points sampling~\cite{Liu2012d}. Note that this procedure effectively models an infinite plane, while guarantying a continuous energy spectrum of the system Hamiltonian which is essential~\cite{Giuliani2008} for properly introducing dissipation effects when calculating nonequilibrium expectation values in quantum statistical mechanics.  The NEGF formalism provides the nonequilibrium density matrix for steady-state transport, $\hat{\rho(k_y)}$, from which the expectation value of the relevant operator $\hat{O}$ is obtained via  $O(k_y) = \langle \hat{O} \rangle = \mathrm{Tr}\, [\hat{\rho(k_y)} \hat{O}]$ at a single value of $k_y$, while its total is an integral over the first Brillouin zone (BZ), $O=\frac{W}{2\pi} \int dk_y \, O(k_y)$, where $W$ is the width of the nanoribbon.

In Fig.~\ref{Fig:REE_and_UREE}, we show the calculated SHE, REE, and UREE efficiencies, $\Theta_{\textrm{SHE}}$, $\alpha_{\textrm{REE}}$, and $\alpha_{\textrm{UREE}}$, as function of the twist angle and Fermi level for the different graphene/TMDC heterostructures, employing the model Hamiltonian parameters from Table~\ref{tab:fit}. We find that graphene/WSe$_2$ has in general both the largest range and highest values of spin conversion efficiencies, due to the highest values and variations of proximity SOC upon twisting. In addition, the large tunability of the Rashba phase angle is responsible for a pronounced UREE for WSe$_2$ and changes sign at a twist angle of around $20^{\circ}$. In all cases, the UREE follows the REE according to $\alpha_{\textrm{UREE}}=\alpha_{\textrm{REE}}\tan(\varphi)$, i. e., a modulation by the Rashba phase angle.

\begin{figure*}[htb]
     \includegraphics[width=.9\textwidth]{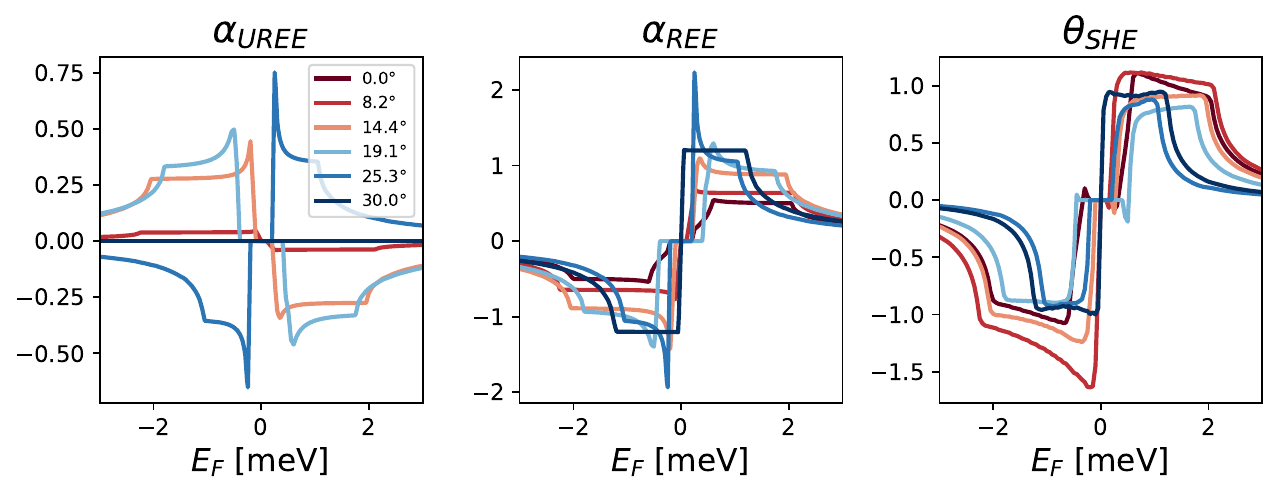}
     \caption{NEGF-computed conversion efficiencies, $\Theta_{\textrm{SHE}}$, $\alpha_{\textrm{REE}}$ and $\alpha_{\textrm{UREE}}$, as function of the Fermi level for selected twist angles for the graphene/WSe$_2$ heterostructure.
    }\label{Fig:plateaus}
    \end{figure*}

Fig.~\ref{Fig:plateaus} shows the REE and UREE efficiencies for a set of twist angles, as a function of the Fermi energy, for graphene/WSe$_2$. The overall behaviour of these curves can simply be understood via the band structure of the corresponding twisted heterostructure. Below the band gap, no states contribute to transport, but as the Fermi energy increases, different cases need to be considered. In the first case, there is no Mexican hat in the band structure and only Rashba-type SOC present, see for example Fig.~\ref{low_energy_bands_MoSe2}(c) for a twist angle of $30^{\circ}$. Once the Fermi energy crosses the first spin-split subband, which is characterized by spin-momentum locking, a plateau in REE emerges~\cite{Offidani2017:PRL}. The plateau is maintained within the Rashba pseudo-gap, followed by an algebraic decay, once the second subband is reached, which contributes with opposite spin-momentum locking.
In the second case, when there is additionally a valley-Zeeman SOC present, as is the case for example in Fig.~\ref{low_energy_bands_MoSe2}(a) for a twist angle of $0^{\circ}$, the REE and UREE efficiencies spike before reaching the plateau.
In the third case, a Mexican hat develops, see for example Fig.~\ref{low_energy_bands_MoSe2}(b), due to proximity SOC that is larger than the pseudospin-asymmetry gap (inverted band structure) \cite{Gmitra2016:PRB}. Instead of directly reaching the plateau or a spike as the Fermi energy increases, the REE and UREE efficiencies now ramp up slowly but still reach a plateau once the Mexican hat is overcome. The analysis from this point is identical to before.

\section{Conclusions}	
\label{Sec:Summary}

In conclusion, we have performed extensive first-principles calculations to reveal the twist-angle and gate dependence of proximity-induced SOC in graphene/TMDC heterostructures. By employing a symmetry-based Hamiltonian, we have extracted orbital and spin-orbit parameters that capture the proximitized low energy Dirac bands. Our results show that the magnitude and the interplay of valley-Zeeman and Rashba SOC can be tuned via twisting, gating, encapsulation, and the interlayer distance. In particular, when twisting from 0$^{\circ}$ to 30$^{\circ}$, the induced valley-Zeeman SOC decreases almost linearly to zero for W-based TMDCs, 
while for Mo-based TMDCs it exhibits a maximum at around 15--20$^{\circ}$ before going to zero. The induced Rashba SOC stays rather constant upon twisting, and acquires a phase angle $\varphi \neq 0$, due to symmetry breaking, for twist angles different from 0$^{\circ}$ and 30$^{\circ}$. 
Within our investigated electric field limits of $\pm 2$~V/nm, mainly the Rashba SOC can be tuned by about 50\%. The interlayer distance provides a giant tunability, since the proximity-induced SOC can be increased by a factor of 2--3, when reducing the distance by only about 10\%. In TMDC-encapsulated graphene, both twist angles are important to control the interference of the individual proximity-induced SOCs, allowing to precisely tailor the valley-Zeeman SOC, while the Rashba SOC becomes suppressed. 

Based on our effective Hamiltonian with fitted parameters, we made specific predictions for experimentally measurable quantities such as spin lifetime anisotropy and charge-to-spin conversion efficiencies. The spin lifetime anisotropy, as well as the charge-to-spin conversion efficiencies are highly tunable by our investigated control knobs and serve as guidance for experimental measurements. 
Our results highlight the important impact of the twist angle, gating, interlayer distance, and encapsulation when employing van der Waals heterostructures in experiments.

\acknowledgments
K.~Z. and J.~F.  were supported by the Deutsche Forschungsgemeinschaft (DFG, German Research Foundation) SFB 1277 (Project No. 314695032), SPP 2244 (Project No. 443416183), the European Union Horizon 2020 Research and Innovation Program under contract number 881603 (Graphene Flagship) and FLAGERA project 2DSOTECH. B.~K.~N. was supported by the US National Science Foundation through the University of Delaware Materials Research Science and Engineering Center, DMR-2011824. The authors thank T. Naimer, E. Icking, and A. Ferreira for fruitful discussions. 
    
\footnotetext[1]{See Supplemental Material, including Refs.~\cite{David2019:arxiv, Naimer2021:arxiv, Kochan2017:PRB, zawadzkiSymmetriesBoundaryConditions2017, keldysh1964,mahfouziHowConstructProper2013, ozakiContinuedFractionRepresentation2007, sanchoHighlyConvergentSchemes1985,mackinnonOneParameterScalingLocalization1981,lewenkopfRecursiveGreenFunction2013a, grothKwantSoftwarePackage2014,gauryNumericalSimulationsTimeresolved2014, weisse2006kernel, santospiresLandauerTransportQuasisteady2020,joaoKITEHighperformanceAccurate2020} where we summarize structural information and fit results in tabular form for the investigated heterostructures. We also analyze the origin of proximity SOC and give details on the real space transport calculations. }

\bibliography{references}

\cleardoublepage


\section*{Supplementary material}

# Supplemental Material:
# Twist- and gate-tunable proximity spin-orbit coupling, spin relaxation anisotropy, and charge-to-spin conversion in heterostructures of graphene and transition-metal dichalcogenides

Klaus Zollner,[1, *] Simão M. João,[2] Branislav K. Nikolić,[3] and Jaroslav Fabian[1]

[1]*Institute for Theoretical Physics, University of Regensburg, 93040 Regensburg, Germany*
[2]*Department of Materials, Imperial College London,
South Kensington Campus, London SW7 2AZ, United Kingdom*
[3]*Department of Physics and Astronomy, University of Delaware, Newark, DE 19716, USA*

## I. STRUCTURAL INFORMATION

TABLE S1. Structural information for the graphene/MoSe$_2$ (graphene/WSe$_2$) heterostructures. We list the twist angle $\vartheta$ between the layers, the supercell tuples $(n,m)$ for graphene and $(n',m')$ for the TMDC, the number of atoms (NoA) in the heterostructure supercell, the number $n_k$ for the $k$-point sampling, the lattice constant and biaxial strain $\varepsilon$ applied to graphene, the relaxed interlayer distance $d_{\text{int}}$, the rippling of the graphene layer $\Delta z_{\text{grp}}$, and the calculated dipole of the MoSe$_2$ (WSe$_2$) structures. The lattice constants of MoSe$_2$ and WSe$_2$ are set to 3.28 Å for all twist angles.

| $\vartheta$ [°] | $(n,m)$ | $(n',m')$ | NoA | $n_k$ | $a_{\text{grp}}$ [Å] | $\varepsilon_{\text{grp}}$ [%] | $d_{\text{int}}$ [Å] | $\Delta z_{\text{grp}}$ [pm] | dipole [debye] |
|---|---|---|---|---|---|---|---|---|---|
| 0.0000 | (0,4) | (0,3) | 59 | 12 | 2.4600 | 0 | 3.4093 (3.3637) | 1.9344 (2.2799) | 0.6054 (0.6156) |
| 2.6802 | (3,5) | (2,4) | 182 | 6 | 2.4794 | 0.7886 | 3.4160 (3.3675) | 1.6146 (1.9063) | 1.8490 (1.8757) |
| 3.8858 | (7,2) | (5,2) | 251 | 3 | 2.5032 | 1.7561 | 3.4217 (3.3767) | 1.3909 (1.6122) | 2.4952 (2.5084) |
| 5.2087 | (1,3) | (1,2) | 47 | 18 | 2.4069 | -2.1585 | 3.3655 (3.3193) | 3.1936 (3.5297) | 0.5047 (0.5178) |
| 8.2132 | (3,5) | (3,3) | 179 | 6 | 2.4348 | -1.0244 | 3.4047 (3.3592) | 1.6455 (1.9513) | 1.8815 (1.9058) |
| 10.8934 | (2,2) | (1,2) | 45 | 18 | 2.5051 | 1.8333 | 3.4263 (3.3834) | 0.7984 (0.9861) | 0.4424 (0.4431) |
| 12.2163 | (7,2) | (6,0) | 242 | 3 | 2.4043 | -2.2642 | 3.3975 (3.3525) | 1.5354 (1.7975) | 2.6027 (2.6547) |
| 14.3916 | (3,4) | (1,4) | 137 | 9 | 2.4711 | 0.4512 | 3.4216 (3.3765) | 0.6920 (0.7975) | 1.3786 (1.3942) |
| 19.1066 | (1,2) | (0,2) | 26 | 24 | 2.4794 | 0.7886 | 3.4247 (3.3796) | 0.3646 (0.4384) | 0.2563 (0.2580) |
| 22.4987 | (7,2) | (3,4) | 245 | 3 | 2.4375 | -0.9146 | 3.4138 (3.3696) | 0.2918 (0.3495) | 2.5023 (2.5403) |
| 25.2850 | (8,0) | (4,3) | 239 | 3 | 2.4939 | 1.3781 | 3.4296 (3.3859) | 0.1669 (0.1939) | 2.3054 (2.3125) |
| 30.0000 | (0,7) | (3,3) | 179 | 6 | 2.4348 | -1.0244 | 3.4131 (3.3692) | 0.1807 (0.2110) | 1.8204 (1.8420) |

TABLE S2. Structural information for the graphene/MoS$_2$ (graphene/WS$_2$) heterostructures. We list the same information as in Table S1. The lattice constants of MoS$_2$ and WS$_2$ are set to 3.15 Å for all twist angles.

| $\vartheta$ [°] | $(n,m)$ | $(n',m')$ | NoA | $n_k$ | $a_{\text{grp}}$ [Å] | $\varepsilon_{\text{grp}}$ [%] | $d_{\text{int}}$ [Å] | $\Delta z_{\text{grp}}$ [pm] | dipole [debye] |
|---|---|---|---|---|---|---|---|---|---|
| 1.0445 | (4,5) | (3,4) | 233 | 3 | 2.4535 | -0.2642 | 3.3643 (3.3090) | 2.9759 (3.4459) | 2.1582 (2.2144) |
| 3.0045 | (1,4) | (1,3) | 81 | 12 | 2.4784 | 0.7480 | 3.3731 (3.3221) | 2.0840 (2.5539) | 0.7318 (0.7403) |
| 6.5868 | (3,2) | (2,2) | 74 | 12 | 2.5034 | 1.7642 | 3.3834 (3.3335) | 1.3572 (1.8227) | 0.6554 (0.6553) |
| 8.9483 | (0,7) | (1,5) | 191 | 6 | 2.5055 | 1.8496 | 3.3886 (3.3363) | 1.0278 (1.4295) | 1.6749 (1.6864) |
| 12.8385 | (3,5) | (1,5) | 191 | 6 | 2.5055 | 1.8496 | 3.3786 (3.3399) | 0.6556 (0.8256) | 1.6562 (1.6724) |
| 14.4649 | (5,1) | (3,2) | 119 | 9 | 2.4661 | 0.2480 | 3.3797 (3.3273) | 0.6684 (0.8666) | 1.0518 (1.0751) |
| 16.1021 | (0,8) | (2,5) | 245 | 3 | 2.4590 | -0.0407 | 3.3783 (3.3139) | 0.5488 (0.7226) | 2.1838 (2.2716) |
| 22.4109 | (6,1) | (3,3) | 167 | 6 | 2.4961 | 1.4675 | 3.3926 (3.3418) | 0.1818 (0.2535) | 1.4186 (1.4332) |
| 24.7913 | (1,4) | (-1,4) | 81 | 12 | 2.4784 | 0.7480 | 3.3878 (3.3363) | 0.1463 (0.1921) | 0.6899 (0.6999) |
| 27.6385 | (1,5) | (3,2) | 119 | 9 | 2.4661 | 0.2480 | 3.3846 (3.3312) | 0.1214 (0.1594) | 1.0246 (1.0390) |
| 29.2649 | (5,3) | (1,5) | 191 | 6 | 2.5055 | 1.8496 | 3.3950 (3.3462) | 0.0977 (0.1141) | 1.6023 (1.6116) |
| 30.0000 | (3,3) | (4,0) | 102 | 9 | 2.4249 | -1.4268 | 3.3712 (3.3166) | 0.1456 (0.1781) | 0.9085 (0.9323) |

---

* klaus.zollner@physik.uni-regensburg.de

## II. BAND OFFSETS

TABLE S3. The calculated position of the Dirac point with respect to the TMDC valence (conduction) band edge, $E_D - E_V$ ($E_D - E_C$), as defined in Fig. 2(a), for the different TMDCs and twist angles.

| | MoSe$_2$ (WSe$_2$) | | | MoS$_2$ (WS$_2$) | |
|---|---|---|---|---|---|
| $\vartheta$ [°] | $E_D - E_V$ [eV] | $E_D - E_C$ [eV] | $\vartheta$ [°] | $E_D - E_V$ [eV] | $E_D - E_C$ [eV] |
| 0.0000 | 0.6093 (0.2347) | -0.8525 (-1.1695) | 1.0445 | 1.3417 (0.9496) | -0.3741 (-0.7488) |
| 2.6802 | 0.5462 (0.1720) | -0.9143 (-1.2321) | 3.0045 | 1.2622 (0.8765) | -0.4527 (-0.8221) |
| 3.8858 | 0.4769 (0.1078) | -0.9843 (-1.2977) | 6.5868 | 1.1862 (0.8035) | -0.5295 (-0.8946) |
| 5.2087 | 0.7824 (0.4056) | -0.6792 (-0.9983) | 8.9483 | 1.1848 (0.7980) | -0.5313 (-0.9012) |
| 8.2132 | 0.6883 (0.3146) | -0.7731 (-1.0910) | 12.8385 | 1.1896 (0.8038) | -0.5275 (-0.8960) |
| 10.8934 | 0.4722 (0.1066) | -0.9888 (-1.2986) | 14.4649 | 1.3133 (0.9249) | -0.4039 (-0.7755) |
| 12.2163 | 0.7897 (0.4157) | -0.6721 (-0.9897) | 16.1021 | 1.3364 (0.9435) | -0.3812 (-0.7565) |
| 14.3916 | 0.5789 (0.2077) | -0.8831 (-1.1978) | 22.4109 | 1.2297 (0.8437) | -0.4885 (-0.8581) |
| 19.1066 | 0.5593 (0.1933) | -0.9030 (-1.2124) | 24.7913 | 1.2863 (0.9009) | -0.4325 (-0.8005) |
| 22.4987 | 0.6930 (0.3215) | -0.7693 (-1.0838) | 27.6385 | 1.3239 (0.9394) | -0.3950 (-0.7623) |
| 25.2850 | 0.5207 (0.1521) | -0.9410 (-1.2542) | 29.2649 | 1.2033 (0.8179) | -0.5160 (-0.8843) |
| 30.0000 | 0.7042 (0.3337) | -0.7585 (-1.0721) | 30.0000 | 1.4588 (1.0690) | -0.2603 (-0.6327) |

## III.  VERTICAL AND LATERAL SHIFTS — MODEL PARAMETERS

TABLE S4. Fit parameters of the model Hamiltonian for the graphene/MoSe$_2$ heterostructures for different angles and for lateral ($x$ and $y$) and vertical ($z$) shifts. The vertical shifts tune the interlayer distance by $\Delta d$, while lateral shifts $x$ and $y$ are in fractions of the supercell lattice vectors. We summarize the twist angle $\vartheta$, the Fermi velocity $v_\text{F}$, the staggered potential gap $\Delta$, the sublattice-resolved intrinsic SOC parameters $\lambda_\text{I}^\text{A}$ and $\lambda_\text{I}^\text{B}$, the Rashba SOC parameter $\lambda_\text{R}$, the phase angle $\varphi$, the position of the Dirac point, $E_\text{D}$, with respect to the Fermi level, and the position of the Dirac point with respect to the TMDC valence (conduction) band edge, $E_D - E_V$ ($E_D - E_C$).

| $\vartheta$ [°] | $\Delta d$ [Å] | $(x,y)$ | $\Delta$ [meV] | $v_\text{F}/10^5[\frac{\text{m}}{\text{s}}]$ | $\lambda_\text{I}^\text{A}$ [meV] | $\lambda_\text{I}^\text{B}$ [meV] | $\lambda_\text{R}$ [meV] | $\varphi$ [°] | $E_\text{D}$ [meV] | $E_D - E_V$ [eV] | $E_D - E_C$ [eV] |
|---|---|---|---|---|---|---|---|---|---|---|---|
| 0.0000 | -0.3 | (0,0) | 1.0981 | 8.2040 | 0.7005 | -0.6360 | 0.8270 | 0 | 3.4731 | 0.4266 | -1.0305 |
| 0.0000 | -0.2 | (0,0) | 0.8369 | 8.2408 | 0.4916 | -0.4538 | 0.5580 | 0 | 3.6012 | 0.4910 | -0.9680 |
| 0.0000 | -0.1 | (0,0) | 0.6465 | 8.2837 | 0.3417 | -0.3204 | 0.3741 | 0 | 3.2502 | 0.5523 | -0.9079 |
| 0.0000 | 0.0 | (0,0) | 0.4917 | 8.2538 | 0.2422 | -0.2258 | 0.2550 | 0 | 1.8970 | 0.6093 | -0.8525 |
| 0.0000 | 0.1 | (0,0) | 0.3752 | 8.3331 | 0.1626 | -0.1575 | 0.1686 | 0 | 2.4345 | 0.6609 | -0.8010 |
| 0.0000 | 0.0 | (2/9,1/9) | 0.4785 | 8.2867 | 0.2331 | -0.2229 | 0.2573 | 0 | 2.7824 | 0.6105 | -0.8501 |
| 19.1066 | -0.3 | (0,0) | 0.1284 | 7.4888 | 1.5851 | -1.6211 | 1.1323 | 1.8564 | -1.9789 | 0.3949 | -1.0635 |
| 19.1066 | -0.2 | (0,0) | 0.1311 | 7.6695 | 1.1478 | -1.1788 | 0.7685 | 2.7330 | -1.8678 | 0.4493 | -1.0103 |
| 19.1066 | -0.1 | (0,0) | 0.1178 | 7.8162 | 0.8305 | -0.8568 | 0.5228 | 3.6440 | -2.1709 | 0.5041 | -0.9576 |
| 19.1066 | 0.0 | (0,0) | 0.1163 | 8.0073 | 0.5627 | -0.5827 | 0.3326 | 4.7154 | 1.0680 | 0.5593 | -0.9030 |
| 19.1066 | 0.1 | (0,0) | 0.0764 | 7.9685 | 0.4584 | -0.4770 | 0.2574 | 5.5749 | -2.3473 | 0.6028 | -0.8599 |
| 19.1066 | 0.0 | (1/3,1/6) | 0.2492 | 8.0107 | 0.5833 | -0.6091 | 0.3462 | 4.5436 | -0.1139 | 0.5579 | -0.9040 |
| 30.0000 | -0.3 | (0,0) | 0.0211 | 8.0369 | 0.0151 | -0.0221 | 0.8275 | 0 | 0.0603 | 0.5305 | -0.9291 |
| 30.0000 | -0.2 | (0,0) | 0.0122 | 8.1618 | 0.0056 | -0.0118 | 0.5486 | 0 | -0.9400 | 0.5889 | -0.8721 |
| 30.0000 | -0.1 | (0,0) | 0.0073 | 8.2489 | 0.0018 | -0.0071 | 0.3626 | 0 | -1.0264 | 0.6473 | -0.8147 |
| 30.0000 | 0.0 | (0,0) | 0.0040 | 8.3109 | 0.0013 | -0.0055 | 0.2398 | 0 | 0.2514 | 0.7042 | -0.7585 |
| 30.0000 | 0.1 | (0,0) | 0.0022 | 8.3555 | 0.0003 | -0.0039 | 0.1599 | 0 | -1.1215 | 0.7531 | -0.7097 |
| 30.0000 | 0.0 | (1/6,1/3) | 0.0075 | 8.3019 | 0.0035 | -0.0007 | 0.2415 | 0 | -0.9588 | 0.7027 | -0.7597 |

TABLE S5.  Same as Table S4, but for the graphene/WSe$_2$ heterostructures.

| $\vartheta$ [°] | $\Delta d$ [Å] | $(x,y)$ | $\Delta$ [meV] | $v_\text{F}/10^5[\frac{\text{m}}{\text{s}}]$ | $\lambda_\text{I}^\text{A}$ [meV] | $\lambda_\text{I}^\text{B}$ [meV] | $\lambda_\text{R}$ [meV] | $\varphi$ [°] | $E_\text{D}$ [meV] | $E_D - E_V$ [eV] | $E_D - E_C$ [eV] |
|---|---|---|---|---|---|---|---|---|---|---|---|
| 0.0000 | -0.3 | (0,0) | 1.2839 | 8.0790 | 3.3416 | -3.1994 | 1.6442 | 0 | -2.5342 | 0.0527 | -1.3469 |
| 0.0000 | -0.2 | (0,0) | 1.0131 | 8.1581 | 2.3642 | -2.2891 | 1.1265 | 0 | 0.2431 | 0.1178 | -1.2830 |
| 0.0000 | -0.1 | (0,0) | 0.7761 | 8.2121 | 1.6673 | -1.6297 | 0.7727 | 0 | -0.0829 | 0.1785 | -1.2244 |
| 0.0000 | 0.0 | (0,0) | 0.5878 | 8.2500 | 1.1722 | -1.1572 | 0.5303 | 0 | 1.2931 | 0.2347 | -1.1695 |
| 0.0000 | 0.1 | (0,0) | 0.4500 | 8.2751 | 0.8221 | -0.8187 | 0.3649 | 0 | 1.3808 | 0.2865 | -1.1183 |
| 0.0000 | 0.0 | (2/9,1/9) | 0.5513 | 8.2679 | 1.1353 | -1.1228 | 0.5137 | 0 | -3.9876 | 0.2333 | -1.1714 |
| 19.1066 | -0.3 | (0,0) | 0.1246 | 7.5802 | 1.1452 | -1.2077 | 2.4147 | -21.5031 | 6.3883 | 0.0445 | -1.3574 |
| 19.1066 | -0.2 | (0,0) | 0.1200 | 7.7572 | 0.9431 | -1.0057 | 1.6603 | -20.8729 | 6.1628 | 0.0946 | -1.3077 |
| 19.1066 | -0.1 | (0,0) | 0.1062 | 7.8433 | 0.7521 | -0.8105 | 1.1507 | -20.2483 | 5.4331 | 0.1444 | -1.2603 |
| 19.1066 | 0.0 | (0,0) | 0.0876 | 7.8914 | 0.5899 | -0.6420 | 0.8215 | -19.6123 | 2.2178 | 0.1933 | -1.2124 |
| 19.1066 | 0.1 | (0,0) | 0.0862 | 8.0208 | 0.4426 | -0.4884 | 0.5490 | -19.0509 | -1.2116 | 0.2336 | -1.2230 |
| 19.1066 | 0.0 | (1/3,1/6) | 0.2048 | 7.9445 | 0.5886 | -0.6707 | 0.8108 | -19.7728 | 1.5774 | 0.1926 | -1.2136 |
| 30.0000 | -0.3 | (0,0) | -0.0717 | 8.0332 | -0.0736 | 0.0667 | 1.7661 | 0 | -1.3089 | 0.1679 | -1.2344 |
| 30.0000 | -0.2 | (0,0) | -0.0383 | 8.1632 | -0.0438 | 0.0390 | 1.2503 | 0 | -0.2899 | 0.2244 | -1.1808 |
| 30.0000 | -0.1 | (0,0) | -0.0218 | 8.2547 | -0.0270 | 0.0228 | 0.8812 | 0 | -0.5389 | 0.2788 | -1.1263 |
| 30.0000 | 0.0 | (0,0) | -0.0093 | 8.3185 | -0.0165 | 0.0128 | 0.6197 | 0 | 1.1670 | 0.3337 | -1.0721 |
| 30.0000 | 0.1 | (0,0) | -0.0129 | 8.3640 | -0.0105 | 0.0072 | 0.4359 | 0 | 0.0510 | 0.3838 | -1.0225 |
| 30.0000 | 0.0 | (1/6,1/3) | -0.0292 | 8.2714 | -0.0138 | 0.0100 | 0.6200 | 0 | -0.6410 | 0.3344 | -1.0714 |

## IV. TRANSVERSE ELECTRIC FIELD — MODEL PARAMETERS

TABLE S6. Fit parameters of the model Hamiltonian for the graphene/MoSe$_2$ heterostructures for different angles and applied transverse electric field. We summarize the twist angle $\vartheta$, the electric field amplitude, the Fermi velocity $v_\mathrm{F}$, the staggered potential gap $\Delta$, the sublattice-resolved intrinsic SOC parameters $\lambda_\mathrm{I}^\mathrm{A}$ and $\lambda_\mathrm{I}^\mathrm{B}$, the Rashba SOC parameter $\lambda_\mathrm{R}$, the phase angle $\varphi$, the position of the Dirac point, $E_\mathrm{D}$, with respect to the Fermi level, and the position of the Dirac point with respect to the TMDC valence (conduction) band edge, $E_D - E_V$ ($E_D - E_C$).

| $\vartheta$ [°] | field [V/nm] | $\Delta$ [meV] | $v_\mathrm{F}/10^5[\frac{\mathrm{m}}{\mathrm{s}}]$ | $\lambda_\mathrm{I}^\mathrm{A}$ [meV] | $\lambda_\mathrm{I}^\mathrm{B}$ [meV] | $\lambda_\mathrm{R}$ [meV] | $\varphi$ [°] | $E_\mathrm{D}$ [meV] | $E_D - E_V$ [eV] | $E_D - E_C$ [eV] |
|---|---|---|---|---|---|---|---|---|---|---|
| 0.0000 | -2.0569 | 0.4685 | 8.2770 | 0.2477 | -0.2108 | 0.3176 | 0 | 0.9127 | 0.3508 | -1.1106 |
| 0.0000 | -1.0284 | 0.4795 | 8.2908 | 0.2443 | -0.2216 | 0.2817 | 0 | -0.4579 | 0.4764 | -0.9844 |
| 0.0000 | 0.0000 | 0.4917 | 8.2538 | 0.2422 | -0.2258 | 0.2550 | 0 | 1.8970 | 0.6093 | -0.8525 |
| 0.0000 | 1.0284 | 0.5112 | 8.2791 | 0.2277 | -0.2254 | 0.2284 | 0 | 0.3842 | 0.7353 | -0.7255 |
| 0.0000 | 2.0569 | 0.4964 | 8.2182 | 0.2250 | -0.2323 | 0.2172 | 0 | 0.3507 | 0.8642 | -0.5961 |
| 19.1066 | -2.0569 | 0.0672 | 7.9629 | 0.5954 | -0.6141 | 0.4519 | -0.7047 | 1.0541 | 0.3021 | -1.1598 |
| 19.1066 | -1.0284 | 0.0847 | 7.9939 | 0.5910 | -0.6129 | 0.3979 | 1.7762 | 0.2231 | 0.4297 | -1.0319 |
| 19.1066 | 0.0000 | 0.1163 | 8.0073 | 0.5627 | -0.5827 | 0.3326 | 4.7156 | 1.0680 | 0.5593 | -0.9030 |
| 19.1066 | 1.0284 | 0.1288 | 8.0129 | 0.5939 | -0.6207 | 0.3059 | 8.2678 | 0.3496 | 0.6877 | -0.7734 |
| 19.1066 | 2.0569 | 0.1518 | 8.0119 | 0.6185 | -0.6498 | 0.2710 | 12.7598 | -0.1338 | 0.8191 | -0.6418 |
| 30.0000 | -2.0569 | 0.0071 | 8.3101 | 0.0028 | -0.0064 | 0.3138 | 0 | -0.1813 | 0.4461 | -1.0168 |
| 30.0000 | -1.0284 | 0.0053 | 8.3128 | 0.0019 | -0.0059 | 0.2735 | 0 | -0.5350 | 0.5739 | -0.8889 |
| 30.0000 | 0.0000 | 0.0040 | 8.3109 | 0.0013 | -0.0055 | 0.2398 | 0 | 0.2514 | 0.7042 | -0.7585 |
| 30.0000 | 1.0284 | 0.0031 | 8.3035 | 0.0006 | -0.0054 | 0.2115 | 0 | 0.1435 | 0.8327 | -0.6294 |
| 30.0000 | 2.0569 | 0.0025 | 8.2883 | 0.0002 | -0.0053 | 0.1866 | 0 | -0.0753 | 0.9610 | -0.5015 |

TABLE S7. Same as Table S6, but for the graphene/WSe$_2$ heterostructures.

| $\vartheta$ [°] | field [V/nm] | $\Delta$ [meV] | $v_\mathrm{F}/10^5[\frac{\mathrm{m}}{\mathrm{s}}]$ | $\lambda_\mathrm{I}^\mathrm{A}$ [meV] | $\lambda_\mathrm{I}^\mathrm{B}$ [meV] | $\lambda_\mathrm{R}$ [meV] | $\varphi$ [°] | $E_\mathrm{D}$ [meV] | $E_D - E_V$ [eV] | $E_D - E_C$ [eV] |
|---|---|---|---|---|---|---|---|---|---|---|
| 0.0000 | -2.0569 | 0.5411 | 8.2389 | 1.1792 | -1.0921 | 0.6196 | 0 | -1.9323 | -0.0030 | -1.4081 |
| 0.0000 | -1.0284 | 0.5667 | 8.2450 | 1.1796 | -1.1320 | 0.5738 | 0 | 0.7371 | 0.1072 | -1.2972 |
| 0.0000 | 0.0000 | 0.5878 | 8.2500 | 1.1722 | -1.1572 | 0.5303 | 0 | 1.2931 | 0.2347 | -1.1695 |
| 0.0000 | 1.0284 | 0.6073 | 8.2522 | 1.1645 | -1.1731 | 0.4950 | 0 | -0.9790 | 0.3602 | -1.0438 |
| 0.0000 | 2.0569 | 0.6256 | 8.2546 | 1.1594 | -1.1876 | 0.4652 | 0 | -0.7261 | 0.4845 | -0.9195 |
| 19.1066 | -2.0569 | 0.0684 | 7.9566 | 0.3418 | -0.3821 | 0.9212 | -23.0902 | -7.5931 | -0.0235 | -1.4293 |
| 19.1066 | -1.0284 | 0.0641 | 7.9215 | 0.4441 | -0.4889 | 0.8698 | -21.8125 | -2.6814 | 0.0585 | -1.3471 |
| 19.1066 | 0.0000 | 0.0876 | 7.8914 | 0.5899 | -0.6420 | 0.8215 | -19.6123 | 2.2178 | 0.1933 | -1.2124 |
| 19.1066 | 1.0284 | 0.1142 | 7.9118 | 0.7122 | -0.7730 | 0.7474 | -17.2804 | -0.2436 | 0.3158 | -1.0893 |
| 19.1066 | 2.0569 | 0.1410 | 7.9332 | 0.8416 | -0.9138 | 0.7124 | -14.5302 | -3.3181 | 0.4357 | -0.9696 |
| 30.0000 | -2.0569 | -0.0134 | 8.3040 | -0.0184 | 0.0161 | 0.6820 | 0 | 0.8303 | 0.0824 | -1.3232 |
| 30.0000 | -1.0284 | -0.0124 | 8.3136 | -0.0170 | 0.0142 | 0.6434 | 0 | -0.0597 | 0.2058 | -1.2001 |
| 30.0000 | 0.0000 | -0.0093 | 8.3185 | -0.0165 | 0.0128 | 0.6197 | 0 | 1.1670 | 0.3337 | -1.0721 |
| 30.0000 | 1.0284 | -0.0184 | 8.3206 | -0.0169 | 0.0121 | 0.6120 | 0 | 0.8132 | 0.4594 | -0.9466 |
| 30.0000 | 2.0569 | -0.0190 | 8.3196 | -0.0181 | 0.0119 | 0.6194 | 0 | -0.0919 | 0.5862 | -0.8190 |

## V. ORIGIN OF PROXIMITY SOC

Let us first investigate the MoSe$_2$ monolayer band structures, as an example, to find out about the magnitude of proximity SOC. According to Ref. [1], the interlayer coupling can be effectively described by tunneling matrix elements from graphene orbitals to TMDC bands at particular $k$ points for different twist angles. From this analysis, one finds that the proximity-induced valley-Zeeman SOC can be estimated as:

$$\lambda_{\mathrm{VZ}} \propto \sum_b \frac{|t_b|^2 \Delta_{s,b}}{(\Delta E_b)^2 - (\Delta_{s,b})^2}, \tag{S1}$$

where $\Delta E_b$ is the energy difference between TMDC band $b$ and the Dirac point without taking into account SOC, $t_b$ is the band tunneling strength, and $\Delta_{s,b} = E_{\uparrow,b} - E_{\downarrow,b}$ is the spin splitting of the TMDC band. It is then obvious that $\lambda_{\mathrm{VZ}}$ depends on the particular $k$ point within the TMDC Brillouin zone, to which the Dirac states couple to, since all these quantities depend on $k$. It is straightforward, to calculate the valley-Zeeman SOC from the monolayer TMDC band dispersions, having knowledge about $t_b$.

In Fig. S1, we summarize our MoSe$_2$ band structure analysis. Depending on the twist angle, the Dirac states fold back to different locations within the TMDC Brillouin zone, as indicated by the dashed lines and black dots for three selected twist angles in Fig. S1(a). We also show the atomic character of the TMDC bands, see Fig. S1(b), which certainly influences the band tunneling strength. Remember that graphene resides above the TMDC and the interlayer coupling depends also on the distance between C atoms and metal/chalcogen atoms. In Fig. S1(c), we show the spin splittings, $\Delta_{s,b}$, of three individual TMDC bands, that are probably most relevant for the coupling to graphene, since they are energetically closest to the Dirac states. Note that the spin splitting can be positive or negative, depending on the energetic order of the spin-split TMDC bands. In Fig. S1(d), we show the valley-Zeeman SOC as calculated from perturbation theory, assuming $t_b = 1$. The maximum valley-Zeeman SOC is expected at the TMDC $K$ point, where the first valence band has a giant spin splitting. However, due to lattice mismatch a coupling of the graphene Dirac states directly to the TMDC valleys is not possible. As we can see for 30°, the Dirac states are folded at a $k$-point along the Γ-M high-symmetry line of the TMDC. Along this line, the TMDC bands are not spin split. Even though graphene couples to the TMDC across the vdW gap, the absent splitting within the TMDC bands prohibits a finite proximity-induced valley-Zeeman SOC in graphene, according to Eq. (S1). This is in agreement with the actual DFT calculation results for 30°. However, comparing 0° and 19.1° in Fig. S1(d), the predicted valley-Zeeman SOC for 0° would be much larger compared to 19.1°. This contradicts the DFT results in Fig. 5 of the main paper, where valley-Zeeman SOC shows a maximum at 19.1° for MoSe$_2$. Moreover, for 19.1° the predicted valley-Zeeman SOC is negative. Comparing to the extracted parameters in Table I, the DFT predicts a positive valley-Zeeman SOC, but a negative one for other twist angles. Therefore it is likely that a sign change could appear upon twisting.

Certainly, the band tunneling strength is not the same for all bands and $k$ points. Therefore, here we also employ the following approach for calculating $t_b$ from the monolayer TMDC dispersion. From the band structure in Fig. S1(b), we have knowledge about the atomic projections. In addition, from the relaxed graphene/MoSe$_2$ heterostructures, we know about the interlayer distances of C atoms to the individual atomic layers within the TMDC. Hence, we assume:

$$t_b = \sum_\alpha 100 \cdot P_\alpha \cdot e^{-d_\alpha}, \tag{S2}$$

where $P_\alpha \in [0;1]$ is the projection onto atom $\alpha = \{\mathrm{Mo}, \mathrm{Se}_1, \mathrm{Se}_2\}$, for given band $b$ and $k$-point, which is weighted by an exponential function taking into account the interlayer distances $d_\alpha$ between graphene and the TMDC atomic layers. In Ref. [1], $t_b$ is actually calculated from knowledge about orbital amplitudes of the TMDC band structure, assuming constant interlayer hopping amplitudes, and taking only the closest chalcogen layer into account. We believe that our approach is somewhat similar and also well justified. Our calculated band tunneling strengths along the high-symmetry paths are summarized in Fig. S1(e). Since the first Se layer in MoSe$_2$ is closest to the graphene, the tunneling strength is large for $k$-points where the Se content is large within a particular band. In Fig. S1(f), we show the valley-Zeeman SOC as calculated from perturbation theory and taking into account our calculated band tunneling strengths. Indeed, this leads to an enhanced valley-Zeeman SOC for 19.1°, compared to the case of $t_b = 1$. Still, the predicted valley-Zeeman SOC for 0° would be larger compared to 19.1° in contradiction to our DFT results. From our monolayer TMDC band structure analysis, one can certainly extract some information about the coupling mechanism, but the predicted valley-Zeeman SOC does not reflect the actual DFT results, most likely due to the limited set of bands that we include in our analysis. We performed the same analysis also for the WSe$_2$ monolayer, see Fig. S2, but the results are similar to MoSe$_2$. However, one can already find pronounced differences. For example, comparing the predicted valley-Zeeman SOC for 0° [Fig. S1(f) and Fig. S2(f)], WSe$_2$ should give a much larger value compared to MoSe$_2$, which is consistent with the DFT results. For 19.1° the predicted valley-Zeeman SOC for WSe$_2$ is smaller than for 0°, which is also consistent with our DFT data.

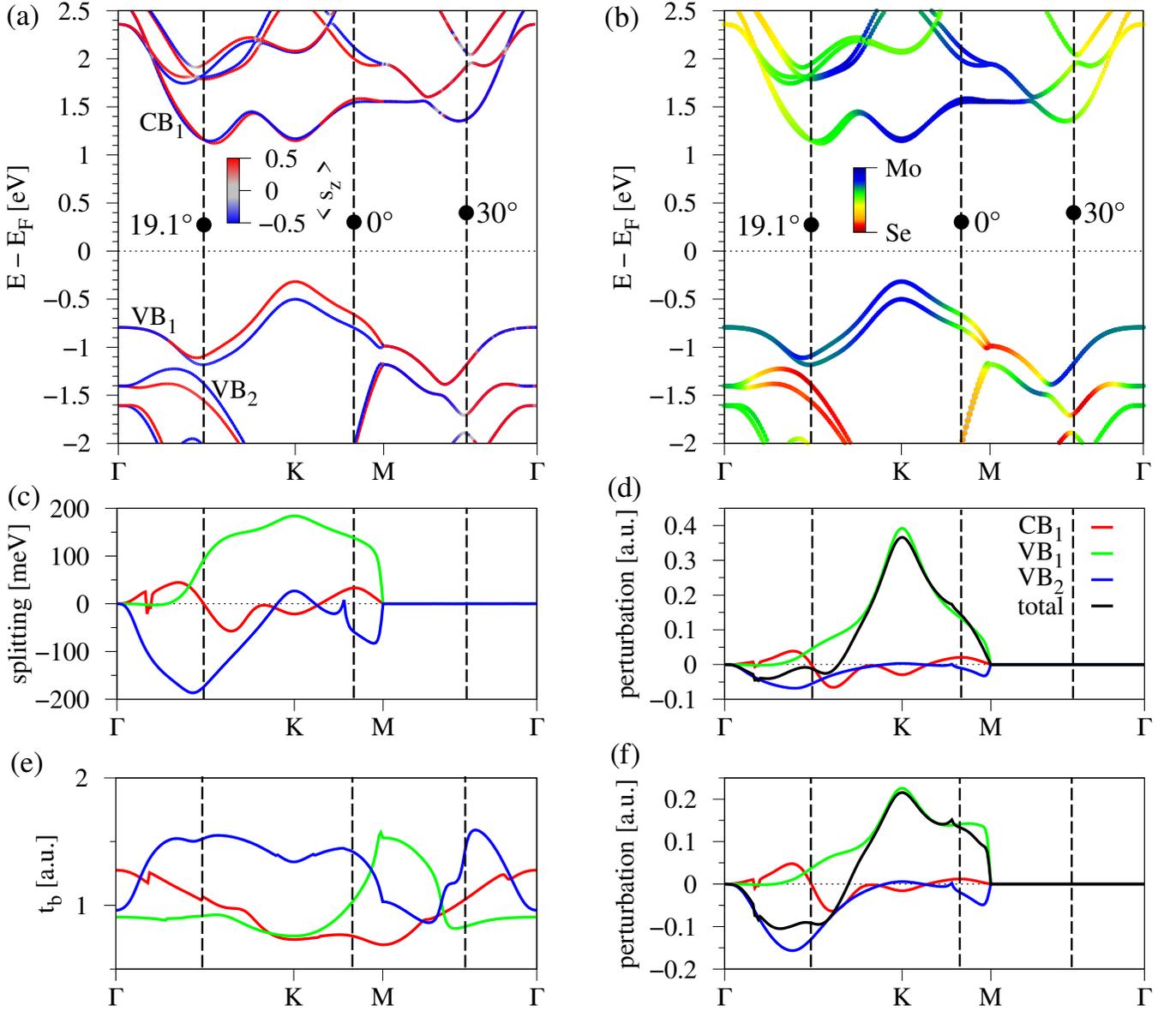

FIG. S1. (a) DFT-calculated band structure of monolayer MoSe$_2$ with lattice constant of $a = 3.28$ Å along the high-symmetry path Γ-K-M-Γ. The color of the lines corresponds to the $s_z$ spin expectation value. The vertical dashed lines indicate the $k$-points, to which the Dirac states couple to, according to the backfolding. The black dots are the locations of the Dirac point for the different twist angles from Table 3. (b) Same as left, but the color code indicates the contribution of the Mo and Se atomic character of the bands. (c) The spin splittings $\Delta_{s,b}$ of the bands VB$_1$, VB$_2$, and CB$_1$, extracted from the band structure in (a). (d) The valley-Zeeman SOC calculated from perturbation theory, evaluating Eq. (S1). We set the Dirac point energy to the 0° twist angle case to calculate $\Delta E_b$, while $\Delta_{s,b}$ is shown in (c). We set $t_b = 1$ for simplicity. The $\sum_b$ is evaluated for the individual bands and for all 3 bands we consider. (e) The calculated band tunneling strength $t_b$, evaluating Eq. (S2). (f) Same as (d), but taking into account the results for $t_b$ from (e).

Another way to find out about the origin of proximity-induced SOC is by carefully analyzing the heterostructure dispersion. In Fig. S3, we do that for the case of graphene/MoSe$_2$ and a twist angle of 0°. Especially from the projected band structure, see Fig. S3(b), we find that the graphene Dirac states couple to TMDC high-energy conduction and valence bands, as indicated by the anticrossings (greenish and yellowish colors). The lowest TMDC conduction bands do not seem to contribute to the interlayer coupling. Looking at the full density of states (DOS), see Fig. S3(e), anticrossings appear whenever the Se content is large, for example at about 1.5 (−1.8) eV above (below) the Fermi level. This is reasonable, since the interlayer coupling happens predominantly at the interface between C and Se atoms. Analyzing the low energy Dirac bands, we find that only 0.3% of Mo and 0.4% of Se content contribute there,

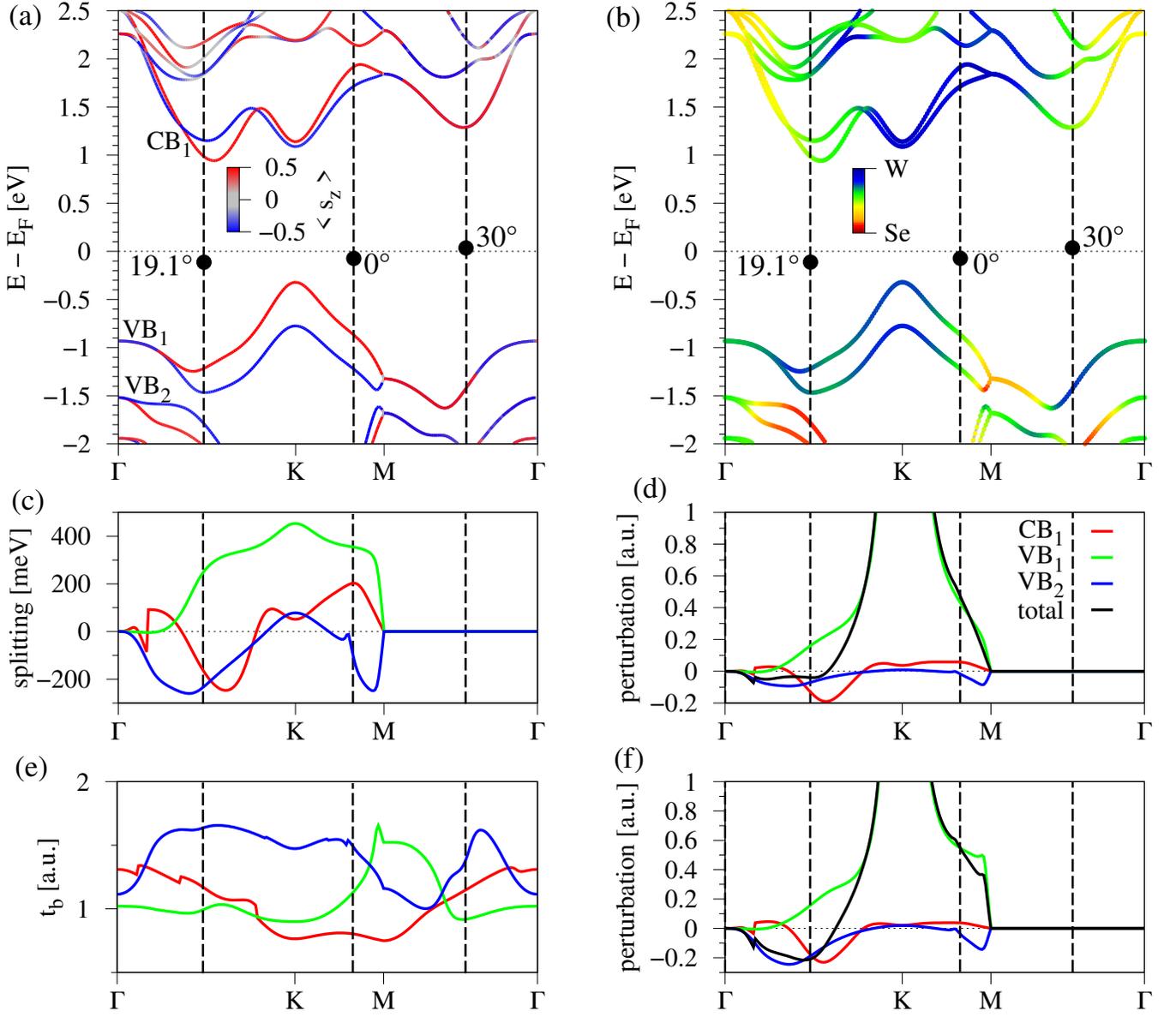

FIG. S2. Same as Fig. S1, but for monolayer WSe$_2$.

leading to a sizable spin splitting of the bands, see Fig. S3(c,d). From the integrated local DOS in real space, see Fig. S4, we find that the whole TMDC is contributing to the Dirac bands, but predominantly interfacial Se $p$ and Mo $d_{xz} + d_{yz}$ orbitals.

In the case of 19.1°, the situation is similar and Dirac states also couple predominantly at higher energies, see Fig. S6. However, the low energy bands have a much larger Mo and Se content compared to the 0° case. In fact, the contribution has doubled, which explains the much larger proximity SOC for 19.1°. The origin may be the coupling to the second highest TMDC valence band (VB$_2$ in Fig. S1), which is almost exclusively formed by Se atoms and has a giant spin splitting at the 19.1° backfolding $k$-point. From the integrated local DOS in real space, see Fig. S5, we find that also Mo $d_{z^2}$ orbitals contribute, which is a significant difference compared to the 0° case.

In Fig. S7, we analyze the 30° geometry. The formerly dominant $s_z$ spin polarization of nearly the whole band structure has vanished, which originates from the arising mirror plane symmetry. Still, the Dirac states couple predominantly to high energy TMDC states. The low energy Dirac bands have nearly the same Mo and Se contribution as for the 19.1° case, but due to symmetry considerations a valley-Zeeman SOC is prohibited.

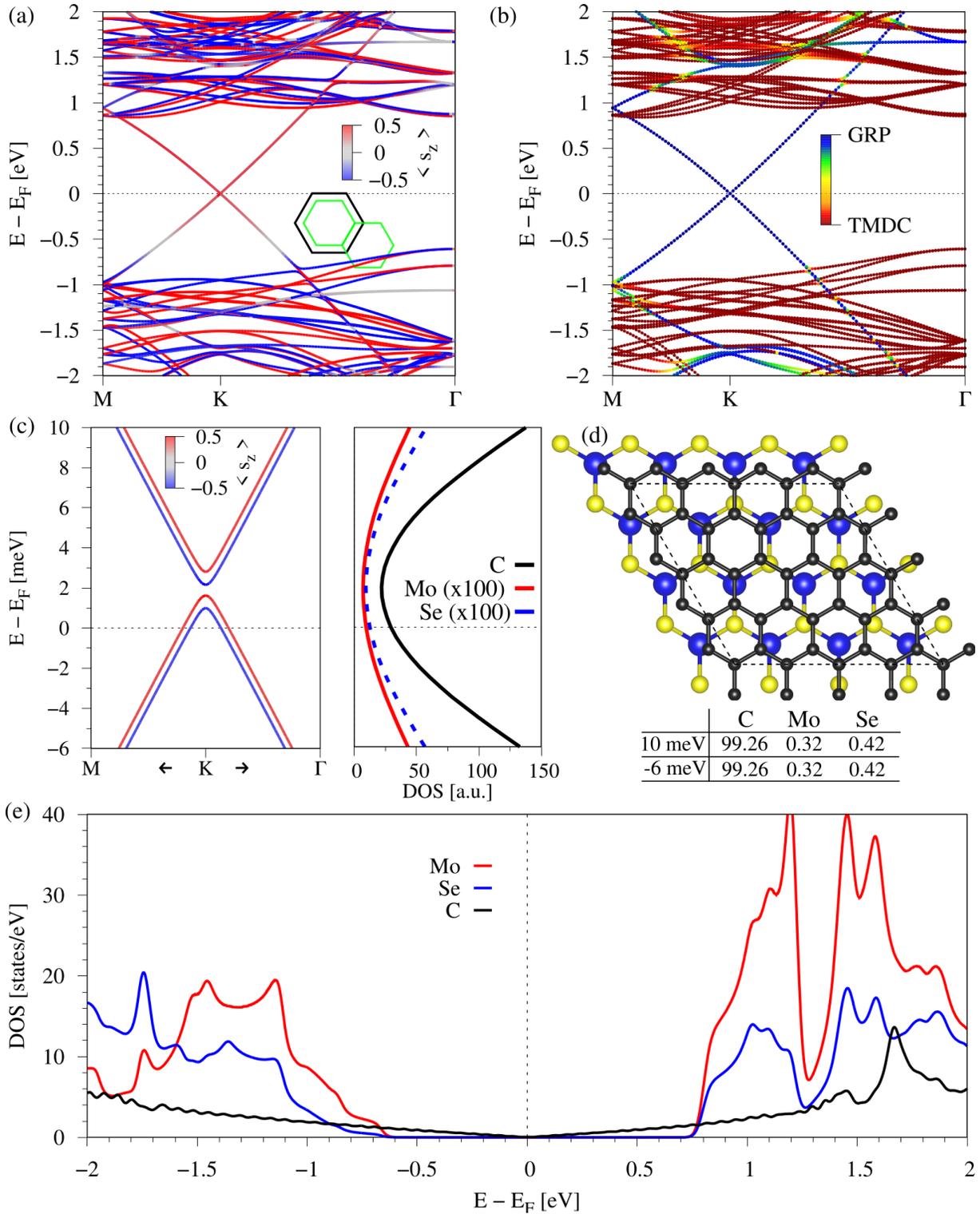

FIG. S3. (a) DFT-calculated band structure of the graphene/MoSe$_2$ heterostructure along the high-symmetry path M-K-Γ for a twist angle of 0°. The color of the lines corresponds to the $s_z$ spin expectation value. The inset shows the backfolding of the graphene Dirac point at $K$. The black (green) hexagon represents the graphene (TMDC) Brillouin zone. (b) Same as (a), but the color code shows the contribution of the individual monolayers to the bands, i. e., the bands appear dark-reddish (dark-blueish) when only TMDC (graphene) orbitals contribute. (c) Zoom to the calculated low-energy Dirac bands near the Fermi level around the $K$ point, corresponding to the band structure in (a). We also show the corresponding atom resolved density of states (DOS). The contribution of Mo and Se is multiplied by a factor of 100 for better visualization. (d) Top view of the heterostructure geometry (black = C, blue = Mo, yellow = Se), where the dashed lines indicate the unit cell. The Table lists the atomic decomposition (in percent) of the DOS shown in (c) at the given energies. (e) The atom resolved DOS.

9

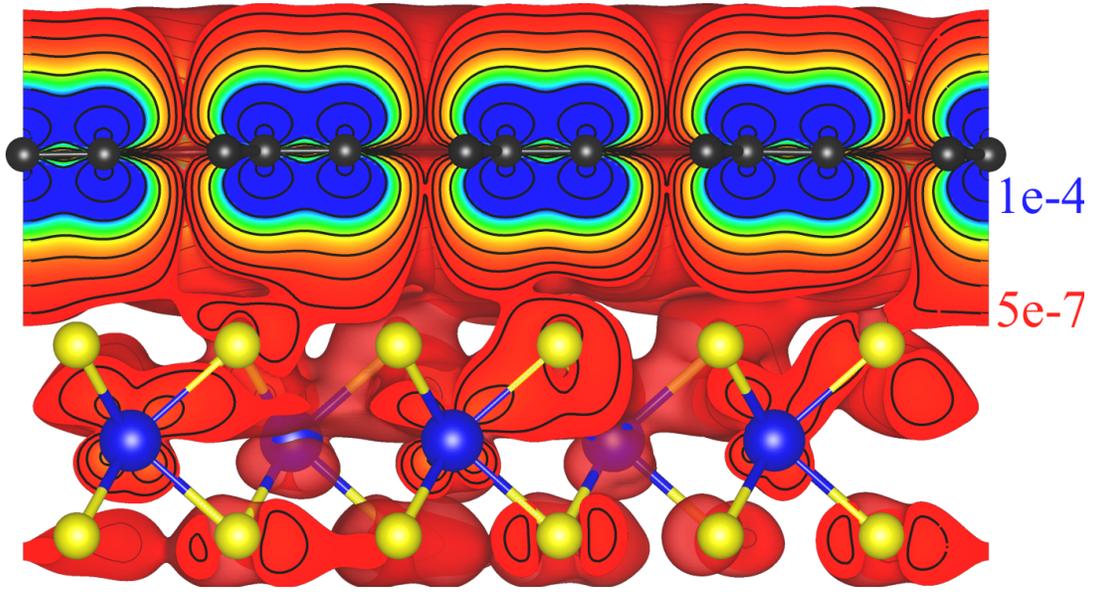

FIG. S4. DFT-calculated integrated local density of states of the $0°$ graphene/MoSe$_2$ heterostructure. The figure is a side view of the cut along the longer diagonal of the unit cell in Fig. S3(d). We take into account only states in an energy window of $\pm 5$ meV around the Dirac point from the low energy dispersion in Fig. S3(c). The colors correspond to isovalues between $1 \times 10^{-4}$ (blue) and $5 \times 10^{-7}$ (red) e/Å$^3$.

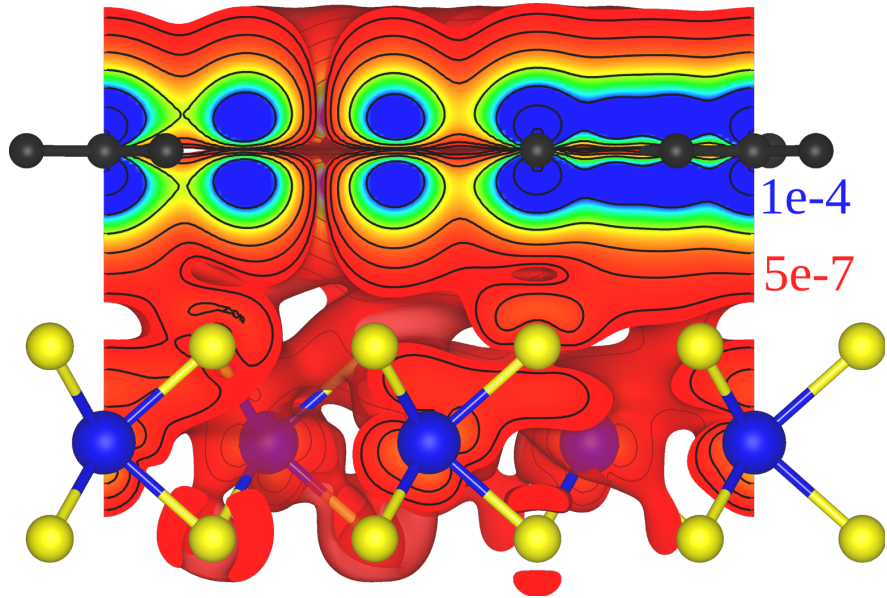

FIG. S5. DFT-calculated integrated local density of states of the $19.1°$ graphene/MoSe$_2$ heterostructure. The figure is a side view of the cut along the longer diagonal of the unit cell in Fig. S6(d). We take into account only states in an energy window of $\pm 5$ meV around the Dirac point from the low energy dispersion in Fig. S6(c). The colors correspond to isovalues between $1 \times 10^{-4}$ (blue) and $5 \times 10^{-7}$ (red) e/Å$^3$.

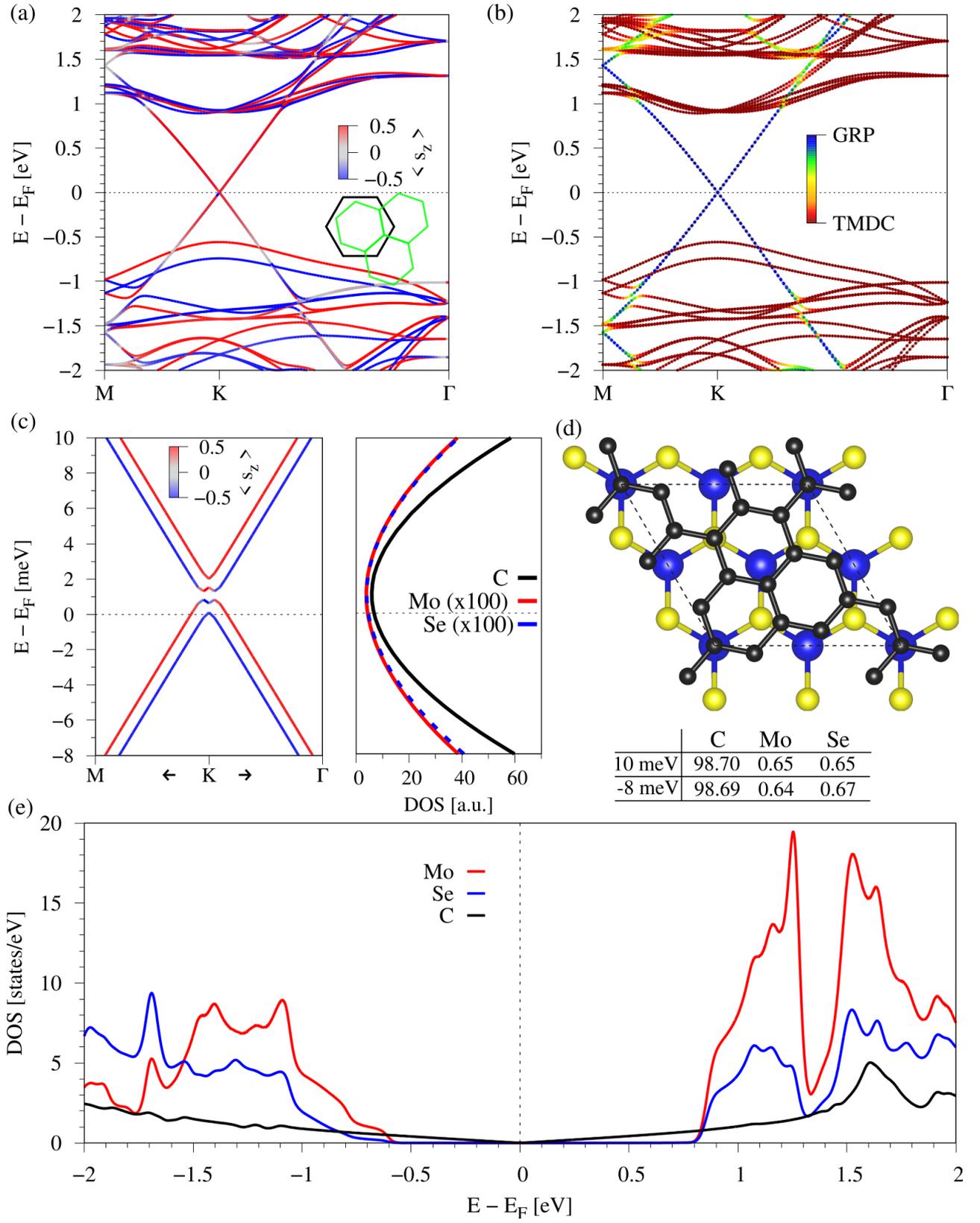

FIG. S6. Same as Fig. S3, but for the graphene/MoSe$_2$ heterostructure with a twist angle of 19.1°.

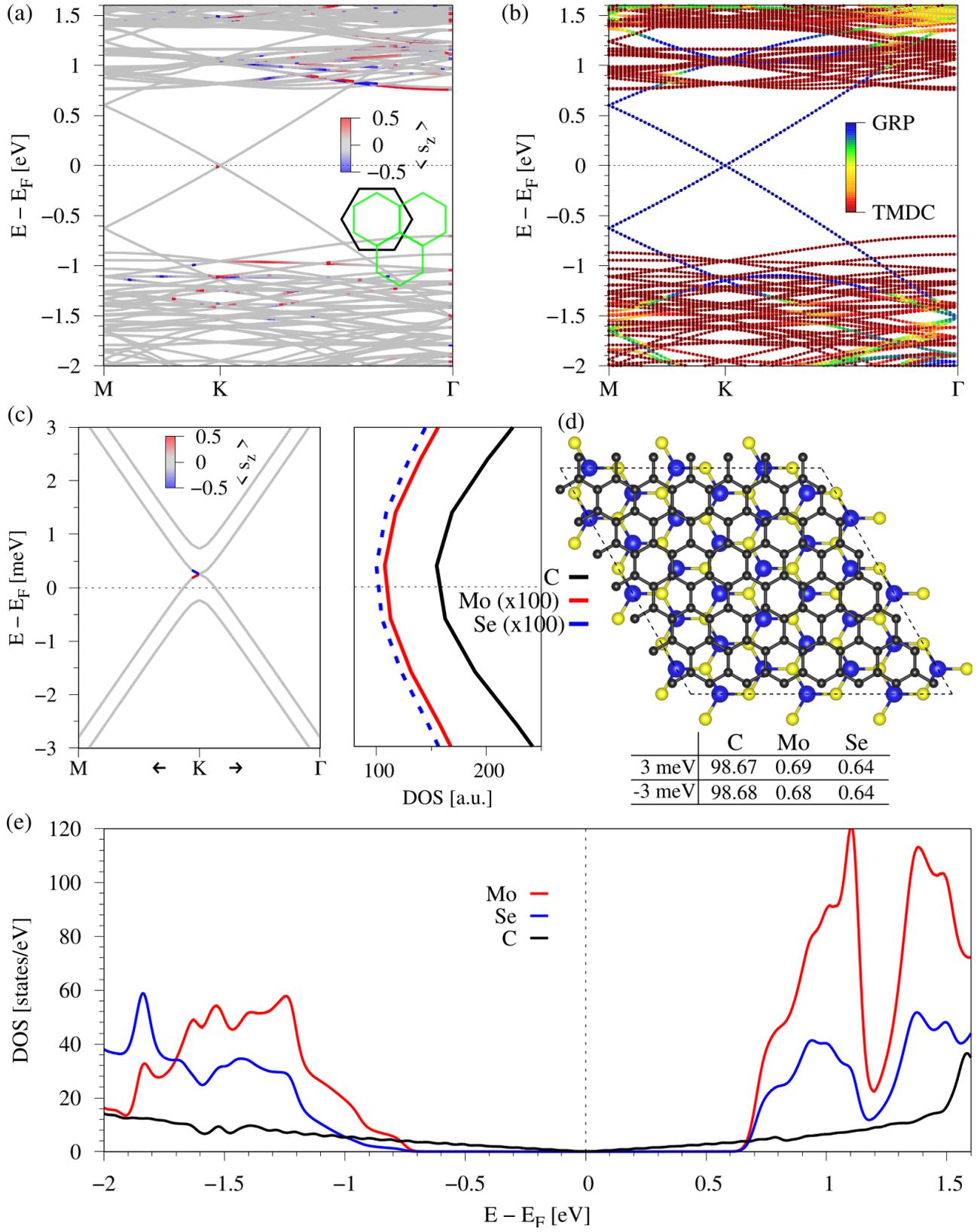

FIG. S7. Same as Fig. S3, but for the graphene/MoSe$_2$ heterostructure with a twist angle of 30°.

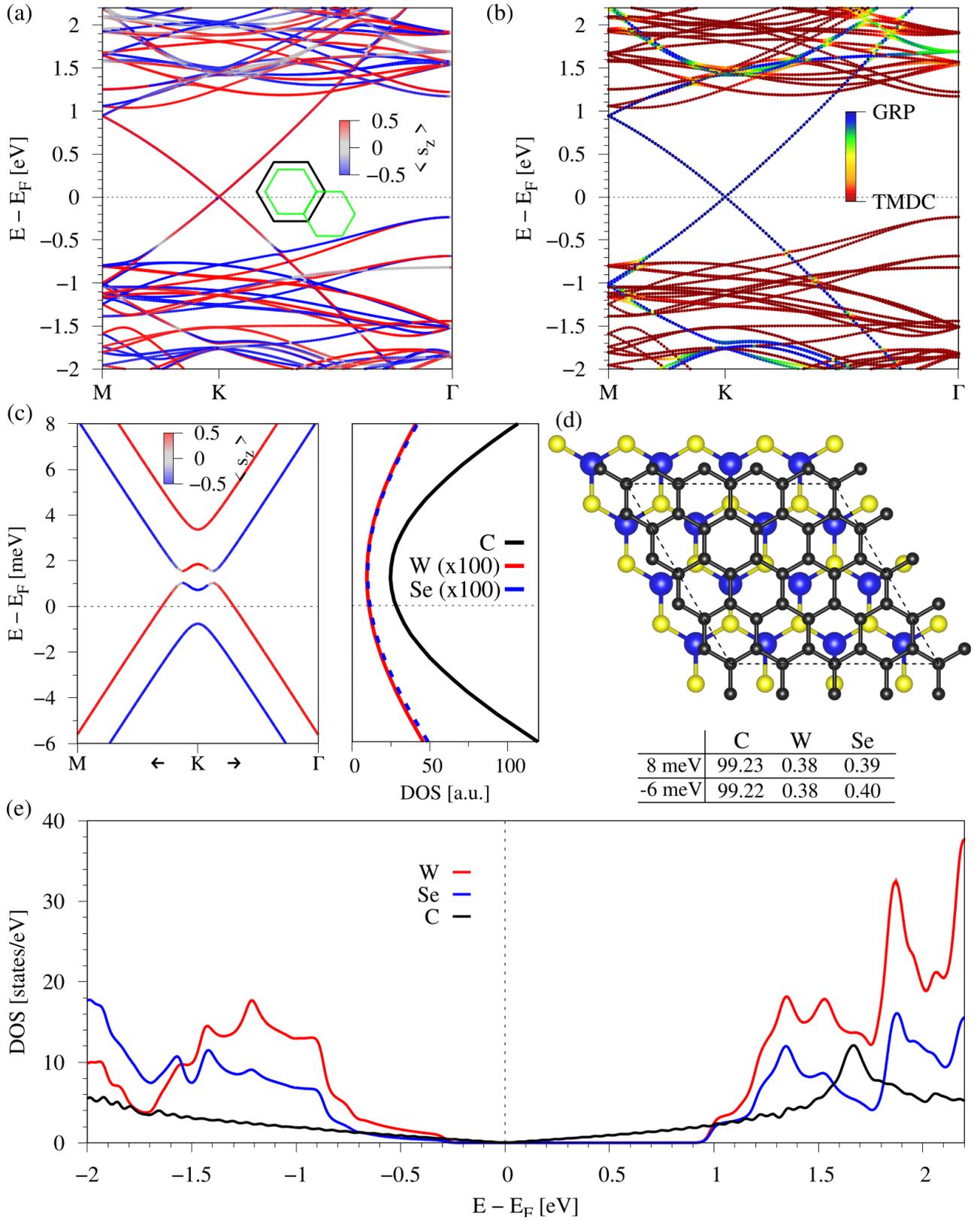

FIG. S8. Same as Fig. S3, but for the graphene/WSe$_2$ heterostructure with a twist angle of 0°.

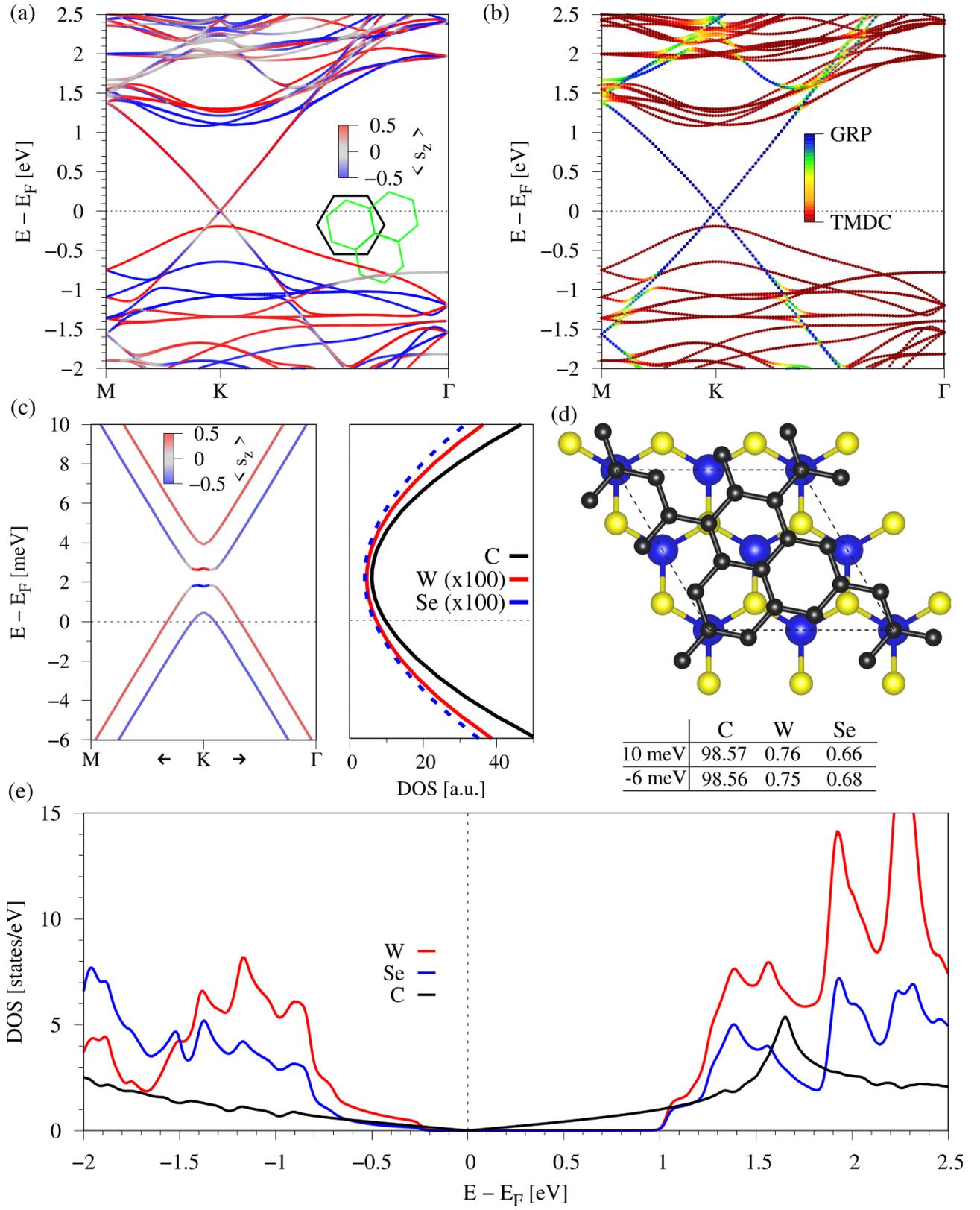

FIG. S9. Same as Fig. S3, but for the graphene/WSe$_2$ heterostructure with a twist angle of 19.1°.

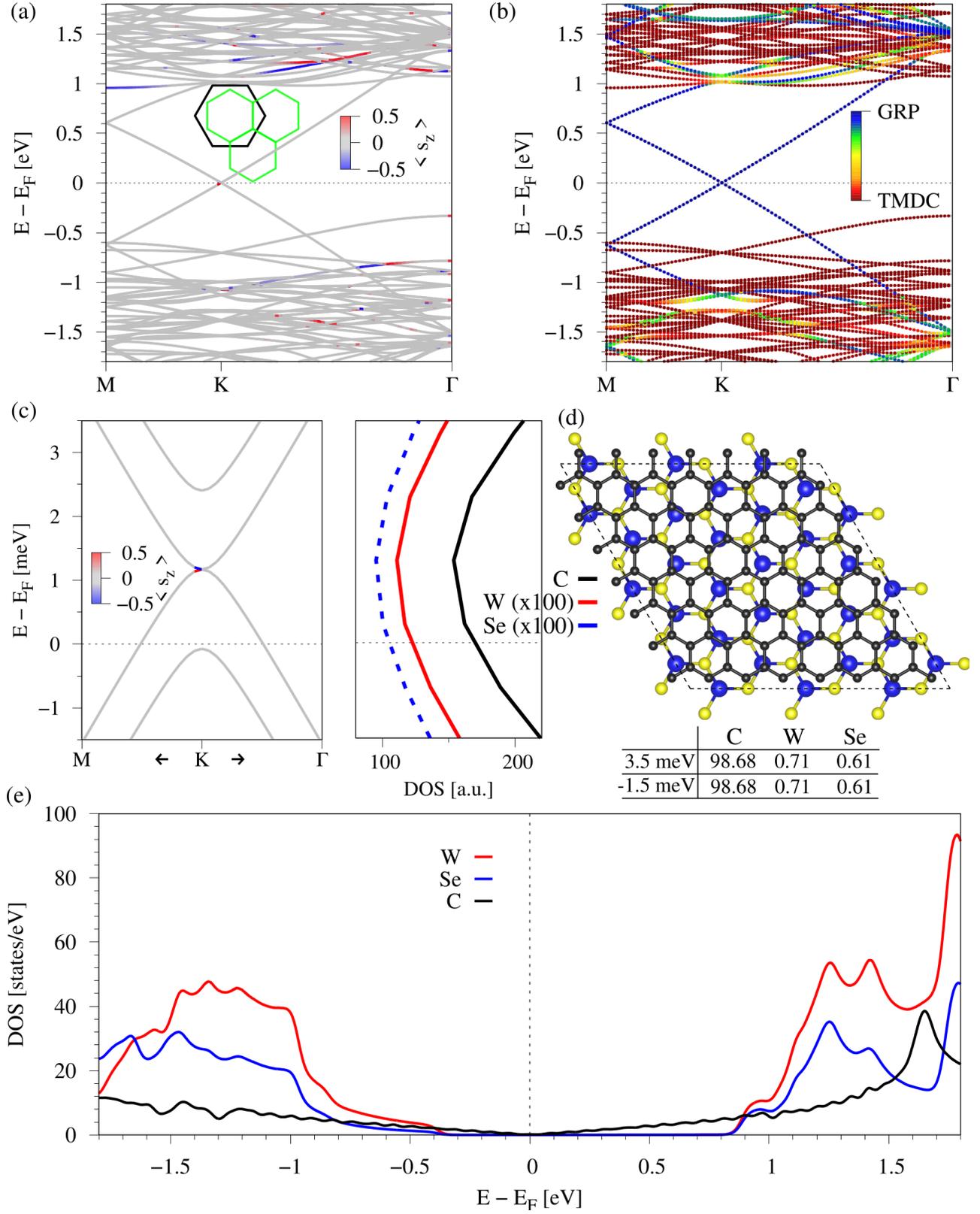

FIG. S10. Same as Fig. S3, but for the graphene/WSe$_2$ heterostructure with a twist angle of 30°.

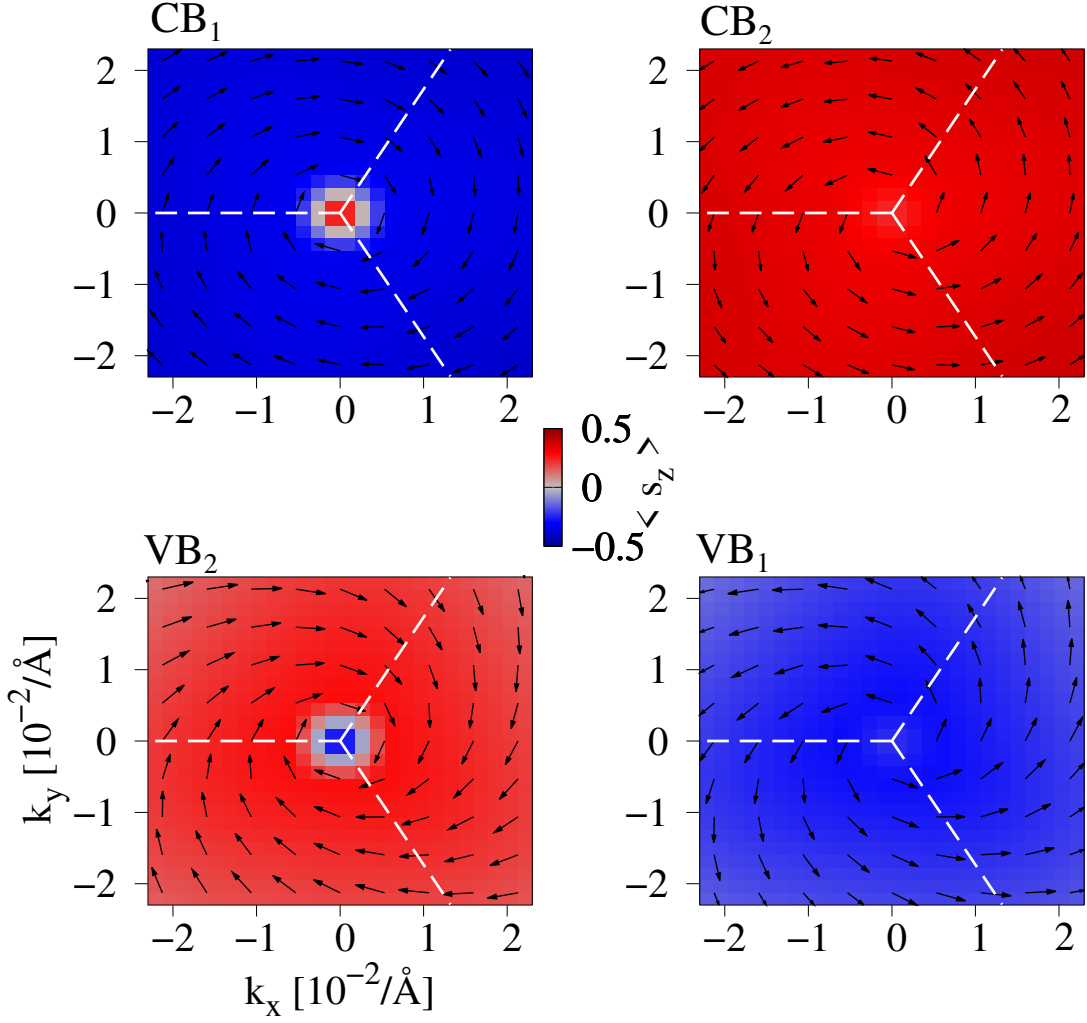

FIG. S11. The calculated spin-orbit fields of the low energy Dirac bands of the graphene/WSe$_2$ heterostructure with a twist angle of 19.1°, corresponding to the dispersion in Fig. S9(c). The color represents the $s_z$ spin expectation value, while the arrows represent $s_x$ and $s_y$ spin expectation values. The dashed white lines represent the edges of the hexagonal Brillouin zone, with the $K$ point at the center. Especially looking at in-plane spins (arrows) along the $k_y = 0$ line emphasizes the presence of the Rashba phase angle $\varphi \approx -19.6°$. This is in contrast to the conventional Rashba field in Fig. 4(f) of the main text, where in-plane spins are always perpendicular to the momentum.

## VI. ATOMIC CONTRIBUTIONS TO PROXIMITY SOC

From Fig. S4, we can see that the C $p_z$ orbitals predominantly couple to the Se $p$ orbitals, which mediate the coupling to Mo $d_{xz} + d_{yz}$ orbitals. However, it is not clear which atomic type gives the dominant contribution for proximity SOC. Therefore, we investigated the impact of artificially turning off SOC on different atoms. In Table S8, Table S9, Table S10, and Table S11 we summarize the fit results for different twist angles and different TMDCs. Turning off SOC on the TMDC, the spin splitting of the TMDC bands vanishes, along with the proximity SOC and we recover the pristine graphene dispersion. Since the TMDC band splittings are reduced, it is not surprising that also the band offsets, of the Dirac states with respect to the TMDC band edges, are modified. For all considered TMDCs and twist angles the valley-Zeeman and Rashba SOC contributions of different atoms nearly perfectly add up, i. e., summing the second and third row fit parameters of a certain structure gives the fit results of the first row. Surprisingly, for the MoSe$_2$ and WSe$_2$ 0° structures we find that the transition-metal and the chalcogen atoms provide the opposite sign for the valley-Zeeman SOC. Moreover, the contribution of the transition-metal atoms dominate over the chalcogen ones. Similarly for the 5.2° structures. For the 19.1° structures, both TMDC atoms provide the same sign for valley-Zeeman SOC, but now the chalcogen atom contribution dominates over the transition-metal one. Furthermore, the Rashba phase angles are opposite in sign. For the 30° structures, again the transition-metal gives the dominant contribution. In the case of MoS$_2$ and WS$_2$, the transition-metal contribution is always dominant compared to the chalcogen atom contribution.

TABLE S8. Fit parameters of the model Hamiltonian for selected graphene/MoSe$_2$ heterostructures, where we artificially turned off SOC on Mo and Se.

| $\vartheta$ [°] | SOC on | $\Delta$ [meV] | $v_\mathrm{F}/10^5[\frac{\mathrm{m}}{\mathrm{s}}]$ | $\lambda_\mathrm{I}^\mathrm{A}$ [meV] | $\lambda_\mathrm{I}^\mathrm{B}$ [meV] | $\lambda_\mathrm{R}$ [meV] | $\varphi$ [°] | $E_\mathrm{D}$ [meV] | $E_D - E_V$ [eV] | $E_D - E_C$ [eV] |
|---|---|---|---|---|---|---|---|---|---|---|
| 0.0000 | C, Mo, Se | 0.4917 | 8.2538 | 0.2422 | -0.2258 | 0.2550 | 0 | 1.8970 | 0.6093 | -0.8525 |
| 0.0000 | C, Mo | 0.4893 | 8.2959 | 0.3267 | -0.3415 | 0.0984 | 0 | 2.8116 | 0.6388 | -0.8553 |
| 0.0000 | C, Se | 0.4591 | 8.2952 | -0.0992 | 0.1207 | 0.1402 | 0 | -0.3492 | 0.6739 | -0.8495 |
| 5.2087 | C, Mo, Se | -1.1162 | 8.5072 | -0.2920 | 0.2166 | 0.2448 | -1.3751 | 1.9400 | 0.7824 | -0.6792 |
| 5.2087 | C, Mo | -1.1374 | 8.5179 | -0.3639 | 0.2915 | 0.1446 | 5.8614 | 0.1243 | 0.8106 | -0.6833 |
| 5.2087 | C, Se | -1.1348 | 8.5431 | 0.0804 | -0.0907 | 0.0946 | -11.8488 | 0.7973 | 0.8470 | -0.6769 |
| 19.1066 | C, Mo, Se | 0.1163 | 8.0073 | 0.5627 | -0.5827 | 0.3326 | 4.7154 | 1.0680 | 0.5593 | -0.9030 |
| 19.1066 | C, Mo | 0.1201 | 8.0673 | 0.1673 | -0.1979 | 0.1431 | -31.7591 | 4.4229 | 0.5974 | -0.8970 |
| 19.1066 | C, Se | 0.1214 | 8.0611 | 0.3887 | -0.3920 | 0.2484 | 29.2495 | -0.4601 | 0.6250 | -0.9001 |
| 30.0000 | C, Mo, Se | 0.0040 | 8.3109 | 0.0013 | -0.0055 | 0.2398 | 0 | 0.2514 | 0.7042 | -0.7585 |
| 30.0000 | C, Mo | 0.0013 | 8.3106 | 0.0001 | -0.0027 | 0.1388 | 0 | -0.1966 | 0.7358 | -0.7589 |
| 30.0000 | C, Se | 0.0007 | 8.3103 | 0.0004 | -0.0029 | 0.0996 | 0 | 0.1012 | 0.7712 | -0.7537 |

TABLE S9. Fit parameters of the model Hamiltonian for selected graphene/WSe$_2$ heterostructures, where we artificially turned off SOC on W and Se.

| $\vartheta$ [°] | SOC on | $\Delta$ [meV] | $v_\mathrm{F}/10^5[\frac{\mathrm{m}}{\mathrm{s}}]$ | $\lambda_\mathrm{I}^\mathrm{A}$ [meV] | $\lambda_\mathrm{I}^\mathrm{B}$ [meV] | $\lambda_\mathrm{R}$ [meV] | $\varphi$ [°] | $E_\mathrm{D}$ [meV] | $E_D - E_V$ [eV] | $E_D - E_C$ [eV] |
|---|---|---|---|---|---|---|---|---|---|---|
| 0.0000 | C, W, Se | 0.5878 | 8.2500 | 1.1722 | -1.1572 | 0.5303 | 0 | 1.2931 | 0.2347 | -1.1695 |
| 0.0000 | C, W | 0.5914 | 8.2516 | 1.2134 | -1.2472 | 0.3036 | 0 | 0.6422 | 0.2691 | -1.1564 |
| 0.0000 | C, Se | 0.5905 | 8.2629 | -0.0623 | 0.1038 | 0.2002 | 0 | -0.1399 | 0.4424 | -1.2144 |
| 5.2087 | C, W, Se | -1.3110 | 8.3911 | -1.1868 | 1.0555 | 0.5979 | -1.3293 | 1.6139 | 0.4056 | -0.8727 |
| 5.2087 | C, W | -1.3320 | 8.4949 | -1.2198 | 1.0480 | 0.4016 | 3.5317 | 5.1394 | 0.4435 | -0.9820 |
| 5.2087 | C, Se | -1.3388 | 8.5136 | 0.0633 | -0.0491 | 0.1678 | -12.3129 | 0.5941 | 0.6116 | -1.0455 |
| 19.1066 | C, W, Se | 0.0876 | 7.8914 | 0.5899 | -0.6420 | 0.8215 | -19.6123 | 2.2178 | 0.1933 | -1.2124 |
| 19.1066 | C, W | 0.0649 | 7.9964 | 0.1751 | -0.2249 | 0.5557 | -29.1750 | 5.1493 | 0.2297 | -1.1965 |
| 19.1066 | C, Se | 0.0696 | 7.8868 | 0.4175 | -0.4232 | 0.3247 | 10.7430 | 0.0455 | 0.3956 | -1.2639 |
| 30.0000 | C, W, Se | 0.0093 | 8.3185 | -0.0165 | 0.0128 | 0.6197 | 0 | 1.1670 | 0.3337 | -1.0721 |
| 30.0000 | C, W | -0.0053 | 8.3184 | -0.0113 | 0.0093 | 0.4719 | 0 | -1.4438 | 0.3676 | -1.0588 |
| 30.0000 | C, Se | -0.0029 | 8.3187 | -0.0042 | 0.0017 | 0.1496 | 0 | 0.7681 | 0.5414 | -1.1176 |

TABLE S10. Fit parameters of the model Hamiltonian for selected graphene/MoS$_2$ heterostructures, where we artificially turned off SOC on Mo and S.

| $\vartheta$ [°] | SOC on | $\Delta$ [meV] | $v_\text{F}/10^5[\frac{\text{m}}{\text{s}}]$ | $\lambda_\text{I}^\text{A}$ [meV] | $\lambda_\text{I}^\text{B}$ [meV] | $\lambda_\text{R}$ [meV] | $\varphi$ [°] | $E_\text{D}$ [meV] | $E_D - E_V$ [eV] | $E_D - E_C$ [eV] |
|---|---|---|---|---|---|---|---|---|---|---|
| 6.5868 | C, Mo, S | 0.4420 | 8.0126 | 0.2445 | -0.2647 | 0.0854 | 21.1421 | 0.2847 | 1.1862 | -0.5295 |
| 6.5868 | C, Mo | 0.4545 | 8.0373 | 0.2537 | -0.2762 | 0.0589 | 33.2029 | -2.5371 | 1.1924 | -0.5286 |
| 6.5868 | C, S | 0.4523 | 8.0361 | -0.0148 | 0.0153 | 0.0292 | -3.6326 | -2.1989 | 1.2503 | -0.5314 |
| 14.4649 | C, Mo, S | 0.3765 | 8.1134 | 0.3053 | -0.3565 | 0.1245 | 15.0688 | 1.1699 | 1.3133 | -0.4039 |
| 14.4649 | C, Mo | 0.3857 | 8.1285 | 0.2882 | -0.3415 | 0.0987 | 18.2143 | -2.0370 | 1.3184 | -0.4044 |
| 14.4649 | C, S | 0.3874 | 8.1770 | 0.0132 | -0.0132 | 0.0300 | 6.2109 | -1.5367 | 1.3781 | -0.4054 |
| 27.6385 | C, Mo, S | -0.0002 | 8.1439 | -0.0410 | 0.0373 | 0.0843 | 32.8878 | 0.1104 | 1.3239 | -0.3950 |
| 27.6385 | C, Mo | -0.0024 | 8.1707 | -0.0220 | 0.0183 | 0.0643 | 28.9229 | -2.6086 | 1.3305 | -0.3939 |
| 27.6385 | C, S | -0.0102 | 8.2031 | -0.0191 | 0.0161 | 0.0214 | 44.2954 | -1.0879 | 1.3910 | -0.3943 |

TABLE S11. Fit parameters of the model Hamiltonian for selected graphene/WS$_2$ heterostructures, where we artificially turned off SOC on W and S.

| $\vartheta$ [°] | SOC on | $\Delta$ [meV] | $v_\text{F}/10^5[\frac{\text{m}}{\text{s}}]$ | $\lambda_\text{I}^\text{A}$ [meV] | $\lambda_\text{I}^\text{B}$ [meV] | $\lambda_\text{R}$ [meV] | $\varphi$ [°] | $E_\text{D}$ [meV] | $E_D - E_V$ [eV] | $E_D - E_C$ [eV] |
|---|---|---|---|---|---|---|---|---|---|---|
| 6.5868 | C, W, S | 0.6485 | 8.0248 | 0.7849 | -0.8638 | 0.2337 | 16.8965 | 1.6459 | 0.8035 | -0.8946 |
| 6.5868 | C, W | 0.6327 | 8.0261 | 0.8424 | -0.8997 | 0.2079 | 21.8240 | 3.3981 | 0.8141 | -0.8903 |
| 6.5868 | C, S | 0.6720 | 8.0120 | -0.0070 | 0.0094 | 0.0428 | -5.3514 | -1.3558 | 1.0089 | -0.9381 |
| 14.4649 | C, W, S | 0.4676 | 8.1248 | 0.5635 | -0.6826 | 0.3678 | -1.3235 | 0.3962 | 0.9249 | -0.7755 |
| 14.4649 | C, W | 0.4759 | 8.1623 | 0.5315 | -0.6532 | 0.3231 | -0.2807 | 1.0492 | 0.9342 | -0.7716 |
| 14.4649 | C, S | 0.4707 | 8.1924 | 0.0209 | -0.0223 | 0.0438 | -1.0199 | 0.8989 | 1.1341 | -0.8146 |
| 27.6385 | C, W, S | 0.0025 | 8.2009 | 0.0059 | -0.0113 | 0.2410 | 18.7310 | 1.8203 | 0.9394 | -0.7623 |
| 27.6385 | C, W | 0.0030 | 8.1990 | 0.0284 | -0.0334 | 0.2179 | 18.3814 | -2.0955 | 0.9440 | -0.7631 |
| 27.6385 | C, S | 0.0027 | 8.2060 | -0.0211 | 0.0184 | 0.0253 | 24.5200 | 0.3798 | 1.1455 | -0.8050 |

## VII. CORRECTING STRAIN-RELATED BAND OFFSET WITH ELECTRIC FIELD?

One particular effect that we would like to address is the fact that the band offset, i. e., the position of the Dirac point within the TMDC band gap, is a linear function of the strain applied to graphene (see main text Fig. 3). In Ref [2] this strain-related band offset was compensated by a transverse electric field in order to extract the zero-strain-like results. How justified is this assumption?

For that purpose, we have considered the 0° twist angle structures of graphene on $MoSe_2$ and $WSe_2$, where no strain was necessary to build the supercells. These structures serve as reference results. In addition, we consider another 0° structure, where graphene is strained by about $-4.8\%$. We compensate the strain-related band offset by a transverse electric field and compare to the reference results. The electric field that is necessary for the correction can be extracted from Table S6 and Table S7 and is about $-2.6$ V/nm.

If we relax the structures, we find that the large amount of strain leads to a sizable rippling of the graphene layer. In fact, the rippling has increased by a factor of 10. This is also the reason, why in Ref. [2] atomic relaxation was neglected and all the twisted structures were kept at fixed $z$. To rule out that the rippling introduces unwanted side effects, we have additionally considered the strained structure and flattened graphene (no rippling) to the average interlayer distance from the fully relaxed strained structure, which is different than the distance obtained for the unstrained sample. Also there, we apply the electric field correction and compare to the reference results. Finally, we consider the flattened graphene samples and change the interlayer distance to the reference cases, as proximity effects are rather sensitive to the interlayer distance.

The comparison of the mentioned cases is summarized in Table S12, Table S13, and Table S14. Comparing the band offsets, the compressive strain pushes the Dirac point closer to the TMDC conduction band edge. The correction of the band offset with the estimated transverse electric field of $-2.6$ V/nm works quite well in the case of the fully relaxed structure. Once we flatten the graphene layer, a larger electric field of about $-3.3$ V/nm is necessary to correct the strain-related band offset.

Most important is how the proximitized Dirac states are affected by the strain and the electric field correction. We find that the large amount of rippling leads to a strongly enhanced sublattice asymmetry, reflected in the parameter $\Delta$, and the opening of a pronounced band gap. Also, the intrinsic SOC parameters are strongly modified by the strain and the rippling. In particular the changes in the $WSe_2$ case are giant and in the range of 0.5–0.8 meV. In contrast, the Rashba SOC, which in the first place originates from the structural $z$-mirror asymmetry and the distortion of graphene $p_z$ orbitals is less affected by the strain. The electric field correction does not help to adjust the results to the reference values.

Once we flatten graphene the sublattice asymmetry nearly vanishes and the Fermi velocity strongly renormalizes, while the intrinsic and Rashba SOC parameters almost match the reference values. With the electric field correction, the Rashba SOC parameters can be brought even closer to the reference value, while intrinsic SOC parameters become less reliable. In particular, the intrinsic SOC values deviate by about 20% (50%) in the case of $WSe_2$ ($MoSe_2$) with flattened graphene and the field correction applied. However, the deviation is also due to the different interlayer distance for the strained and the reference structure.

As a final check, for the flattened case, we tune the interlayer distance to match the one of the reference structure, since the distance strongly affects proximity effects. Without electric field correction, intrinsic SOCs again match the reference values, while the Rashba SOC is underestimated. Similar to before, the field correction makes the intrinsic (Rashba) SOC parameters less (more) reliable.

Based on these findings, we believe that the electric field correction can be partially justified when comparing twisted structures with fixed interlayer distance and no structural relaxation which could lead to rippling, in particular at large strain. However, the presented results for the 0° structure may not be representative for the other twist angles.

TABLE S12. Structural information for the graphene/$MoSe_2$ (graphene/$WSe_2$) heterostructures. We list the same information as in Table S1. Red indicates the structure with more strain. Blue is the same as red, but with flat graphene. Green is the same as blue, but the interlayer distance is the same as for the reference structure.

| $\vartheta$ [°] | (n,m) | (n',m') | NoA | $n_k$ | $a_{grp}$ [Å] | $\varepsilon_{grp}$ [%] | $d_{int}$ [Å] | $\Delta z_{grp}$ [pm] | dipole [debye] |
|---|---|---|---|---|---|---|---|---|---|
| 0.0000 | (0,4) | (0,3) | 59 | 12 | 2.4600 | 0 | 3.4093 (3.3637) | 1.9344 (2.2799) | 0.6054 (0.6156) |
| <span style="color:red">0.0000</span> | (0,7) | (0,5) | 173 | 9 | 2.3429 | -4.7602 | 3.3749 (3.3380) | 19.3940 (19.2038) | 2.0499 (2.0348) |
| <span style="color:blue">0.0000</span> | (0,7) | (0,5) | 173 | 9 | 2.3429 | -4.7602 | 3.3749 (3.3380) | 0 (0) | 1.9444 (1.9558) |
| <span style="color:green">0.0000</span> | (0,7) | (0,5) | 173 | 9 | 2.3429 | -4.7602 | 3.4093 (3.3637) | 0 (0) | 1.8183 (1.8901) |

TABLE S13. The calculated position of the Dirac point with respect to the TMDC valence (conduction) band edge, $E_D - E_V$ ($E_D - E_C$), as defined in Fig. 2(a), for the graphene/MoSe$_2$ (graphene/WSe$_2$) heterostructures. Red indicates the structure with more strain. Blue is the same as red, but with flat graphene. Green is the same as blue, but the interlayer distance is the same as for the reference structure.

| $\vartheta$ [°] | E-field [V/nm] | $E_D - E_V$ [eV] | $E_D - E_C$ [eV] |
| --- | --- | --- | --- |
| 0.0000 | 0 | 0.6093 (0.2347) | -0.8525 (-1.1695) |
| 0.0000 | 0 | 0.9154 (0.5518) | -0.5453 (-0.8502) |
| 0.0000 | -2.57 | 0.6131 (0.2518) | -0.8481 (-1.1507) |
| 0.0000 | 0 | 1.0203 (0.6495) | -0.4407 (-0.7531) |
| 0.0000 | -3.34 | 0.6056 (0.2421) | -0.8562 (-1.1613) |
| 0.0000 | 0 | 1.0434 (0.6653) | -0.4179 (-0.7375) |
| 0.0000 | -3.34 | 0.6170 (0.2482) | -0.8448 (-1.1642) |

TABLE S14. Fit parameters of the model Hamiltonian. We list the same information as in Table I of the main text. Red indicates the structure with more strain. Blue is the same as red, but with flat graphene. Green is the same as blue, but the interlayer distance is the same as for the reference structure.

| TMDC | $\vartheta$ [°] | E-field [V/nm] | $\Delta$ [meV] | $v_\text{F}/10^5 [\frac{\text{m}}{\text{s}}]$ | $\lambda_\text{I}^\text{A}$ [meV] | $\lambda_\text{I}^\text{B}$ [meV] | $\lambda_\text{R}$ [meV] | $\varphi$ [°] | $E_\text{D}$ [meV] |
| --- | --- | --- | --- | --- | --- | --- | --- | --- | --- |
| MoSe$_2$ | 0.0000 | 0 | 0.4917 | 8.2538 | 0.2422 | -0.2258 | 0.2550 | 0 | 1.8970 |
| | 0.0000 | 0 | 2.4419 | 8.2928 | -0.0353 | -0.3081 | 0.1935 | 0 | 0.2123 |
| | 0.0000 | -2.57 | 1.9394 | 8.3885 | 0.0853 | -0.1075 | 0.2088 | 0 | -0.0015 |
| | 0.0000 | 0 | 0.0422 | 8.8716 | 0.2767 | -0.2634 | 0.1886 | 0 | -0.3451 |
| | 0.0000 | -3.34 | 0.0487 | 8.9140 | 0.3347 | -0.3226 | 0.2500 | 0 | -0.2220 |
| | 0.0000 | 0 | 0.0340 | 8.8793 | 0.2424 | -0.2307 | 0.1647 | 0 | 0.4358 |
| | 0.0000 | -3.34 | 0.0399 | 8.9070 | 0.2952 | -0.2847 | 0.2178 | 0 | -0.0377 |
| WSe$_2$ | 0.0000 | 0 | 0.5878 | 8.2500 | 1.1722 | -1.1572 | 0.5303 | 0 | 1.2931 |
| | 0.0000 | 0 | 2.3658 | 8.2914 | 0.3939 | -0.5162 | 0.4881 | 0 | -0.0413 |
| | 0.0000 | -2.57 | 1.6894 | 8.2439 | 0.5374 | -0.0649 | 0.4747 | 0 | -0.5142 |
| | 0.0000 | 0 | 0.0269 | 8.8884 | 1.3056 | -1.2737 | 0.4416 | 0 | 0.8119 |
| | 0.0000 | -3.34 | 0.0326 | 8.9017 | 1.4471 | -1.4127 | 0.5486 | 0 | 1.0723 |
| | 0.0000 | 0 | 0.0239 | 8.8938 | 1.1849 | -1.1559 | 0.3986 | 0 | -0.7790 |
| | 0.0000 | -3.34 | 0.0295 | 8.9121 | 1.3126 | -1.2815 | 0.4950 | 0 | -1.1163 |

## VIII. REAL SPACE TRANSPORT CALCULATIONS

The Rashba-Edelstein effect (REE) and the unconventional REE (UREE) were evaluated using a real-space equivalent of our effective low energy Hamiltonian $\mathcal{H}$ within the Keldysh formalism on a graphene nanoribbon. The expectation values were evaluated using the Recursive Green's function method (RGFM).

### S1. Hamiltonian

The Hamiltonian used for the transport calculations is the real space tight-binding equivalent of $\mathcal{H} = \mathcal{H}_0 + \mathcal{H}_\Delta + \mathcal{H}_I + \mathcal{H}_R$ [3]. The first term

$$\mathcal{H}_0 = -t \sum_{\langle ij \rangle} \sum_\sigma c_{i\sigma}^\dagger c_{j\sigma}$$

is the graphene Hamiltonian, where $\langle \cdots \rangle$ denotes a sum over nearest neighbors and $c_{i\sigma}$ ($c_{i\sigma}^\dagger$) annihilates (creates) an electron in site $i$ with spin $\sigma$. The next term

$$\mathcal{H}_\Delta = \Delta \sum_{i \in A} \sum_\sigma c_{i\sigma}^\dagger c_{i\sigma} - \Delta \sum_{i \in B} \sum_\sigma c_{i\sigma}^\dagger c_{i\sigma}$$

arises from sublattice asymmetry. The first (second) sum is over all sites belonging to sublattice A (B). The third term

$$\mathcal{H}_I = \sum_{S=A,B} \frac{i\lambda_I^S}{3\sqrt{3}} \sum_{\langle\langle i,j \rangle\rangle} \sum_\sigma v_{ij} [s_z]_{\sigma\sigma} c_{i\sigma}^\dagger c_{j\sigma}$$

contains the sublattice-resolved intrinsic spin-orbit coupling, where $\langle\langle \cdots \rangle\rangle$ denotes a sum over next-nearest neighbors and the first sum separates the contributions coming from either sublattice. $v_{ij} = +1$ ($-1$) if the electron takes a left (right) turn along the lattice to get to the next-nearest neighbor. The last term

$$\mathcal{H}_R = \frac{2i\lambda_R}{3} e^{-i\frac{\varphi}{2} s_z} \sum_{\langle ij \rangle} \sum_{\sigma\sigma'} c_{i\sigma}^\dagger c_{j\sigma} (\mathbf{s}_{\sigma\sigma'} \times \mathbf{d}_{ij}) \cdot \hat{\mathbf{z}} \, e^{i\frac{\varphi}{2} s_z}$$

is the Rashba SOC with an additional phase angle $\varphi$. $\mathbf{d}_{ij}$ is the unit vector connecting site $i$ to $j$ and $\mathbf{s}_{\sigma\sigma'} = \left([s_x]_{\sigma\sigma'}, [s_y]_{\sigma\sigma'}, [s_z]_{\sigma\sigma'}\right)$ is a vector of Pauli matrices' components.

### S2. Geometry and symmetries

The honeycomb lattice is set to an armchair nanoribbon geometry, consisting of a central sample (of width $W$ unit cells and length $L$ unit cells) attached to two infinite leads made of the same material with the same width. In these simulations, we used $W = 6$ and $L = 10$. Twisted boundary conditions [4] are imposed to allow for $k$-point sampling along the transverse nanoribbon direction. An additional phase $0 < \phi < 2\pi$ is added to the hoppings crossing the periodic boundary conditions and the final expectation value is averaged over $\phi$. Then, even if $W$ is small, the infinite-width limit can be retrieved with enough sampling over $\phi$. Effectively, what this does is to repeat the system along the transverse direction. Therefore, despite $W$ being a rather small number, sampling over $k$ yields the same result as an infinitely wide lattice.

Transport properties are calculated with respect to operators $\mathcal{A}$ defined in a single slice of the lattice. The nanoribbon can be organized by slices across its cross-section. The Hamiltonian of the nanoribbon is described in terms of the intra-slice Hamiltonian $h$ and the inter-slice Hamiltonian $u$ connecting slice $n$ to $n+1$. Both $u$ and $h$ can be slice-dependent, as long as they are uniform in the leads. A sketch of the sample geometry is shown in Fig. S12.

Due to translation invariance along the longitudinal direction, the operator $\mathcal{A}$ only needs to be nonzero in one of the slices. This simplifies the process of calculating $\langle \mathcal{A} \rangle$ with the RGFM, since only the matrix elements of the Green's functions that connect this slice to the beginning of the leads need to be computed.

Symmetries also play a big role in our numerical results. In the absence of a twist ($\varphi = 0$) in our graphene/TMDC structures, the $\langle s_x \rangle_{\varphi=0}$ response is forbidden and $\langle s_y \rangle_{\varphi=0}$ comes entirely from the Fermi surface. When a twist is introduced, these two mix: $\langle s_y \rangle_\varphi = \langle s_y \rangle_{\varphi=0} \cos(\varphi)$ and $\langle s_x \rangle_\varphi = \langle s_y \rangle_{\varphi=0} \sin(\varphi)$, but they are still determined by $\langle s_y \rangle_{\varphi=0}$ and thus rely only on a Fermi surface calculation, considerably simplifying the numerical procedure. The $\langle s_z \rangle$ response is also forbidden.

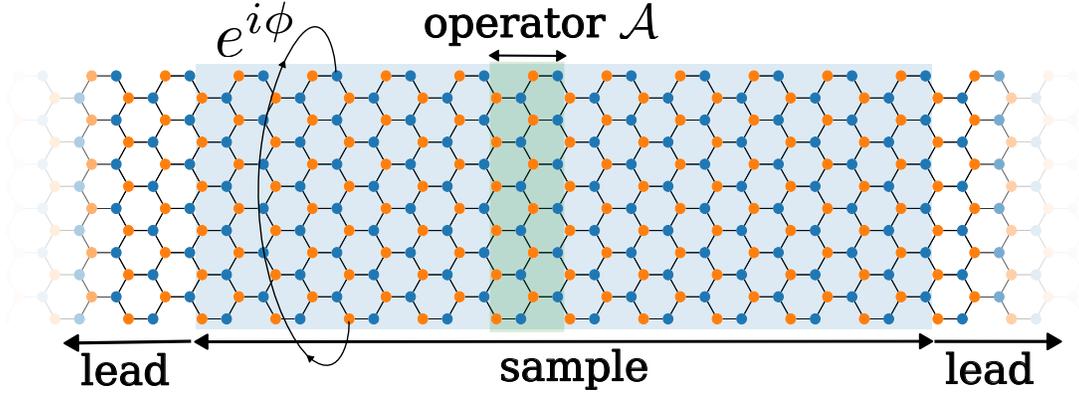

FIG. S12. Lattice geometry used in the transport simulations: graphene nanoribbon sample (blue) with twisted boundary conditions, attached to two identical leads. The operator $\mathcal{A}$ is defined in the green slice.

### S3. Recursive Green's function method

The REE and UREE were calculated using the Recursive Green's function method (RGFM) which is now briefly explained. At $t < 0$, the leads are disconnected from the sample and lie in thermal equilibrium with the corresponding reservoirs at energy $-\Delta V/2$ and $\Delta V/2$. At $t = 0$ the leads are connected to the sample, a transient regime ensues and eventually an equilibrium state is reached, at which point the desired observables are measured.

The expectation value $\langle \mathcal{A} \rangle$ of observable $\mathcal{A}$ is obtained via the Keldysh formalism [5, 6] as the sum of two terms, stemming from the Fermi surface and the Fermi sea:

$$\langle \mathcal{A} \rangle_{\text{surf}} = \frac{i}{2\hbar} \int_{-\infty}^{\infty} d\varepsilon \, (f_R - f_L) \, \text{Tr} \left[ A G^r \left( \Gamma^L - \Gamma^R \right) G^a \right]$$

$$\langle \mathcal{A} \rangle_{\text{sea}} = -\frac{1}{2} \int_{-\infty}^{\infty} d\varepsilon \, (f_R + f_L) \, \text{Tr} \left[ A \left( G^r - G^a \right) \right]$$

where $f_{R(L)}(\varepsilon)$ is the Fermi function of the right (left) lead defined through the Fermi-Dirac distribution $f_{R/L}(\varepsilon) = f(\varepsilon \pm \Delta V/2)$, $G^{r(a)}(\varepsilon) = (\varepsilon - H \pm i0^+)^{-1}$ is the retarded (advanced) Green's function of the whole system and $\Gamma^{L(R)}(\varepsilon) = i \left[ \Sigma^r_{L(R)}(\varepsilon) - \Sigma^{r\dagger}_{L(R)}(\varepsilon) \right]$ is the level-width function defined through the left (right) self-energies of the leads. Finally, the self-energies are defined in terms of the surface Green's function at the left and right leads, respectively: $\Sigma^r_L(\varepsilon) = u g^r_L(\varepsilon) u^\dagger$ and $\Sigma^r_R(\varepsilon) = u^\dagger g^r_R(\varepsilon) u$.

Within this formalism, the full expectation value is

$$\langle \mathcal{A} \rangle = \langle \mathcal{A} \rangle_{\text{surf}} + \langle \mathcal{A} \rangle_{\text{sea}} - \langle \mathcal{A} \rangle_0$$

where $\langle \mathcal{A} \rangle_0$ is the expectation value of $\langle \mathcal{A} \rangle$ at zero bias. The Fermi sea term can be calculated efficiently via Ozaki countour integration [7], while the Fermi surface term simplifies at low bias because $f_R - f_L$ is only nonzero in a very narrow window of energy.

The recursive Green's function method for nanoribbons requires matrix inversions for each slice, so its numerical complexity scales as $W^3 L$, making it very difficult to deal with wide nanoribbons [8–10]. In the case of simple lattices, this complexity can be reduced to $W^2 L$ at the expense of a high memory cost [11, 12], and more recently a real-space method based on the Kernel Polynomial Method (KPM) [13] has been proposed for general systems to reduce this complexity to $WL$ at the expense of introducing stochasticity to the formalism and finite leads [14, 15]. An alternative way of effectively increasing $W$ is through the use of twisted boundary conditions, or $k$-point sampling, in the transverse direction. When the system has translation invariance along the transverse direction, as is the case

here, this approach exactly reproduces the infinite-width case.

---

\clearpage

\end{document}